\documentclass[12pt]{article}
\usepackage{amsfonts,amsthm,amsmath,amssymb,upgreek}
\usepackage{hyperref}
\usepackage[paper=letterpaper,margin=1in]{geometry}
\usepackage{graphicx}
\usepackage{units}
\usepackage{color}
\usepackage[export]{adjustbox}
\usepackage[T1]{fontenc}

\def\Tr{{\rm Tr }}
\def\hat{\widehat}
\newcommand{\be}{\begin{equation}}
\newcommand{\ee}{\end{equation}}

\def\res{\mathop{\text{Res}}}
\def\Ai{\text{Ai}}

\def\s{\text{Sch}}
\def\tilde{\widetilde}

\begin{document}
\thispagestyle{empty}

\vspace*{.5cm}
\begin{center}

{\bf {\LARGE JT gravity as a matrix integral}}\\

\begin{center}

 {\bf Phil Saad,$^{a}$ Stephen H. Shenker,$^{a}$ and Douglas Stanford$^b$}\\
  \bigskip \rm
  
\bigskip
 $^a$ Stanford Institute for Theoretical Physics,\\Stanford University, Stanford, CA 94305\\ \vspace{5pt}
   $^b$Institute for Advanced Study, Princeton, NJ 08540

\rm
  \end{center}

\vspace{2.5cm}
{\bf Abstract}
\end{center}
\begin{quotation}
\noindent

We present exact results for partition functions of  Jackiw-Teitelboim (JT) gravity on two-dimensional surfaces of arbitrary genus with an arbitrary number  of boundaries. The boundaries are of the type relevant in the NAdS${}_2$/NCFT${}_1$ correspondence. We show that the partition functions correspond to the genus expansion of a certain matrix integral. A key fact is that Mirzakhani's recursion relation for Weil-Petersson volumes maps directly onto the Eynard-Orantin ``topological recursion'' formulation of the loop equations for this matrix integral.

The matrix integral provides a (non-unique) nonperturbative completion of the genus expansion, sensitive to the underlying discreteness of the matrix eigenvalues. In matrix  integral descriptions of noncritical strings, such effects are due to an infinite number of disconnected worldsheets connected to D-branes.    In JT gravity, these effects can be reproduced by a sum over an infinite number of disconnected geometries -- a type of D-brane logic applied to spacetime.

\end{quotation}

\setcounter{page}{0}
\setcounter{tocdepth}{2}
\setcounter{footnote}{0}
\newpage

\tableofcontents

\pagebreak

\section{Introduction}
Studies of the  Sachdev-Ye-Kitaev (SYK) model \cite{Sachdev:1992fk,KitaevTalks,Kitaev:2017awl}  have led to a resurgence of interest in Jackiw-Teitelboim (JT) gravity \cite{Jackiw:1984je,Teitelboim:1983ux,Almheiri:2014cka}. This is because the low energy dynamics of the SYK model is described by the  1D Schwarzian theory \cite{Kitaev:2017awl,Maldacena:2016hyu} which in turn is the boundary description of bulk  2D JT gravity \cite{Jensen:2016pah,Maldacena:2016upp,Engelsoy:2016xyb}. In particular, the low energy limit of the thermal partition function of the SYK model $\langle Z(\beta) \rangle  = \langle \Tr e^{-\beta H_{\text{SYK}}} \rangle $ is approximated by the  Schwarzian theory on the circle, which is dual to JT gravity on the Euclidean disk.  

Several studies have pointed to the importance in JT gravity of surfaces of other topologies \cite{Maldacena:2018lmt,Harlow:2018tqv,Saad:2018bqo}. In particular,  recent work on random matrix statistics in SYK \cite{Saad:2018bqo} showed that the ``ramp'' region of the spectral form factor $\langle Z(\beta +iT)Z(\beta -iT) \rangle$ is described by the ``double cone''  geometry with the topology of the cylinder.   This, as well as other hints described in \cite{Saad:2018bqo},  led us to study JT gravity on surfaces of arbitrary topology.  

The main result in this paper is the calculation of the  the Euclidean JT gravity partition functions  $\langle Z(\beta_1)...Z(\beta_n)\rangle_{\text{conn.}}$ corresponding to surfaces with $n$ Schwarzian boundaries and arbitrary numbers of handles.  These  can be  computed from a certain double-scaled matrix integral that we will discuss.   The topological expansion of this integral gives the sum over handles in JT gravity.  The matrix integral supplies a (non-unique) nonperturbative completion of this expansion, which among other things gives an explanation of the ``plateau'' in  the spectral form factor.  
 
We now give a brief summary of this paper.

In \hyperref[secMatrix]{{\bf section two}} we review the topological expansion of Hermitian matrix integrals. We start by describing a conventional integral over $L\times L$ Hermitian matrices with potential $V(H)$:
\be\label{matrixint}
\mathcal{Z} = \int dH \, e^{-L \Tr V(H)}.
\ee
We will think of the random matrix $H$ as the Hamiltonian of the boundary theory. A natural observable to consider is the thermal partition function $Z(\beta) = \Tr\, e^{-\beta H}$. In practice, when studying matrix integrals, it is more convenient to work with the ``resolvent'' function $R(E) = \Tr\frac{1}{E-H}$. The two are related by a simple integral transform.

For a rather general $V(H)$, the matrix ensemble correlation functions $\langle Z(\beta_1)...Z(\beta_n)\rangle_{\text{conn.}}$ and $\langle R(E_1)...R(E_n)\rangle_{\text{conn.}}$ are of a very constrained type. In particular, a set of equations known as the ``loop equations'' \cite{migdal1983loop} determine all orders in the $1/L^2$ expansion of these quantities in terms of a function $\rho_0(E)$. This function is simply the leading density of eigenvalues that characterizes the infinite $L$ theory.\footnote{For example, in the Gaussian matrix model, where the potential is chosen to be $V(H) = \frac{H^2}{2}$, this function would be the Wigner semicircle distribution $\rho_0(E) = \frac{1}{2\pi}\sqrt{4-E^2}$. This has been normalized so that the integral is one; the actual density of eigenvalues would be $L \cdot \rho_0(E)$.} So if one knows $\rho_0(E)$, correlation functions are fixed to all orders in $1/L^2$. And, in fact, this is a practical tool: the procedure of implementing the loop equations was streamlined by Eynard \cite{eynard2004all} into simple recursion relation, which is part of the more general theory of ``topological recursion'' \cite{eynard2007invariants}.

The type of matrix ensemble that will be relevant to JT gravity is not quite of the type described in (\ref{matrixint}). It is, formally, a matrix integral in which $\rho_0(E)$ is not normalizable. This type of problem is known as a ``double scaled'' matrix integral \cite{Brezin:1990rb,Douglas:1989ve,Gross:1989vs}, and it can be understood as a limit of a conventional matrix integral. In this limit, $L$ is taken to infinity, the potential is tuned in a certain way, and one focuses attention near the edge of the eigenvalue distribution, where the density of eigenvalues remains finite. We refer to this density as $e^{S_0}$. The loop equations commute with this limit, and they lead to an expansion of correlation functions in powers of $e^{-2S_0}$.

In \hyperref[sec:JTpert]{{\bf section three}}, we will see that the JT gravity correlation functions $\langle Z(\beta_1)...Z(\beta_n)\rangle_{\text{conn.}}$ are consistent with the matrix integral recursion relation described above, to all orders in $e^{-S_0}$. Let's briefly describe how this arises. JT gravity is a theory of a two dimensional metric $g_{\mu\nu}$ and a dilaton field $\phi$, with Euclidean action
\be\label{eq:JTaction}
I = -\underbrace{\frac{S_0}{2\pi}\left[\frac{1}{2}\int_{\mathcal{M}}\sqrt{g}R + \int_{\partial\mathcal{M}}\sqrt{h}K\right]}_\text{topological term $= S_0\,\chi(\mathcal{M})$} -\bigg[\underbrace{\frac{1}{2}\int_{\mathcal{M}}\sqrt{g}\phi(R+2)}_\text{sets $R = -2$} +\underbrace{\int_{\partial\mathcal{M}}\sqrt{h}\phi (K-1)}_\text{gives action for boundary}\bigg].
\ee
To study $\langle Z(\beta_1)...Z(\beta_n)\rangle$ in this theory, we impose boundary conditions that the manifold $\mathcal{M}$ should have $n$ boundaries, with regulated lengths $\beta_1/\epsilon,...,\beta_n/\epsilon$ and with $\phi = \gamma/\epsilon$ at each boundary. We regard $\epsilon$ as a ``holographic renormalization'' parameter that should be taken to zero. With boundary conditions of this type, geometries of various topologies will contribute to the JT path integral. The first term in (\ref{eq:JTaction}) is proportional to the Euler characteristic of $\mathcal{M}$, and it implies that different topologies will be weighted by $(e^{S_0})^\chi$. For example, with a single boundary, the first three topologies are
\be
\includegraphics[width=.7\textwidth,valign = c]{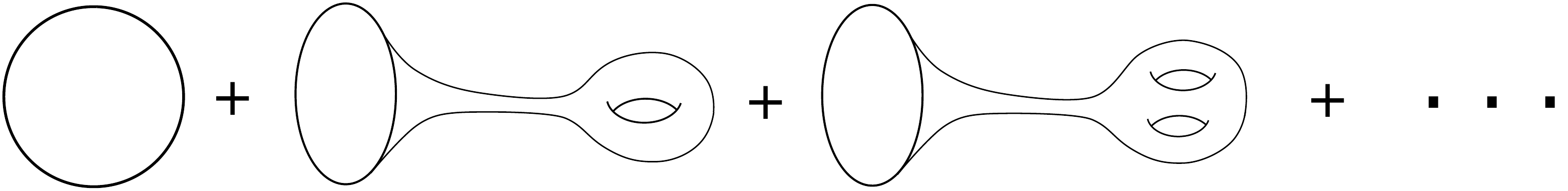}\label{trumpetsExamples}
\ee
A surface with one boundary and $g$ handles, of genus $g$, has Euler characteristic $\chi = 1-2g$, so the weighting is $e^{(1-2g)S_0}$.

The sum over topologies is of the same form as a perturbative string expansion with  ``string coupling''  $g_s = e^{-S_0}$.  Our perspective is different, though.  In string perturbation theory these diagrams represent the perturbative  splitting and joining of closed strings, the amplitude determined by the perturbative coupling $g_s$.    In the JT context these diagrams represent the nonperturbative splitting and joining of closed JT ``baby universes,'' the amplitude controlled by $e^{-S_0}$, nonperturbative in the gravitational coupling $G_N \sim \frac{1}{S_0}$.   Such an expansion in multiple ``baby universes'' is sometimes referred to as a ``third quantized'' description.

Our basic claim is that the JT gravity answer for the geometries (\ref{trumpetsExamples}), and their generalizations with $n$ boundaries, are precisely consistent with the recursion relations of a double-scaled matrix integral. To check this, a first step is to compute the analog of $\rho_0(E)$. For this, one has to compute the leading (genus zero) contribution to $\langle Z(\beta)\rangle$, and then interpret it as a Laplace transform of a density of states $e^{S_0}\rho_0(E)$. A nice feature of JT gravity is that this can be evaluated exactly. One can think about doing the JT path integral in two steps, where we first integrate over the dilaton $\phi$. If we do this integral along a contour parallel to the imaginary axis, the dilaton functions as a Lagrange multiplier setting $R = -2$ everywhere. This implies that $\mathcal{M}$ is a rigid constant-negative-curvature surface. At genus zero with one boundary, this is simply a piece of the hyperbolic disk. Since the boundary conditions only fix the length of the boundary, we have to integrate over the ``cutout shape'' of the geometry within the hyperbolic disk \cite{Maldacena:2016upp}. The action for this integral comes from the final extrinsic curvature term in (\ref{eq:JTaction}). The resulting path integral is the partition function of the Schwarzian theory, which has been computed by several different methods \cite{Cotler:2016fpe,Bagrets:2017pwq,Stanford:2017thb,Mertens:2017mtv,Kitaev:2018wpr,Yang:2018gdb} and it leads to the expression\footnote{Up to a multiplicative ambiguity that can be absorbed into $S_0$. In this paper, unless stated otherwise, we take $\sqrt{x}$ to be positive for real positive $x$ and branched along the negative real $x$ axis.}

\be\label{eq:rho0}
\rho_0(E) = \frac{\gamma}{2\pi^2}\sinh(2\pi\sqrt{2\gamma E}).
\ee
Below, we will often set our units of energy so that $\gamma = \frac{1}{2}$.

A next step is to compute higher genus contributions to correlation functions such as $\langle Z(\beta_1)...Z(\beta_n)\rangle$, and see if they satisfy the matrix integral recursion relation of \cite{eynard2004all}, with the leading density (\ref{eq:rho0}) as input. For example, suppose we are trying to compute $\langle Z(\beta)\rangle$ in JT gravity. Then we we should consider geometries with one asymptotic region and a boundary of length $\beta/\epsilon$. The first three topological classes of orientable geometries\footnote{See \cite{wittenStanford} for treatment of the nonorientable case relevant for other matrix ensembles.} are the ones sketched in (\ref{trumpetsExamples}). The constraint $R = -2$ coming from the integral over $\phi$ simplifies the path integral within each topological class significantly. However, there is still some work to do. For each geometry, we have to do a path integral over the cutout shape of the regulated boundary, with the Schwarzian action induced by the final extrinsic curvature term in (\ref{eq:JTaction}). For all but the first (genus zero) case, we also have to do a finite dimensional integral over some moduli associated to the surface. It is convenient to divide this into two steps.\footnote{A related approach was proposed in \cite{Blommaert:2018iqz}.} First we fix the length of the minimal geodesic that circles the neck of the geometry to be $b$. This geodesic separates a hyperbolic ``trumpet'' from a Riemann surface with boundary:
\be\label{wormholepicIntro}
\includegraphics[width=.3\textwidth,valign =c]{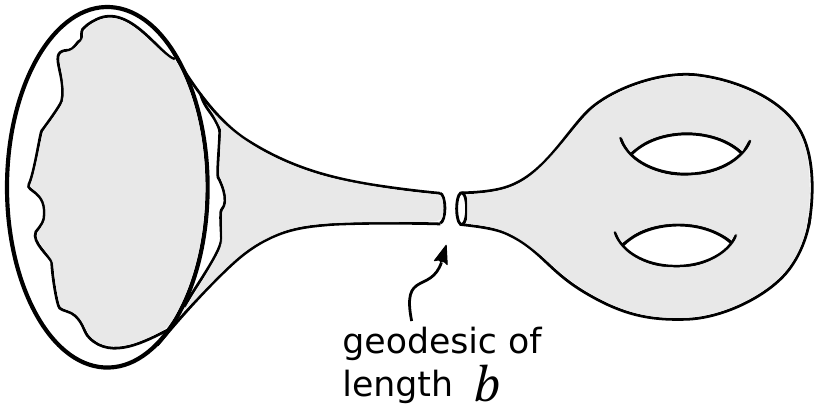}
\ee
Holding $b$ fixed, we integrate over all of the moduli of the Riemann surface, and also over the wiggles at the boundary of the trumpet. Finally, we integrate over $b$. As we will explain in detail in section (\ref{sec:JTpert}), this leads to an expression
\be
\langle Z(\beta)\rangle \simeq e^{S_0}Z_{\text{Sch}}^{\text{disk}}(\beta) + \sum_{g = 1}^\infty  e^{(1-2g)S_0}\int_0^\infty b \,db \ V_{g,1}(b)Z_{\text{Sch}}^{\text{trumpet}}(\beta,b).\label{eq:ZofbetaIntro}
\ee
Here the first term comes from the disk topology, and the others are from genus $g$. The $Z_{\text{Sch}}$ expressions come from doing the path integral over the wiggly boundary of the disk and the trumpet. These are explicit functions
\be
Z_{\text{Sch}}^{\text{disk}}(\beta) = \frac{\gamma^{3/2}e^{\frac{2\pi^2\gamma}{\beta}}}{(2\pi)^{1/2}\beta^{3/2}}, \hspace{20pt} Z_{\text{Sch}}^{\text{trumpet}}(\beta,b) = \frac{\gamma^{1/2}e^{-\frac{\gamma b^2}{2\beta}}}{(2\pi)^{1/2}\beta^{1/2}}.
\ee
The only non-explicit part of this formula is the factor of $V_{g,1}(b)$. This is the Weil-Petersson volume of the moduli space of hyperbolic Riemann surfaces with genus $g$ and one geodesic boundary of length $b$. The ``$\simeq$'' symbol will be commented on below.

If we were studying instead $\langle Z(\beta_1)...Z(\beta_n)\rangle_{\text{conn.}}$, we would have a similar integral to do involving $V_{g,n}(b_1,...,b_n)$, the Weil-Petersson volume of the moduli space of a genus $g$ surface with $n$ geodesic boundaries of lengths $b_1,...,b_n$. These volumes were shown by Mirzakhani \cite{mirzakhani2007simple} to satisfy a recursion relation.\footnote{The first general procedure for calculating Weil-Petersson volumes used topological gravity techniques and the associated matrix integral constructions \cite{Witten:1990hr,kontsevich1992,1999math......2051M}. For a recent discussion see \cite{ Dijkgraaf:2018vnm}.} This was later shown by Eynard and Orantin \cite{eynard2007weil} to be closely related to the ``topological recursion'' that determines the genus expansion of a matrix integral. As we described above, the matrix integral recursion relation depends as input on a leading expression for the density of states $\rho_0(E)$. It turns out that if one plugs in the leading answer computed from the disk contribution to JT gravity (\ref{eq:rho0}), then the two recursion relations agree with one another. To summarize: to all orders in $e^{-S_0}$, and for arbitrary correlators $\langle Z(\beta_1)...Z(\beta_n)\rangle$, the JT gravity answer coincides precisely with the expected answer if we say that $Z(\beta) = \Tr(e^{-\beta H})$ and $H$ is drawn from a double-scaled random matrix ensemble with leading density of states given by (\ref{eq:rho0}).

In \hyperref[sec:minimalString]{{\bf section four}}, we briefly review an older connection between 2D gravity and random matrices \cite{David:1984tx,Ambjorn:1985az,Kazakov:1985ds,Kazakov:1985ea,Brezin:1990rb,Douglas:1989ve,Gross:1989vs}~\footnote{For reviews see \cite{Ginsparg:1993is,DiFrancesco:1993cyw} and for some more recent work see \cite{Seiberg:2004at}.}, the $c<1$ ``minimal string.'' In this body of work, it was argued that Liouville gravity coupled to a $(2,p)$ minimal model (where $p$ is odd) is related to a matrix integral with \cite{Moore:1991ir}
\be
\rho_0(E) = \sinh\left[\frac{p}{2}\text{arccosh}\left(1 + \frac{E}{\kappa}\right)\right] = \sqrt{E}\left(a_0 + a_1 E + a_2 E^2 + ... + a_{\frac{p-1}{2}}E^{\frac{p-1}{2}}\right).
\ee
In the limit $p\rightarrow \infty$, this agrees with (\ref{eq:rho0}) after rescaling $E$. Since both JT gravity and the minimal string are supposed to be dual to matrix integrals, matching $\rho_0(E)$ implies that their partition functions agree on surfaces with arbitrarily many boundaries and handles. This suggests that JT gravity is the large $p$ limit of the $(2,p)$ minimal string. Further work on this is in progress \cite{seibergStanford}.

From the perspective of a matrix integral, the $e^{-S_0}$ expansion is a perturbative expansion. Matrix integrals also contain nonperturbative effects, and in \hyperref[sec:nonperturbative]{{\bf section five}}, we review these and give a tentative JT gravity interpretation. One can motivate the need for these effects from Eq.~(\ref{eq:ZofbetaIntro}). This was written with a ``$\simeq$'' symbol because the sum over genus is an asymptotic series, due to the $(2g)!$ growth of the volume of moduli space with genus $g$ \cite{penner1992,Shenker:1990uf,mirzakhani2011towards}. To make sense of this formula, one would like to find a nonperturbative completion of this series.  This $(2g)!$ behavior is generic in string theory \cite{Shenker:1990uf} and the $e^{-1/g_s}$ nonperturbative effects it points to are due to D-branes \cite{Polchinski:1994fq}.  In the context of matrix integrals, these effects are related to the dynamics of single eigenvalues and there are procedures for computing them. In the Liouville-minimal string context these effects have been connected to two types of branes, ZZ branes and FZZT branes \cite{Fateev:2000ik,Teschner:2000md,Zamolodchikov:2001ah,McGreevy:2003kb,Martinec:2003ka,Klebanov:2003km,Seiberg:2003nm,Kutasov:2004fg,Maldacena:2004sn,Seiberg:2003nm,Seiberg:2004at}.\footnote{There has been an important parallel line of development investigating nonperturbative ``brane'' effects in the topological string.  See, for example, \cite{Dijkgraaf:2002fc,Aganagic:2003qj,Dijkgraaf:2018vnm}.}  In JT gravity we can imitate this connection and give these matrix phenomena a spacetime interpretation. In doing so, we will again encounter two different types of nonperturbative effect. 

The first is called, in the matrix literature, a ``one-eigenvalue instanton'' \cite{Neuberger:1980qh,Ginsparg:1990as,David:1990sk,Shenker:1990uf}.\footnote{For a review see \cite{Marino:2012zq}.} This represents a configuration in the matrix integral where one out of the $L$ eigenvalues has been displaced away from the rest. In JT gravity, this can be interpreted as a type of D-brane effect which in Liouville theory is called a ZZ brane. In JT gravity there is a similar construction where the spacetime plays the role of the worldsheet. The spacetime is allowed to end at a new type of boundary associated to the location of the eigenvalue. By the usual rules of D-branes \cite{Polchinski:1994fq}, we can allow other disconnected spacetimes to also end at the same boundary condition. These disconnected spacetimes exponentiate, adding up to a tiny prefactor that makes the contribution proportional to $e^{-\text{(const.)}e^{S_0}}$. In fact, as we will see, this contribution also comes with a prefactor of $i$, indicating that in the interpretation of JT gravity as a matrix integral, it is one which is expanded about a metastable point. Such a model can be defined nonperturbatively on an appropriate contour, but it doesn't have the reality properties of a conventional matrix integral.

The second type of effect corresponds to adding a ``probe brane,'' and is analogous to the FZZT brane in the Liouville context.  These branes are characterized by a parameter $E$, and they correspond go an insertion of $\det(E-H)$ in the matrix integral. (An insertion of $1/\det(E-H)$ will be referred to as a ``ghost brane.'') In the JT gravity interpretation, this object involves an infinite set of disconnected spacetimes. To see this, one can write 
\begin{align}
\det(E-H) &= \exp(\Tr \log (E-H))  \label{detintro}
\end{align}
Each trace corresponds to a boundary of the spacetime, so after expanding out the exponential, this object involves an infinite number of boundaries. At leading order in the genus expansion, each trace is associated to its own disconnected geometry with the topology of a disk, with a boundary condition that depends on $E$. As in the ZZ brane case, this is formally equivalent to a D-brane calculation \cite{Polchinski:1994fq}.  

These probe branes can be related to more familiar quantities like the resolvent using
\be\label{resolvintro}
\Tr \frac{1}{E-H} =  \partial_E \frac{\det(E-H)}{\det(E'-H)}\bigg|_{E'\rightarrow E}.
\ee
This expresses the resolvent as a ``dipole'' of a brane and ghost brane. It is a trivial equality for matrices, but in the semiclassical approximation to the matrix integral, it is quite surprising. In particular, the infinite number of disconnected spacetimes do not quite cancel out of the RHS. These are interpreted as a nonperturbative contribution to $R(E)$.

One of the motivations for this work was to understand the origin of random matrix statistics in the SYK model \cite{You:2016ldz,Garcia-Garcia:2016mno,Cotler:2016fpe,Saad:2018bqo} from the gravitational perspective.   The eigenvalue pair correlation function $\langle \rho(E) \rho(E')\rangle$ can be extracted from the two resolvent correlator $\langle R(E) R(E') \rangle$.   This in turn can be extracted from a four determinant expectation value similar to \eqref{resolvintro}, and this can be related to a FZZT brane  type expression in JT gravity.   These contain effects that are nonperturbative in the ``string'' coupling, of order $e^{ie^{S_0}}$, which explain the rapidly oscillating part of the sine kernel formula for the pair correlator.   This oscillating behavior, a D-brane effect,  is responsible for the ``plateau'' in the spectral form factor which occurs at exponentially late times, $t \sim e^{S_0}$.\footnote{Essentially the same method was used in the semiclassical quantum chaos literature to derive the plateau from a sum over periodic orbits \cite{berry1990rule,keating1992semiclassical,keating2007resummation,2009NJPh...11j3025M,haake2010quantum}.}

Our analysis of these nonperturbative effects is not rigorous so it is important to make a check.  In principle, Borel resummation of the perturbative series gives another way to determine nonperturbative effects (up to a choice of Borel contour). Nonperturbative effects are associated to singularities in the Borel plane, and the closest singularity to the origin encodes the large-order asymptotics of the perturbative series. We can therefore use the nonperturbative effects described above to make a prediction for the large genus behavior of the perturbative series.\footnote{For a review see \cite{Marino:2012zq}.} In particular, this gives a prediction for the large genus behavior of the volumes of moduli space. For $V_{g,0}$ and for $V_{g,1}(b)$ with $b\ll g$, this prediction agrees with the Zograf conjecture \cite{zograf2008large}.\footnote{Parts of this conjecture have been established rigorously \cite{1999math......2051M,mirzakhani2013growth,mirzakhani2011towards}.} For $V_{g,1}(b)$ at general $b/g$, this method provides a new conjecture, which appears consistent with extrapolation to $g = \infty$ of results up to $g = 20$ kindly provided to us by Peter Zograf.

In \hyperref[sec:Discussion]{{\bf section six}} we discuss some open questions.

\section{Review of the genus expansion in matrix integrals}\label{secMatrix}
For the purposes of this paper, a matrix integral means an integral over Hermitian $L\times L$ matrices $H$ with a weighting that is determined by a potential function $V(H)$:
\be
\mathcal{Z} = \int dH \, e^{-L \Tr\, V(H)}, \hspace{20pt} H  = L\times L \text{ Hermitian matrix}.
\ee
We use the curly ``$\mathcal{Z}$'' to refer to the matrix integral partition function, not to be confused with the partition function $Z(\beta) = \Tr\,e^{-\beta H}$, which is an observable in this matrix ensemble. Expectation values of such observables are given by e.g.
\be\label{eq:expectationvalue}
\langle Z(\beta_1)...Z(\beta_n)\rangle = \frac{1}{\mathcal{Z}}\int dH\, e^{-L\Tr\,V(H)}Z(\beta_1)...Z(\beta_n).
\ee
Matrix ensembles of this type have been widely studied. They are solvable in the infinite $L$ limit, in the $1/L$ expansion, and in some cases beyond.  For reviews, see \cite{DiFrancesco:1993cyw,Marino:2004eq,eynard2015random}. In this section, we will review the $1/L$ expansion, so that we can recognize the same structure when it appears in the context of JT gravity.

In the matrix integral literature, a quantity that is commonly studied is the matrix resolvent
\be\label{eq:defres}
R(E) = \Tr\,\frac{1}{E -H} = \sum_{j = 1}^L \frac{1}{E - \lambda_j}.
\ee 
Here $E$ is an arbitrary complex number. For a fixed matrix $H$, this function is a sum of poles in $E$ corresponding to the eigenvalues $\lambda_j$. But after taking the expectation value over matrices, the poles are smeared into a branch cut. To specify the resolvent one has to indicate which side of the cut one is studying; the discontinuity across the real axis given by
\be\label{eq:discres}
R(E+i\epsilon) - R(E-i\epsilon) = -2\pi i\rho(E)
\ee
where $\rho(E)$ is the density of eigenvalues
\be
\rho(E) = \sum_{j = 1}^L \delta(E-\lambda_j).
\ee
For a fixed matrix, $\rho(E)$ is a discrete sum of delta functions, but its expectation value over an ensemble of matrices is a smooth function. From either $\rho(E)$ or $R(E)$, one can compute $Z(\beta) = \Tr\, e^{-\beta H}$ by an integral transform, and vice versa.

Correlation functions of resolvents have a $1/L$ expansion of the form
\be\label{bwu}
\langle R(E_1)...R(E_n)\rangle_{\text{conn.}} \simeq  \sum_{g = 0}^\infty \frac{R_{g,n}(E_1,...,E_n)}{L^{2g+n-2}}.
\ee
Of course, one also has a similar expansion for correlation functions of partition functions $Z(\beta)$ or eigenvalue densities $\rho(E)$. The existence of such an expansion follows from the analysis of matrix integral perturbation theory \cite{Brezin:1977sv} in terms of `t Hooft double line diagrams \cite{tHooft:1973alw}. The parameter $g$ is called the genus, because it is the genus of the double-line diagrams that contribute to a given term. The $\simeq$ is because the series is asymptotic, and the ``conn.'' subscript means that we take the connected part, or cumulant.

In the next three subsections, we will illustrate several techniques for studying matrix integrals by showing how to compute $R_{0,1},R_{0,2},R_{1,1}$.

\subsection{Computing \texorpdfstring{$R_{0,1}$}{R01}}

At infinite $L$, matrix integrals of the type described above simplify dramatically. The eigenvalues of the matrix become very finely spaced, and from the perspective of coarse-grained observables like correlation functions of resolvents, they can be approximated as a smooth density. In the large $L$ limit, this smooth density is non-fluctuating, or ``self-averaging'' which means that it is the same for all typical matrices drawn from the ensemble.

We will refer to the unit-normalized density of eigenvalues in the large $L$ limit as
\be\label{unitnorm}
\rho_0(E) = \lim_{L\to\infty}\frac{1}{L}\langle \rho(E)\rangle.
\ee
We will focus on the simplest type of problem (the so-called ``one-cut'' matrix models) where the density is supported in a single interval on the energy axis between $a_\pm$. As an example, $\rho_0(E)$ might look like the following:
\be\label{fig:rho0example}
\includegraphics[width=.33\textwidth,valign =c]{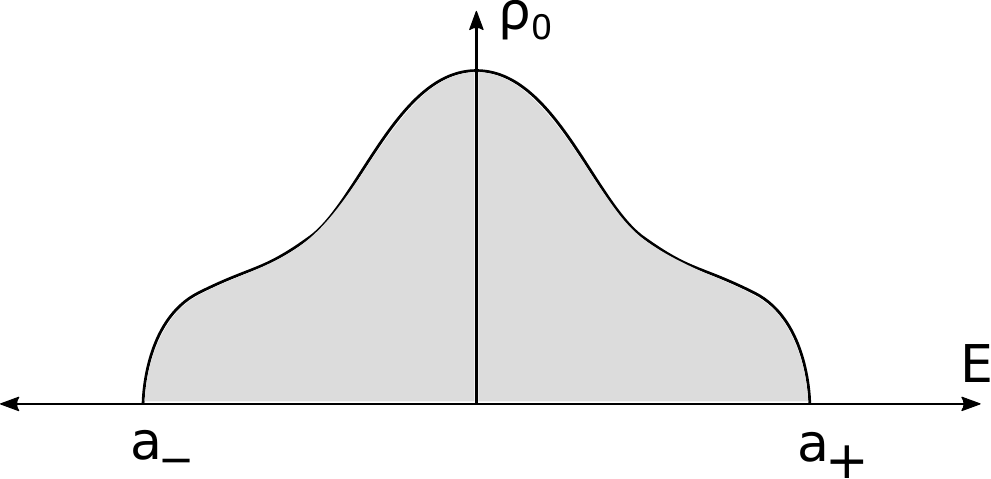}
\ee
Note that in (\ref{unitnorm}), $\rho_0$ is normalized so that its integral is one. It is related to the leading term $R_{0,1}$ in the $1/L$ expansion of the expectation value of the resolvent (\ref{bwu}), by
\be
R_{0,1}(E) = \int_{a_-}^{a_+} d\lambda \frac{\rho_0(\lambda)}{E-\lambda}
\ee
and an inverse expression
\be\label{bxn}
\rho_0(E) = -\frac{1}{2\pi i}\left(R_{0,1}(E+i\epsilon) - R_{0,1}(E-i\epsilon)\right).
\ee

There are several ways to determine $\rho_0$ or $R_{0,1}$, starting from the matrix potential $V(H)$. One way begins by writing the matrix integral in terms of an integral over the matrix eigenvalues,
\be
\mathcal{Z} = C_L \int d^{L}\lambda \prod_{i<j}(\lambda_i-\lambda_j)^2 e^{-L \sum_{j = 1}^L V(\lambda_j)}.
\ee
Here the numerical factor $C_L$ and the Vandermonde determinant $\prod_{i<j}(\lambda_i-\lambda_j)^2$ arise from integrating out the unitary matrix that diagonalizes $H$. The leading density $\rho_0(E)$ can be derived from a mean-field analysis (valid at large $L$) of the statistical mechanics problem for these eigenvalues. The idea is as follows: the effective potential that a single eigenvalue feels is a combination of the bare potential and a term coming from the Vandermonde factor:
\be\label{effpot}
V_{\text{eff}}(\lambda_j) = L V(\lambda_j) - \sum_{i\neq j}\log\left[(\lambda_i-\lambda_j)^2\right].
\ee
After making a continuum approximation to the contribution from all of the other eigenvalues
\be
\sum_{i\neq j}\log\left[(\lambda_i-\lambda_j)^2\right] = L\int d\lambda \rho_0(\lambda)\log\left[(\lambda-\lambda_j)^2\right]
\ee
the equilibrium condition is that each eigenvalue should sit at a stationary point of the effective potential \cite{Brezin:1977sv}:
\be
V'_{\text{eff}}(E) = 0 \hspace{20pt} \implies\hspace{20pt} V'(E) = 2\int d\lambda \frac{\rho_0(\lambda)}{E-\lambda}.\label{eq:equilibrium}
\ee
Here the integral should be defined by a principal value prescription, analogous to leaving out the $i = j$ term from the sum (\ref{effpot}). Taking this prescription into account (\ref{eq:equilibrium}) can be written as
\be\label{Avg}
R_{0,1}(E+i\epsilon) + R_{0,1}(E-i\epsilon) = V'(E).
\ee

To use (\ref{Avg}) to get an expression for $R_{0,1}(E)$, one can use the following trick \cite{migdal1983loop}. Let's imagine for a moment that we knew the end points $a_\pm$, and let's define a function
\be
\sigma(x) = (x-a_+)(x-a_-).
\ee
We will define $\sqrt{\sigma(x)}$ as a function with a cut between $x = a_\pm$, so that in particular for large $x$, $\sqrt{\sigma(x)}\sim x$. Then, for complex $E$ away from the cut,
\begin{align}
R_{0,1}(E) &= \oint_{E} \frac{d\lambda}{2\pi i}\frac{R_{0,1}(\lambda)}{\lambda-E}\sqrt{\frac{\sigma(E)}{\sigma(\lambda)}}=-\int_{\mathcal{C}} \frac{d\lambda}{2\pi i}\frac{R_{0,1}(\lambda)}{\lambda-E}\sqrt{\frac{\sigma(E)}{\sigma(\lambda)}}\notag\\
&=-\frac{1}{2}\int_{\mathcal{C}} \frac{d\lambda}{2\pi i}\frac{V'(\lambda)}{\lambda-E}\sqrt{\frac{\sigma(E)}{\sigma(\lambda)}}.\label{eq:leadingrho}
\end{align}
Here $\mathcal{C}$ is a contour surrounding the interval $[a_-,a_+]$.
In the first equality we used the residue theorem. In the second we deformed the contour past infinity\footnote{It is clear from the definition that $R_{0,1}(\lambda) \approx \frac{1}{\lambda}$ at infinity so this is OK.} to surround the cut on the real axis. In the third equality we used that $\sqrt{\sigma(\lambda)}$ takes the opposite sign on the two sides of the cut. This sign change is canceled by the sign of $d\lambda$ on the two sides, so the net effect is that $R_{0,1}(\lambda)$ can be replaced by its average value on the two sides (\ref{Avg}).

As a final step, we need to determine the endpoints $a_\pm$. To do this, one can impose that $R_{0,1}(E) \sim \frac{1}{E}$ for large $E$. By expanding (\ref{eq:leadingrho}) for large $E$, one finds that this implies
\be\label{endpoints}
0 = \int_{\mathcal{C}}\frac{d\lambda}{2\pi i} \frac{V'(\lambda)}{\sqrt{\sigma(\lambda)}}, \hspace{20pt} 2 = \int_{\mathcal{C}} \frac{d\lambda}{2\pi i}\frac{\lambda V'(\lambda)}{\sqrt{\sigma(\lambda)}},
\ee
which gives two conditions for the two endpoints $a_\pm$. After solving for the endpoints using these equations, (\ref{eq:leadingrho}) becomes an explicit equation for the resolvent $R_{0,1}(E)$, and we can take its discontinuity (\ref{bxn}) to get the density of eigenvalues $\rho_0(E)$.

\subsection{Computing \texorpdfstring{$R_{0,2}$}{R02}}
A useful trick for computing $R_{0,2}(E_1,E_2)$ is the following \cite{ambjorn1990multiloop,ambjorn1995matrix}. If we write the potential as a power series
\be
V(M) = \sum_{n = 0}^\infty v_n M^n
\ee
then it is straightforward to check that an appropriate sum of derivatives of $R_{0,1}$ gives $R_{0,2}$:
\be
R_{0,2}(E_1,E_2) = -\sum_{n = 0}^\infty \frac{1}{E_2^{n+1}}\partial_{v_n}R_{0,1}(E_1).
\ee
Since we have a formula (\ref{eq:leadingrho}) for $R_{0,1}$ in terms of the potential, we can differentiate it with respect to the parameters $v_n$ and get $R_{0,2}$. In taking the derivative of (\ref{eq:leadingrho}), the term where $\partial_{v_n}$ hits the potential can be simplified using
\be
\sum_{n = 0}^\infty \frac{1}{E^{n+1}}\partial_{v_n}V'(\lambda) = \sum_{n = 0}^\infty \frac{n \lambda^{n-1}}{E^{n+1}} = \frac{1}{(\lambda - E)^2}.
\ee
One also needs to know the derivative of the endpoints, $\partial_{v_n}a_\pm$. These can be determined by taking derivatives of (\ref{endpoints}). For example, in the symmetric case with $a_\pm = \pm a$, one finds after some algebra that
\begin{align}
R_{0,2}(E_1,E_2) 
=\frac{1}{2(E_1-E_2)^2}\left(\frac{E_1E_2 - a^2}{\sqrt{\sigma(E_1)}\sqrt{\sigma(E_2)}} - 1\right).\label{ryt}
\end{align}
Somewhat miraculously, this depends on the potential only through the endpoints $\pm a$. This is a very general feature of one-cut matrix integrals \cite{ambjorn1990multiloop,brezin1993universality}.

\subsection{Computing \texorpdfstring{$R_{1,1}$}{R11}}
In order to compute $R_{1,1}$, we will introduce the machinery of the ``loop equations,'' which make it possible to systematically compute all of the $R_{g,n}$. The starting point is the equation
\be
0=\int d^L\lambda \frac{\partial}{\partial\lambda_a}\left[\frac{1}{E-\lambda_a}R(E_1)...R(E_k)\prod_{i<j}(\lambda_i-\lambda_j)^2 e^{-L \sum_{j}V(\lambda_j)}\right].
\ee

To get $R_{1,1}$, one only needs the first of these equations, with $k = 0$, so no extra resolvents are inserted in the integrand. Evaluating the derivative explicitly in this case, one gets
\be
0 = \left\langle \frac{1}{(E-\lambda_a)^2} + \frac{1}{E-\lambda_a}\sum_{j\neq a}\frac{2}{\lambda_a-\lambda_j} - \frac{L V'(\lambda_a)}{E-\lambda_a}\right\rangle.
\ee
After symmetrizing the second term under $\lambda_a\leftrightarrow \lambda_j$, combining it with the first term, and then summing over $a$, this equation can be rewritten as
\begin{align}\label{eq:imposeatleading}
\left\langle \left(\Tr\frac{1}{E-H}\right)^2 - L\Tr\frac{V'(H)}{E-H}\right\rangle=0.
\end{align}
This is an exact equation, but it can be expanded in powers of $1/L$. At leading order in the expansion, we find an equation that can be written (after dividing by $L^2$ and adding terms involving $V'(E)$ to both sides of the equation)
\be\label{single}
\left(R_{0,1}(E)-\frac{V'(E)}{2}\right)^2 = \frac{V'(E)^2}{4} - \left\langle \frac{1}{L}\Tr\frac{V'(E)-V'(H)}{E-H}\right\rangle_0
\ee
where the zero subscript means the large $L$ limit. If the potential $V(E)$ is an analytic function, then the RHS of this equation is also analytic in $E$ (in fact, if the potential is a polynomial then the RHS is also a polynomial). So if we define the quantity
\be\label{spec}
y = R_{0,1}(E) - \frac{V'(E)}{2},
\ee
then (\ref{single}) defines a hyperelliptic curve $y^2 = f(E)$, where $f$ is the RHS of (\ref{single}). This curve is referred to as the spectral curve of the matrix integral. Concretely, it is a double cover of the complex $E$ plane, with the two sheets differing by $y \rightarrow -y$. We define it so that (\ref{spec}) holds on the physical sheet, away from the cut along the positive real axis. Eq.~(\ref{bxn}) then implies that if we continue to the positive real axis through the upper half plane for $E$, we have
\be
y = -\pi i\rho_0(E)\label{spec2}.
\ee
If we continue through the lower half plane, we would have the opposite sign.

Correlation functions of resolvents have branch points in the energy plane, but they become single-valued functions on the spectral curve. For example, the same loop equation (\ref{eq:imposeatleading}) can be expanded to higher orders in $1/L^2$. And, at next order, one finds
\be
2y(E) R_{1,1}(E) = -R_{0,2}(E,E)-\left\langle\frac{1}{L}\Tr\frac{V'(E) - V'(H)}{E-H}\right\rangle_1\label{yeu}
\ee
where the subscript on the expectation value in the RHS means the next-to-leading (i.e.~$1/L^2$) term. Note that the RHS is a single-valued function of $E$. This implies that $R_{1,1}$ must be a double-valued function, changing sign in the same way as $y$ on the two sheets of the spectral curve. So $R_{1,1}$ can be defined as a single-valued function on the spectral curve. 

An alternative way to get a single-valued function related to the resolvents is to multiply them by factors of $\sqrt{\sigma(E)}$, which changes sign in a compensating way between the two sheets. So, for example $R_{1,1}(E)\sqrt{\sigma(E)}$ is a single-valued function in the original energy plane. A very important point is that $R_{1,1}(E)\sqrt{\sigma(E)}$ is singular only at the endpoints of the eigenvalue distribution $E = a_\pm$. This follows from the representation 
\be
R_{1,1}(E) = \int_{a_-}^{a_+} d\lambda \frac{\rho_1(\lambda)}{E-\lambda}
\ee
in terms of the $1/L^2$ correction $\rho_1$ to the density of eigenvalues, which is itself analytic away from the endpoints of the cut. It is obvious from this definition that $R_{1,1}$ is singular only when $E$ coincides with an endpoint $a_\pm$. For example, if $E$ approaches the real axis somewhere away from the endpoints, the contour can be deformed smoothly to avoid a singularity. 

The fact that $R_{1,1}(E)\sqrt{\sigma(E)}$ is single-valued and only singular at $a_\pm$ makes it possible to bypass the fact that we don't know the second term on the RHS of (\ref{yeu}). Following \cite{eynard2004all}, we write a dispersion relation
\begin{align}
R_{1,1}(E)\sqrt{\sigma(E)} &= \oint_E \frac{d\lambda}{2\pi i}\frac{R_{1,1}(\lambda)\sqrt{\sigma(\lambda)}}{\lambda-E}= -\sum_{\pm}\oint_{a_\pm} \frac{d\lambda}{2\pi i}\frac{R_{1,1}(\lambda)\sqrt{\sigma(\lambda)}}{\lambda-E}\\
&= \sum_{\pm}\oint_{a_\pm} \frac{d\lambda}{2\pi i}\frac{R_{0,2}(\lambda,\lambda)}{\lambda-E}\frac{\sqrt{\sigma(\lambda)}}{2y(\lambda)}.\label{dis}
\end{align}
The first equality uses the residue theorem, and the second equality uses that $R_{1,1}(\lambda)\sqrt{\sigma(\lambda)}$ has singularities only at $a_\pm$.\footnote{Since $R(E) \sim \frac{L}{E} + \frac{\text{Tr}(H)}{E^2} + ...$ for large $E$, a $\frac{1}{E}$ piece appears only in $R_{0,1}(E)$, and in particular $R_{1,1}$ has no pole at infinity.} In going to the third line, we used (\ref{yeu}) and exploited the fact that the unknown term on the RHS of (\ref{yeu}) is not singular at the points $a_\pm$.

Eq.~(\ref{dis}) relates $R_{1,1}$ to $R_{0,2}$, which was computed in (\ref{ryt}). In \cite{eynard2004all,eynard2015random}, it is shown how a similar strategy works for higher order quantities $R_{g,n}$. One expands the loop equations in powers of $1/L$ and gets a set of equations for $R_{g,n}$ involving some unknown correlators that include a factor of $\Tr\frac{V'(E)-V'(H)}{E-H}$. These unknown terms are not singular at $E = a_\pm$, so the dispersion relation method in (\ref{dis}) gives a set of equations purely involving $R_{g,n}$. These equations can be solved recursively. We will write a special case of the resulting recursion relation below, after introducing the idea of double-scaled matrix integrals.

\subsection{Double scaling}
In this paper, we will be interested in an analog of a matrix integral for which the leading density of eigenvalues is of the form
\be\label{hjk}
\rho_{0}^{\text{total}}(E) = \frac{e^{S_0}}{(2\pi)^2}\sinh(2\pi\sqrt{E}), \hspace{20pt} E>0.
\ee
Here the superscript ``total'' is meant to indicate that unlike (\ref{unitnorm}), we are not dividing by $L$ here, so this density of eigenvalues is normalized so that its integral is $L$. Of course, such an equation doesn't make sense, since the integral of the RHS is not normalized at all. However, (\ref{hjk}) can be realized as a limit of an ordinary matrix integral. For finite $a$, we can imagine choosing a potential and a value of $L$ so that\be\label{yro}
\rho_0^{\text{total}}(E) = \frac{e^{S_0}}{(2\pi)^2}\sinh\left(2\pi\sqrt{\frac{a^2-E^2}{2a}}\right), \hspace{20pt} -a<E<a.
\ee
Shifting $E \rightarrow E-a$ and taking $a$ large, we recover (\ref{hjk}). Note that as $a$ becomes large, the ratio $L/e^{S_0}$ also becomes large, in order to keep $\rho_0^{\text{total}}(E)$ normalized to have integral $L$. So the total number of eigenvalues is going to infinity. However, this doesn't mean that the matrix integral becomes trivial. The reason is that in the region of interest, the density of eigenvalues is finite, controlled by $e^{S_0}$, and the $1/L$ expansion will be replaced by an $e^{-S_0}$ expansion. This procedure is an example of what is called ``double scaling,'' and what we are saying here is that (\ref{hjk}) makes sense as a leading density of eigenvalues for a double-scaled matrix integral. In such an integral, correlation functions of resolvents have an expansion of the form
\be\label{bwu3}
\langle R(E_1)...R(E_n)\rangle_{\text{conn.}} \simeq  \sum_{g = 0}^\infty \frac{R_{g,n}(E_1,...,E_n)}{(e^{S_0})^{2g+n-2}}.
\ee
Note that we are risking confusion by writing the coefficients of the $e^{-S_0}$ expansion of a double-scaled matrix integral with the same notation $R_{g,n}$ as the coefficients of the $1/L$ expansion of a conventional matrix integral. Similarly, whereas for a conventional matrix integral one defines $\rho_0 = \frac{1}{L}\rho_0^{\text{total}}$, for a double-scaled theory we will use 
\be\label{total}
\rho_0(E) = e^{-S_0}\rho_0^{\text{total}}(E).
\ee

The double-scaling procedure needed to achieve (\ref{hjk}) would be quite complicated at the level of the potential, since we have to tune it carefully as we take the limit. And, in fact, the potential itself diverges in the limit.\footnote{The quantity that remains finite is the effective potential on a given eigenvalue (\ref{effpot}), which is the sum of the ``bare'' potential and the Vandermonde potential from an increasingly large number of other eigenvalues.} Fortunately, as we will see in a moment, the $e^{-S_0}$ expansion is determined directly by the leading density of eigenvalues itself, without any explicit reference to the potential.\footnote{The same is also true for the $1/L$ expansion of a conventional matrix integral.}

In practice, double-scaled theories are a little simpler than conventional matrix integrals, since they have only a single endpoint of the eigenvalue distribution, which we take to be at $E = 0$. So for example, a formula that will be important below is the expression for $R_{0,2}$. If we start with (\ref{ryt}) and shift $E_1 \rightarrow E_1 - a$ and $E_2\rightarrow E_2 - a$ and finally take $a$ large, one finds a somewhat simpler expression
\be\label{pairdouble}
R_{0,2}(E_1,E_2) = \frac{1}{4z_1z_2(z_1+z_2)^2}, \hspace{20pt} z_i = \sqrt{-E_i}.
\ee
This is valid for any double-scaled theory. But in some cases one can say much more. For example, it will sometimes be helpful to have in mind the simple double-scaled theory
\be\label{sim}
\rho_0^{\text{total}}(E) = \frac{e^{S_0}}{\pi}\sqrt{E}, \hspace{20pt} E>0.
\ee
This theory is dual to topological gravity \cite{Witten:1990hr,kontsevich1992}. For a recent discussion see \cite{ Dijkgraaf:2018vnm}.

The density of eigenvalues (\ref{sim}) can be obtained by starting with the Gaussian matrix integral, for which $\rho_0$ is a semicircle:
\be
\rho_0^{\text{total}}(E) = \frac{e^{S_0}}{\pi}\sqrt{\frac{a^2-E^2}{2a}}, \hspace{20pt} -a<E<a.
\ee
The condition that $\rho_0$ should have unit normalization implies that $L/e^{S_0} = (a/2)^{3/2}$. To take the double-scaled limit, one shifts $E\rightarrow E-a$ and takes $a$ large, recovering (\ref{sim}). In this limit, many quantities can be computed exactly, such as the exact ensemble average of the density of eigenvalues:
\be\label{exactairy}
\langle \rho(E)\rangle = e^{\frac{2S_0}{3}}\Big[\Ai'(\xi)^2 - \xi \Ai(\xi)^2\Big], \hspace{20pt} \xi = -e^{\frac{2S_0}{3}}E.
\ee
Note that unlike the leading expression (\ref{sim}), this eigenvalue distribution is extended along the entire real axis, although it is nonperturbatively small (in $e^{-S_0}$) for fixed negative $E$.

\subsection{Topological recursion}
We will now present the recursion relation of \cite{eynard2004all}, which efficiently determines all of the higher $R_{g,n}$. Instead of working with functions of the energy, it will be convenient to define a new coordinate $z$ by\footnote{The minus sign here is conventional, and we have chosen to define things so that real $z$ corresponds to the ``forbidden'' region of negative $E$, where the large $L$ density of eigenvalues vanishes.}
\be
z^2 = -E.
\ee
In terms of $E$, the functions $R_{g,n}$ and the spectral curve quantity $y$ are double valued. They can be viewed as single-valued functions living on a double cover of the $E$ plane. For a conventional one-cut matrix integral, this double cover is branched over the endpoints $a_-,a_+$. In the double scaled limit, we move one branch point to the origin and the other to infinity, so we have a double cover of the $E$ plane branched over the origin. The coordinate $z$ defined above is a good coordinate on that space.  The variable $x = -E$ is often used.  Then, more precisely, 
the spectral curve is  given by the locus
\be
(x(z), y(z)) \subset \mathbb{C}^2 ~.
\ee
This curve is uniformized by $z$ and so is genus $0$.\footnote{  It can be also be viewed as a degenerate infinite genus curve with degenerations located at each of the infinite number of zeros of $y(z)$.}

Concretely, $R_{g,n}$ and $y$ are single-valued functions of $z$. For example, for the case (\ref{hjk}), using (\ref{spec2}) and (\ref{total}) one has (up to a minus sign that can be fixed once and for all)
\be
y = \frac{\sin(2\pi z)}{4\pi}.
\ee
In the topological gravity case (\ref{sim}), one has $y = z$. For small $z$ the two differ by a factor of two that is conventional; it could have been absorbed into a shift of $S_0$ by $\log(2)$.

We can now present Eynard's recursion relation for the expansion of the multi-resolvent correlators $R_{g,n}$ defined in (\ref{bwu3}). We will write this for the special case of a double-scaled matrix integral, with one branch point at the origin. In order to line up with the notation in \cite{Eynard:2014zxa}, we define functions $W_{g,n}$ by
\be\label{nwy}
W_{g,n}(z_1,...,z_n) = (-1)^n2^nz_1...z_n\,R_{g,n}(-z_1^2,...,-z_n^2) \hspace{20pt}
\ee
except for two special cases where
\be\label{special}
W_{0,1}(z) = 2z\,y(z), \hspace{20pt} W_{0,2}(z_1,z_2) = \frac{1}{(z_1-z_2)^2}.
\ee
These two special cases are the inputs to the recursion relation of \cite{eynard2004all}, which determines all of the other $W_{g,n}$ by:
\begin{align}\label{topRecur}
W_{g,n}(z_1,\overbrace{z_2,\dots,z_n}^{J})&= \res_{z\to 0}\Bigg\{\frac{1}{(z_1^2-z^2)}\,\frac{1}{4y(z)}\,\,\,\Big[ W_{g-1,n+1}(z,-z,J)\\
& \hspace{30pt}+\sum_{I\cup I'=J; h+h'=g}' W_{h,1+|I|}(z,I)\,W_{h',1+|I'|}(-z,I') \Big]\Bigg\}.\notag
\end{align}
This expression is a residue at the origin (the one remaining branch point) because it is derived using a dispersion relation argument as in (\ref{dis}). The only complication relative to the discussion above for the case $(g,n) = (1,1)$ is that, in general, the loop equations involve another term that appears here on the second line. Regarding the notation, $I,I'$ are subsets of the arguments $z_2,...,z_n$ and $|I|$ is the number of elements in subset $I$. Also, $\sum'$ means that two cases are excluded: $(I = J,h = g)$ and $(I' = J,h' = g)$. This recursion relation is an example of the more general structure referred to as Eynard/Orantin topological recursion \cite{eynard2007invariants}. For more on this in the context of matrix integrals, see \cite{eynard2015random}.

Let's now see how this looks in the first few orders for the case (\ref{sim}) where $y=z$. The first two $W_{0,1}$ and $W_{0,2}$ are given by (\ref{special}), and working out the residue formula explicitly for the first four nontrivial cases gives
\begin{align}
W_{0,1} &= 2z_1^2,\hspace{34pt}
W_{0,2} = \frac{1}{(z_1-z_2)^2},\hspace{20pt}\notag
W_{0,3}= \frac{1}{2z_1^2z_2^2z_3^2}\\
W_{1,1}&= \frac{1}{16z_1^4},\hspace{26pt}
W_{1,2}=\frac{5z_1^4+3z_1^2z_2^2+5z_2^4}{32z_1^6z_2^6}\\
W_{2,1} &=\frac{105}{1024z_1^{10}}.\notag
\end{align}
Using (\ref{nwy}) to convert back to $R_{g,n}$ variables, and using (\ref{eq:discres}) to compute the density of eigenvalues, one finds that $R_{1,1}$ and $R_{2,1}$ agree with the expansion of the exact result (\ref{exactairy}).

One can also work out the answer for the case (\ref{hjk}) where $y = \frac{\sin(2\pi z)}{4\pi}$:
\begin{align}
W_{0,1} &= 2z_1\frac{\sin(2\pi z_1)}{4\pi} ,\hspace{15pt}
W_{0,2} = \frac{1}{(z_1-z_2)^2},\hspace{20pt}
W_{0,3}= \frac{1}{z_1^2z_2^2z_3^2}\\
W_{1,1}&= \frac{3+2\pi^2z_1^2}{24z_1^4},\hspace{30pt}
W_{1,2}=\frac{5(z_1^4+z_2^4) + 3z_1^2z_2^2 + 4\pi^2(z_1^4z_2^2+z_2^4z_1^2) + 2\pi^4z_1^4z_2^4}{8z_1^6z_2^6}\notag\\
W_{2,1} &=\left(\frac{105}{128z_1^{10}} + \frac{203\pi^2}{192z_1^8} + \frac{139\pi^4}{192z_1^6} + \frac{169\pi^6}{480z_1^4} + \frac{29\pi^8}{192z_1^2}\right).\notag
\end{align}
In this case, we don't have an exact formula to expand and compare to. However, a fact of central importance in this paper is the following.  These quantities $W_{g,n}$ are the Laplace transforms of the volumes $V_{g,n}(b_1,...,b_n)$ of the moduli space of bordered Riemann surfaces with geodesic boundaries of lengths $b_1,...,b_n$ and genus $g$. The computation of these volumes was difficult even for $V_{1,1}(b)$ \cite{nakanishi2001areas} until the discovery by Mirzakhani of a recursion relation \cite{mirzakhani2007simple} that enables their computation. The first few are\footnote{The tilde on $V_{1,1}$ is due to the fact that for this case, at every point in the moduli space of this surface, there is a $\mathbb{Z}_2$ symmetry, and the volume we reported here is the volume taking this symmetry into account, in other words counting only one of the pair of surfaces related by the $\mathbb{Z}_2$ symmetry.}
\begin{align}\label{WPvolsTable}
V_{0,1} &= \text{undefined},\hspace{48pt}
V_{0,2} = \text{undefined},\hspace{48pt}
V_{0,3}= 1\\
\widetilde{V}_{1,1}&= \frac{1}{48}(b_1^2 + 4\pi^2),\hspace{33pt}
V_{1,2}=\frac{1}{192}(4\pi^2+b_1^2+b_2^2)(12\pi^2+b_1^2+b_2^2)\notag\\
V_{2,1} &=\frac{1}{2211840}(4\pi^2+b_1^2)(12\pi^2+b_1^2)(6960\pi^4+384\pi^2b_1^2+5b_1^4).\notag
\end{align}
One can verify explicitly that for these examples, we have
\be\label{elt}
W_{g,n}(z_1,...,z_n) = \int_0^\infty b_1 db_1 e^{-b_1z_1}...\int_0^\infty b_ndb_n e^{-b_nz_n} \ V_{g,n}(b_1,...,b_n).
\ee
This was noted and proven in general in work by Eynard and Orantin \cite{eynard2007weil}, who showed that after Laplace transform, Mirzakhani's recursion relation takes the form of (\ref{topRecur}) with the spectral curve $y = \frac{\sin(2\pi z)}{4\pi}$. This relation provides a link between the loop equations of a matrix integral and the volumes of the moduli space of curves. In the next section, we will see that it implies that JT gravity is dual to a matrix ensemble.

\section{The genus expansion in JT gravity}\label{sec:JTpert}

Now that we have reviewed the structure of the genus expansion in matrix integrals, we will see how the same thing arises from an analysis of the path integral in JT gravity. This theory is defined by the Euclidean action
\be\label{eq:JTaction2}
I_{JT} = -\frac{S_0}{2\pi}\left[\frac{1}{2}\int_{\mathcal{M}}\sqrt{g}R + \int_{\partial\mathcal{M}}\sqrt{h}K\right] -\left[ \frac{1}{2}\int_{\mathcal{M}}\sqrt{g}\phi(R+2) +\int_{\partial\mathcal{M}}\sqrt{h}\phi (K-1)\right].
\ee
The basic strategy is to compute something like the correlation functions of resolvents studied in the previous section, and show that they satisfy the recursion relation of a double-scaled matrix integral.

More precisely, we will study correlation functions $\langle Z(\beta_1)...Z(\beta_n)\rangle_{\text{conn.}}$, where in the matrix integral interpretation, $Z(\beta) = \Tr(e^{-\beta H})$ and $H$ is the random matrix. This is related to the resolvent by an integral transform
\be
R(E) = -\int_0^\infty d\beta e^{\beta E}Z(\beta),
\ee
which makes sense for $E$ less than the ground state energy (which we will set to be zero). The expression can be continued in $E$ after doing the integral.

We can translate $\langle Z(\beta_1)...Z(\beta_n)\rangle_{\text{conn.}}$ to a bulk gravity computation using the usual NAdS${}_2$/NCFT${}_1$ dictionary \cite{Maldacena:2016upp}. Here we think about the random matrix $H$ as the Hamiltonian of the boundary theory. The way the translation works is as follows. One integrates over 2d geometries with the rule that for each factor of $Z(\beta)$, the geometry must have a boundary with length $\beta/\epsilon$, and where the dilaton has the value $\phi = \gamma / \epsilon$. (One takes $\epsilon \rightarrow 0$ at the end.) In order to compute the connected part of the correlator, we require that the 2d geometry be connected. In what follows we will often take $\gamma = 1/2$, which amounts to a choice of units.

The upshot of this is that to compute $\langle Z(\beta_1)...Z(\beta_n)\rangle_{\text{conn.}}$, we are looking for connected geometries with $n$ boundaries of a specific kind. For a fixed number of boundaries, these are classified topologically\footnote{In this paper we restrict to orientable surfaces, which are appropriate for a Hermitian matrix integral. We are grateful to Edward Witten for comments on nonorientable contributions. See \cite{wittenStanford} for discussion.} by the number of handles (``genus'') $g = 0,1,...$. The first $S_0$ term\footnote{We should note that the presence of the topological term in the low energy limit of the SYK model has only been verified for the disk and the cylinder.} in the JT action (\ref{eq:JTaction2}) gives a factor of $(e^{S_0})^\chi$, where $\chi$ is the Euler character $\chi = 2-2g-n$. So, summing over different topologies, we get an expression
\be\label{zgn}
\langle Z(\beta_1)...Z(\beta_n)\rangle_{\text{conn.}} \simeq \sum_{g = 0}^\infty \frac{Z_{g,n}(\beta_1,...,\beta_n)}{(e^{S_0})^{2g+n-2}}
\ee
where $Z_{g,n}(\beta_1,...,\beta_n)$ is the JT path integral for a given topology with the $S_0$ term left out of the action. As an example of the type of geometry involved, we have
\be\label{threeTrumpets}
Z_{g=2,n=3}(\beta_1,\beta_2,\beta_3) \hspace{10pt}=\hspace{15pt} \includegraphics[width=.27\textwidth,valign =c]{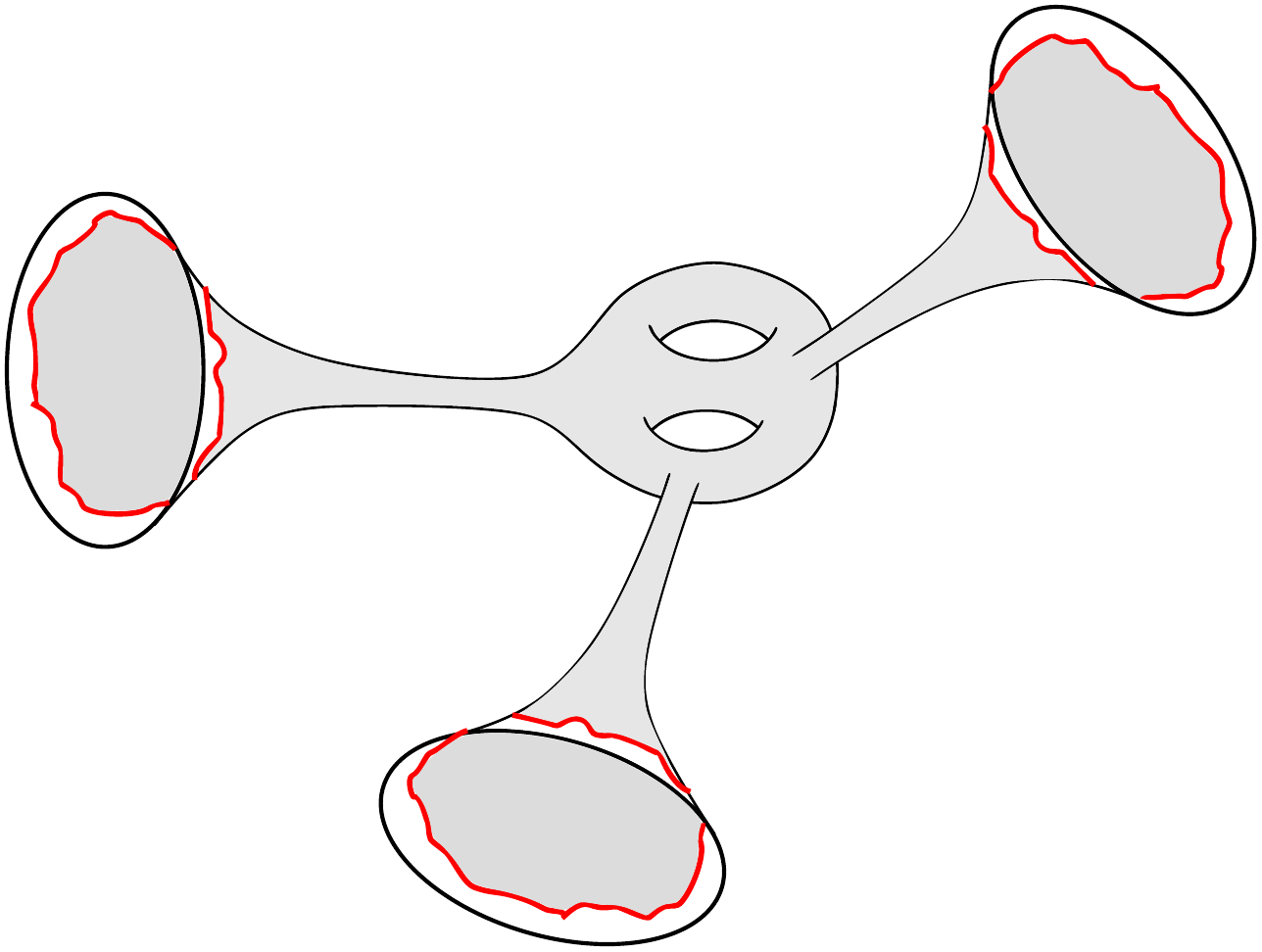}
\ee
where the lengths of the regularized (red) boundaries are fixed to be $\beta_j/\epsilon$, but they are allowed to take wiggly shapes as we discuss in a moment. 

The path integral over such geometries is simplified considerably by the fact that the integral over $\phi$ with the action (\ref{eq:JTaction2}) imposes a constraint that the metric should be constant negative curvature $R = -2$. It is possible to put a metric with $R = -2$ on a topology such as (\ref{threeTrumpets}), and in fact there is an infinite-dimensional space of such geometries, consistent with our boundary conditions. This corresponds to the freedom to make the boundary wiggly, and also a finite-dimensional moduli space associated to the surface itself. The path integral over the metric then becomes an integral over the moduli space, and a path integral over the boundary wiggles. The JT action reduces to the final extrinsic curvature term, and we have
\be\label{nvx}
Z_{g,n}(\beta_1,...,\beta_n) = \int  d(\text{bulk moduli})\int \mathcal{D}(\text{boundary wiggles}) e^{\int_{\partial \mathcal{M}}\sqrt{h}\phi (K-1)}.
\ee
As we will see, the integral over the boundary wiggles is easy to do. The integral over the bulk moduli is difficult to get explicit expressions for, but we will see that Mirzakhani's recursion relation implies that the result satisfies the recursion of a double-scaled matrix integral.

Since our goal is to compute the quantities $Z_{g,n}$ exactly, it will be important to understand the integration measure. As we will see, the measure that follows from the JT path integral is the Weil-Petersson measure on the bulk moduli, and the symplectic measure on the boundary wiggles, with a relation between the normalization of these measures. As a first step we will review the definition of the Weil-Petersson measure for the moduli space, deferring until slightly later an explanation of why it is relevant in JT gravity.

\subsection{The Weil-Petersson symplectic form}
The Weil-Petersson measure is derived from a symplectic form. This form can be explained very concretely in terms of the ``pants'' construction of a hyperbolic Riemann surface. This is based on building up surfaces from an elementary building block, which is a hyperbolic surface with genus zero and three geodesic boundaries, of lengths $b_1,b_2,b_3$. For example, the three shaded pieces here
\be\label{pantsfig}
\includegraphics[width=.45\textwidth,valign =c]{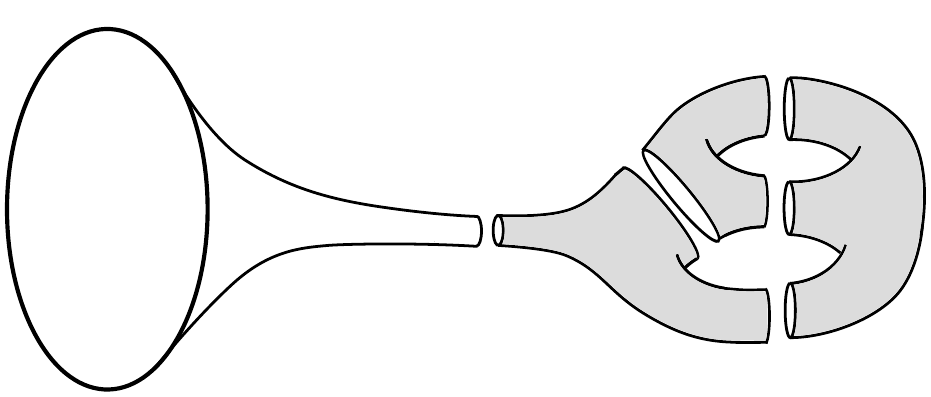}
\ee
can be glued together to give a genus two surface with one geodesic boundary. Such a geometry can be attached to a ``trumpet'' (shown unshaded above) in order to form a geometry with an asymptotic boundary, but for the moment we will focus on the shaded components only. By gluing together several of these three-boundary building blocks, one ends up with a manifold with some genus $g$ and some number $n$ (possibly zero) of leftover geodesic boundaries. In the example, $(g,n) = (2,1)$. The space of such manifolds for fixed values of the external boundary lengths is denoted $\mathcal{M}_{g,n}(b_1,...,b_n)$, and referred to as the moduli space of bordered Riemann surfaces.

A set of coordinates on this space are simply the parameters of the gluing construction, known as Fenchel-Nielsen coordinates. These consist of the lengths $\widetilde{b}_1,...,\widetilde{b}_k$ of the pairs of internal boundaries that were glued together, along with a corresponding set of twist parameters $\tau_1,..,\tau_k$, which represent the proper distance that one boundary is rotated relative to its partner before the two are glued together. So, all together we have $2k$ coordinates, and one can show that $k = 3g+n-3$, where $g$ is the genus and $n$ is the number of fixed leftover boundaries. The Weil-Petersson form is a symplectic form on this space, and in these coordinates it is simply \cite{wolpert1985weil}
\be\label{WPform}
\Omega = \alpha\sum_{i = 1}^{3g+n-3} d\widetilde{b}_i\wedge d\tau_i,
\ee
where $\alpha$ is a numerical constant that depends on convention. In the literature, $\alpha = 1$ is widely used, and we will often use this convention below. The quantity $\frac{1}{k!}\Omega^k$ gives a volume form, and one can integrate this to get the volume of the moduli space $\mathcal{M}_{g,n}(b_1,...,b_n)$.

In practice, a major complication is that the pants decomposition overcounts the moduli space. Because hyperbolic Riemann surfaces have many geodesics on which to cut, one can find different pants decompositions of the same geometrical surface. Although it is not obvious in these coordinates, the form (\ref{WPform}) is invariant under such changes in decomposition. But to correctly compute the volume of the moduli space
\be
V_{g,n}(b_1,...,b_n) = \text{vol}\left(\mathcal{M}_{g,n}(b_1,...,b_n)\right)
\ee
one still needs to restrict the integral to a fundamental region that counts each distinct surface once. This is tricky: see e.g.~\cite{nakanishi2001areas} for an example computation.

In \cite{mirzakhani2007simple}, Mirzakhani showed that these volumes satisfy a recursion relation that makes it possible to compute them in practice (see \cite{Dijkgraaf:2018vnm} for a recent introduction to Mirzakhani's method). We will not show the recursion relation explicitly, but an important fact for this paper is that it is equivalent to the statement that the $W_{g,n}$ defined in (\ref{elt}) satisfy the recursion relation of a double scaled matrix integral, with $y = \frac{\sin(2\pi z)}{4\pi}$ \cite{eynard2007weil}.

\subsection{Optional: the compact case, in the second-order formalism}\label{secondCompact}
We will now begin to explain why the Weil-Petersson measure is relevant for JT gravity. In this section we start with an explanation intended to appeal to readers who have background in string perturbation theory; a more complete explanation will be given later.

To avoid the complication of boundary wiggles for the moment, we will assume that we are integrating over closed manifolds, and in order to have a hyperbolic metric, we assume the genus is $g\ge 2$. The path integral on a closed genus $g$ surface gives a contribution to the logarithm of the matrix integral partition function:
\be
\log\mathcal{Z} \supset (e^{S_0})^\chi\mathcal{F}_g = \int \frac{\mathcal{D}g_{\mu\nu}\mathcal{D}\phi}{\text{Vol}(\text{diff})}\, e^{-I_{JT}[g_{\mu\nu},\phi]}.
\ee
To evaluate the path integral, one can fix to conformal gauge
\be
g_{ab} = e^{2\omega}\hat{g}_{ab}, \hspace{20pt} \sqrt{g}R = \sqrt{\hat{g}}(\hat{R}-2\hat\nabla^2\omega)
\ee
where $\hat{g}$ is a metric with $\hat{R} = -2$. The path integral becomes (see, e.g. eq.~(3.13) of \cite{DHoker:1985een} with $d = 0$ and including the extra JT gravity term)
\be\label{dhoker}
\mathcal{F}_g = \int_{\text{moduli}}d(\text{Weil-Pet.})(\det \hat{P}_1^\dagger \hat{P}_1)^{1/2}\int\mathcal{D}\omega\mathcal{D}\phi \  e^{-26 S_L[\omega]}e^{\int \sqrt{g}\phi(R+2)}.
\ee
The integral over moduli is the integral over metrics $\hat{g}$ with $\hat{R} = -2$. In \cite{DHoker:1985een} it is shown that the path integral reduces to the Weil-Petersson measure on this space, times the Fadeev-Popov determinant for the gauge-fixing to conformal gauge, $(\det \hat{P}_1^\dagger \hat{P}_1)^{1/2}$. The operator $\hat{P}_1$ takes vectors to traceless symmetric tensors
\be
(\hat{P}_1v)_{ab} = \hat{\nabla}_av_b + \hat{\nabla}_bv_a - \hat{g}_{ab}\hat{\nabla}^cv_c,
\ee
and for a conventional choice of metric, its adjoint acts as $(\hat{P}_1^\dagger\alpha)^a = -2g^{ab}\hat\nabla^c \alpha_{bc}$. 
$S_L$ is the Liouville action; it will not play an important role in this theory because the Liouville field will be localized to zero, and the action will give a multiple of the Euler characteristic.

The final term in the action in (\ref{dhoker}) can be written more explicitly as
\be
\int \sqrt{g}\phi(R + 2) = \int \sqrt{\hat{g}} \phi(\hat{R} - 2\hat\nabla^2\omega+2e^{2\omega}) = 2\int \sqrt{\hat{g}}\,\phi(-\hat\nabla^2+2)\omega + O(\omega^2).
\ee
The path integral over $\phi$ imposes a delta function constraint $\delta((-\hat\nabla^2+2)\omega)$ at every point in space. This condition implies $\omega = 0$, but in doing the integral over $\omega$ we get an inverse power of $\det(-\hat\nabla^2+2)$. So the path integral becomes 
\begin{align}\label{ratioOfDets}
\mathcal{F}_g&= e^{(\text{const.})\chi}\int_{\text{moduli}}d(\text{Weil-Pet.)}\frac{(\det \hat P_1^\dagger \hat P_1)^{1/2}}{\det(-\hat\nabla^2+2)}\\
&=e^{(\text{const.})\chi}\int_{\text{moduli}}d(\text{Weil-Pet.)}
\end{align}
In going to the second line, we used that the ratio of determinants in (\ref{ratioOfDets}) for hyperbolic manifolds is proportional to $e^{\# \chi}$. To show this, one can use the constant negative curvature condition and a short calculation to write $\hat P_1^\dagger \hat P_1 = 2(-\hat\nabla_1^2 +1)$, where $\nabla_1^2$ is the Laplacian acting on vectors. One can then write out the vector Laplacian explicitly, using coordinates $ds^2 = \frac{dx^2 + dy^2}{y^2}$. One finds a simple expression after conjugating by a factor of $y$:
\be
y^{-1}\hat\nabla_1^2 y = \left(\begin{array}{cc}D_{-1} - 1  & 0 \\
0 & D_{1} -1\end{array}\right)
\ee
acting on two-component vectors with top component $v^x + iv^y$ and bottom component $v^x-iv^y$. The differential operators $D_n$ are defined, following the conventions in \cite{d1986determinants}, as
\be
D_n \equiv y^2\left(\partial_x^2 + \partial_y^2\right) -2iny \partial_x.
\ee
Note that $D_0$ is simply the scalar Laplacian, so we conclude that
\be
\frac{(\det \tfrac{1}{2}\hat P_1^\dagger \hat P_1)^{1/2}}{\det(-\hat\nabla^2+2)} = \frac{\det^{1/2}(-D_1+2)\det^{1/2}(-D_{-1}+2)}{\det(-D_0+2)} = \frac{\det(-D_1+2)}{\det(-D_0+2)}.
\ee
These determinants differ by at most a factor $(\text{const})^{\chi}$ where $\chi$ is the Euler characteristic, see \cite{d1986determinants,sarnak1987determinants}. Note that we inserted a factor of $\tfrac{1}{2}$ in the first expression here. After regularization, this factor also contributes at most a term proportional to $(\text{const})^{\chi}$.

So we find that, up to a factor $e^{(\text{const.})\chi}$ that can be absorbed into $S_0$, the path integral on a compact $g\ge 2$ surface reduces to the Weil-Petersson volume of the moduli space.\footnote{Note that this volume also depends on an arbitrary normalization of the form in (\ref{WPform}). If we rescale the form by a factor of $\alpha$, we will rescale the answer by a factor of $\alpha^{3g-3} = \alpha^{-\frac{3}{2}\chi}$, so this factor can also be absorbed into $S_0$.} The cancellation of the determinants that made this possible was quite specific to the JT theory. We will see this from another perspective below.

\subsection{First-order formalism}
In order to evaluate (\ref{nvx}), we need the measure both for the bulk moduli and the boundary wiggles. And, moreover, we will need a relation between the normalizations of these measures. In order to derive this, it will be convenient to use the first-order formalism for JT gravity, which is a topological $BF$-type theory.

\subsubsection{Basic setup}
We start with some equations for first order gravity in two dimensions. The basic objects are the one forms $e^a = e^a_\mu dx^\mu$ and the spin connection $\omega^a_{ \ b} = \epsilon^{a}_{ \ b}\omega$ where $\epsilon_{12} = 1$, and $a,b$ indices are raised and lowered with $\delta_{ab}$, so we don't need to distinguish up vs.~down. The no-torsion condition, which determines the spin connection, is
\be\label{notor}
de^a = -\omega^a_{ \ b}\wedge e^b.
\ee
The curvature two form is
\be
R^{a}_{ \ b} = d\omega^{a}_{ \ b} + \omega^{a}_{ \ c}\wedge \omega^c_{ \ b} = d\omega^a_{ \ b}.
\ee
The first equation is the general definition. The second equation is true in two dimensions, since in this case $\omega^a_{ \ c}\wedge \omega^{c}_{ \ b} =-\delta^a_b \omega\wedge\omega = 0$. In general, the curvature two form is related to the Riemann tensor by
\be
R^{a}_{ \ b} = \frac{1}{2}e^a_\mu e^\nu_bR^\mu_{ \ \nu\rho\sigma}dx^\rho\wedge dx^\sigma.
\ee
In two dimensions this simplifies because $R_{\mu\nu\rho\sigma} = \frac{1}{2}R (g_{\mu\rho}g_{\nu\sigma} - g_{\mu\sigma}g_{\nu\rho})$ and we find that
\be
d\omega^a_{ \ b} = \frac{1}{2}Re^a\wedge e_b.
\ee
Now, $e^1\wedge e_2 = e^1\wedge e^2 = \sqrt{g}d^2x$, so
\be
\sqrt{g}d^2xR = 2 d\omega^1_{ \ 2} = 2d\omega.
\ee
We can then write the JT action in the first order formalism as
\begin{align}
\frac{1}{2}\int \sqrt{g}\phi(R+2) &\rightarrow \int \left[\phi (d\omega + e^1\wedge e^2) + \phi_a(de^a + \epsilon^a_{ \ b}\omega\wedge e^b)\right]\\
&=i\int \text{Tr}\left(B F\right).\label{finalBF}
\end{align}
In the first line, we introduced the $\phi^a$ as Lagrange multipliers that enforce the no-torsion condition (\ref{notor}). In the second line we wrote the answer in terms of a matrix of scalars $B$ and a matrix of one forms $A$:
\be
B = -i\left(\begin{array}{cc} -\phi^1 & \phi^2 + \phi \\ \phi^2-\phi & \phi^1 \end{array}\right) \hspace{20pt} A = \frac{1}{2}\left(\begin{array}{cc} -e^1 & e^2 - \omega \\ e^2+\omega & e^1 \end{array}\right)\label{SL2r}
\ee
where the field strength $F$ is defined as
\be
F = DA = dA + A\wedge A.
\ee
The action can now be recognized as an $SL(2,\mathbb{R})$ ``$BF$'' theory. Note that we introduced a factor of $i$ in (\ref{finalBF}), and a compensating factor of $-i$ in (\ref{SL2r}). With this convention, the purely imaginary contour for our Lagrange multipliers $\phi,\phi_1,\phi_2$ becomes a real contour for the matrix $B$. The action (\ref{finalBF}) then has the usual factor of $i$ for a BF theory in Euclidean signature.

The field $B$ enters the action only linearly, so we can integrate it out, getting a constraint that $F = 0$, in other words that we should integrate only over flat connections.\footnote{Actually, in JT gravity we only want to integrate over one topological component of the space of flat $SL(2,\mathbb{R})$ connections (see e.g.~footnote 5 of \cite{Dijkgraaf:2018vnm}). Global aspects of the space of flat connections will not be important for us, because we view the second order Lagrangian as the definition of the theory, and only use the BF description locally, in order to compute the correct measure about a given configuration.} The advantage of the BF theory presentation is that there is a very simple description for the measure on the space of flat connections. This measure arises from the original ultralocal measure for the path integral, after integrating out the nonzero modes. After gauge-fixing and using the Fadeev-Popov procedure, it was shown in \cite{Witten:1991we} for $BF$ theories with compact gauge group that the path integral on orientable surfaces reduces to an integral over flat connections, with the measure induced by the symplectic form on the space of gauge fields
\be\label{sympform}
\Omega(\sigma,\eta) = 2\alpha \int \Tr\left(\sigma\wedge \eta\right).
\ee
Here $\sigma$ and $\eta$ are elements of the tangent space in the space of gauge fields. Concretely, they are one forms that parametrize infinitesimal variations of $A$. The constant $\alpha$ is arbitrary and can be absorbed into a shift of $S_0$, as we will see in detail below. 

An important point in \cite{Witten:1991we} is that this symplectic form is Kahler-compatible with a metric on the space of one-forms, which can be used together with a similar metric on the space of zero-forms to define the gauge-fixed path integral measure. Concretely, Kahler-compatibility of a metric and a symplectic form means that $g(\sigma,\eta) = \Omega(\sigma,J\eta)$, where $J^2 = -1$ and $\Omega(J\sigma,J\eta) = \Omega(\sigma,\eta)$. In the case of compact BF theory, one can take $J = *$, and the corresponding metric on the space of one-forms is
\be
g(\sigma,\eta)= 2\alpha \int \Tr\left(\sigma\wedge *\eta\right).\label{metric}
\ee
Unlike (\ref{sympform}), this formula involves a choice of metric on the underlying surface, which is needed to define $*$, but in the gauge-fixed path integral, this dependence drops out if one uses a similar formula to define the metric on the space of ghost and antighost fields, as explained in \cite{Witten:1991we}. 

In the noncompact case, (\ref{metric}) is not a positive metric on the space of field configurations, since the Lie algebra metric has negative directions. However, following \cite{BarNatan:1991rn}, we can use a non-invariant but positive metric on the Lie algebra in order to define the gauge fixing condition, and also to define the metric in the space of field configurations:
\be
g(\sigma,\eta)= 2\alpha \int \Tr\left(\sigma\wedge *T\eta\right).\label{metric2}
\ee
Here $T$ reverses the sign of the negative-metric component of the Lie algebra.\footnote{We are grateful to Edward Witten for suggesting this.} This metric is also Kahler-compatible with (\ref{sympform}), using $J = *T$. This means that  the analysis in the noncompact case continues to apply, and in particular the induced measure on the space of flat connections is the one that follows from (\ref{sympform}).

\subsubsection{The Weil-Petersson measure from BF theory}
An important fact for us is that, on the space of flat $SL(2,\mathbb{R})$ connections, (\ref{sympform}) is locally the same as the Weil-Petersson form on the moduli space of curves \cite{goldman1984symplectic}. We will explain this with a simple example in a moment, but first we should explain the formula in more detail.

Suppose we are at a point in the space of flat $SL(2,\mathbb{R})$ connections. A symplectic form is a two-form, which is supposed to take as input two vectors in the tangent space to this point. The tangent space can be described (up to gauge transformations) as consisting of gauge fields $\delta A$ such that $A + \epsilon \delta A$ is still flat to linear order in $\epsilon$, i.e.~
\be\label{stillflat}
d(\delta A) + A\wedge \delta A + \delta A \wedge A = 0.
\ee
We can stick two such configurations into (\ref{sympform}) and get a number $\Omega(\delta_1A,\delta_2A)$, so (\ref{sympform}) is a two form in that sense. Note also that it is gauge-invariant on the space of flat connections. To see this, suppose we make a gauge transformation of one of the deformations: $\delta_2 A \rightarrow \delta_2 A + d\Theta + [A,\Theta]$. Then the change in the value of $\Omega(\delta_1 A,\delta_2A)$ is
\be
2\alpha\int\Tr\left(\delta_1 A\wedge (d\Theta + [A,\Theta])\right)
\ee
which vanishes after integrating by parts and using (\ref{stillflat}). (The possible boundary term in this integration by parts will be important below.)

Let's now see concretely how the form (\ref{sympform}) is related to the Weil-Petersson form (\ref{WPform}). To do this we focus on a particular geodesic where two tubes of pants have been glued together. Near this region, we choose coordinates $\rho,x$ so that $\rho$ measures the distance to the joining locus ($\rho<0$ is one tube, and $\rho > 0$ is the other). The metric is
\be
ds^2 = d\rho^2 + \cosh^2(\rho) \left[b dx + \tau\delta(\rho)d\rho\right]^2, \hspace{20pt} x \sim x + 1.
\ee
The funny term involving $\delta(\rho)$ is to implement the twist by distance $\tau$. To see that this works correctly, we can write the expression in brackets as \footnote{It should be clear from context that $\theta(\rho)$ in the following equation is the step function, not an angular coordinate.}
\be
b dx + \tau \delta(\rho) d\rho = dy, \hspace{20pt} y = b x + \tau \theta(\rho).
\ee
The metric is smooth in $y$, which is a continuous coordinate on the geometry. This means that $x$ is actually discontinuous: the two spaces have been glued together with a shift by proper length $\tau$ in the $x$ direction.

In $SL(2,\mathbb{R})$ gauge theory terms, up to a gauge choice $A$ is given by (\ref{SL2r}) with
\be
e^1 = d\rho, \hspace{20pt} e^2 = b \cosh(\rho)dx + \tau \delta(\rho) d\rho, \hspace{20pt} \omega = -b\sinh(\rho)dx.
\ee
We now consider small variations of $\delta_i b,\delta_i\tau$ for $i = 1,2$ and we compute
\begin{align}
\text{Tr}\left(\delta_1 A\wedge \delta_2 A\right) &= \frac{1}{2}\left(\delta_1 e^1\wedge \delta_2 e^1 + \delta_1 e^2\wedge \delta_2 e^2 - \delta_1\omega\wedge \delta_2\omega\right)\\
&= \frac{1}{2}(\delta_1b\,\delta_2\tau - \delta_2 b\,\delta_1\tau)\delta(\rho)dx d\rho.
\end{align}
Integrating over $\rho,x$, the symplectic form  (\ref{sympform}) is
\be
\Omega(\delta_1A,\delta_2A) = \alpha(\delta_1b\,\delta_2\tau - \delta_2 b\,\delta_1\tau).
\ee
This coincides with (\ref{WPform}) including the normalization factor of $\alpha$, which is why we chose the specific normalization in (\ref{sympform}).

\subsubsection{Boundary conditions}
The real advantage to working with the BF formulation is that it is straightforward to use the same symplectic form (\ref{sympform}) to get the measure for the integral over wiggly boundaries. Before we can do that, though, we need to discuss the boundary conditions that we are imposing on the $BF$ theory. Our guiding principle is that we should choose boundary conditions that reproduce the ones from the second order formulation of JT gravity. Essentially, we need to arrange that the boundary conditions should allow a wiggly boundary but not more.

The boundary conditions that are imposed in JT gravity in order to compute the partition function $Z(\beta)$ are
\be\label{constmet}
g_{uu}\big|_{\text{bdy}} = \frac{1}{\epsilon^2}, \hspace{20pt} \phi\big|_{\text{bdy}} = \frac{\gamma}{\epsilon}, \hspace{20pt} \epsilon \rightarrow 0.
\ee
The $u$ coordinate is a rescaled proper length coordinate along the boundary.  It should run from zero to $\beta$, so that the total length of the boundary is $\beta/\epsilon$. The invariant content of (\ref{constmet}) is that that $\phi|_{\text{bdy}} = \frac{\gamma}{\beta}\cdot\text{length}(\text{bdy})$, but these equations also define a preferred coordinate $u$ along the boundary. This coordinate will be convenient below. As emphasized in \cite{Maldacena:2016upp}, this type of boundary condition allows a ``boundary graviton'' mode corresponding to a wiggly boundary. This is typically described by giving the trajectory of the $\epsilon$-regularized boundary in Euclidean AdS${}_2$.\footnote{JT gravity localizes to exact Euclidean AdS${}_2$, but all that is really important for parametrizing the boundary this way is that the geometry is asymptotically Euclidean AdS${}_2$.} To specify this, one can use e.g.~global coordinates
\be\label{global}
ds^2 = d\rho^2 + \sinh^2(\rho)d\theta^2
\ee
and then give a function $\theta(u)$, which specifies the angle in the hyperbolic plane as a function of the boundary proper length coordinate $u$. The function $\rho(u)$ is then fixed by (\ref{constmet}). Instead of parametrizing things this way, we would rather have a formula for the asymptotic behavior of the metric, so that we can read off conditions that can be translated to the first-order formalism. To do this, we can use coordinates $r,u$ where $r$ is a coordinate that measures distance to the boundary (which is taken to $r = \infty$ in the $\epsilon\rightarrow 0$ limit), and $u$ is the coordinate we have already discussed. Then one can show that a wiggly boundary leads to a metric with the large $r$ behavior
\be\label{asymptoticAdS}
ds^2 = dr^2 + \left(\tfrac{1}{4}e^{2r} - \s(u)+...\right) du^2, \hspace{20pt} \s(u) = \text{Sch}\left(\tan\frac{\theta(u)}{2},u\right).
\ee
One way to check this is to note that this metric has $R = -2$ to the order we are working, and then to compute the extrinsic curvature of a constant $r$ surface for large $r$, which is 
\be
K = \frac{1}{2}g^{uu}\partial_r g_{uu} = \frac{\frac{1}{2}\partial_r\left(\tfrac{1}{4}e^{2r} -\s(u)+...\right)}{\tfrac{1}{4}e^{2r} -  \s(u)+...} = 1 + 4e^{-2r}\s(u)+...
\ee
This agrees with the computation of the extrinsic curvature of the parametrized boundary in the fixed metric (\ref{global}), as described in \cite{Maldacena:2016upp}.\footnote{And the extrinsic curvature of the boundary curve specifies it uniquely, up to an $SL(2,\mathbb{R})$ isometry of the entire geometry that is a gauge symmetry in this context.} So, the summary of this discussion is that a to allow wiggly boundaries, we can impose that the metric should asymptotically take the form (\ref{asymptoticAdS}) with an arbitrary function $\s(u)$ that is allowed to vary.

We would now like to translate this condition to the $BF$ theory. First of all, a standard minimal boundary condition for $BF$ theory would be to fix some linear combination of $B$ and $A_u$ to zero
\be\label{BA}
B + ic A_u \big|_{\text{bdy}} = 0, \hspace{20pt} c = \text{const. to be fixed below}.
\ee
The coefficient was taken to be purely imaginary so that the boundary condition is a real one after continuing the Euclidean boundary time $u$ to Lorentzian time. One can check that with this boundary condition, the action with an appropriate boundary term
\be
I = -i\int_\mathcal{M} \Tr(B F) + \frac{i}{2} \int_{\partial\mathcal{M}}\Tr\left(B A\right).
\ee
is stationary with respect to linear variations about a solution. The boundary condition (\ref{BA}) is closely related to a familiar boundary condition in three-dimensional Chern-Simons theory. There, some linear combination of the boundary components of the gauge field are typically set to zero. $BF$ theory is a dimensional reduction of Chern-Simons theory where the gauge field is taken to be independent of the third coordinate $x^3$, and where
\be
A^{(3d)} = A + B\, dx^3.
\ee
From the three-dimensional perspective, the boundary components of $A^{(3d)}$ are $A_u$ and $B$, so (\ref{BA}) is indeed the usual type of condition. Chern-Simons theory with this type of boundary condition leads to a boundary theory that is a chiral WZW model \cite{Witten:1988hf,Elitzur:1989nr}. The analogous statement for $BF$ theory with Lie group $G$ is that the boundary theory is the quantum mechanics of a particle moving on the $G$ manifold \cite{Mertens:2018fds}. However, this isn't quite what we want. We are aiming for the Schwarzian boundary theory, which is a path integral over a quotient of $\text{diff}(S^1)$, whereas the above suggests that we will get a path integral over the larger space $\text{loop}(SL(2,\mathbb{R}))/SL(2,\mathbb{R})$ \cite{Blommaert:2018oro}.

The problem can be fixed \cite{Mertens:2017mtv,Grumiller:2017qao,Gonzalez:2018enk,Blommaert:2018oro} by following the steps \cite{Coussaert:1995zp} used to impose the asymptotic boundary conditions for AdS${}_3$ \cite{Brown:1986nw} in the Chern-Simons formulation \cite{Achucarro:1987vz,Witten:1988hc} (see also \cite{Cordova:2016cmu,Cotler:2018zff} for recent 3d discussion). Basically, we need to impose stronger boundary conditions, namely the asymptotic conditions (\ref{asymptoticAdS}). These conditions can be written in first-order variables as
\be\label{1stconstmet}
e^1 = dr, \hspace{20pt} e^2 = \left(\tfrac{1}{2}e^{r} - \s(u)e^{-r}\right)du, \hspace{20pt} \omega = -\left(\tfrac{1}{2}e^{r} + \s(u)e^{-r}\right)du.
\ee
In terms of the gauge field, this imposes the large $r$ behavior
\begin{align}
A &= \frac{dr}{2} \left(\begin{array}{cc} -1 & 0 \\ 0 & 1 \end{array}\right) + \frac{du}{2} \left(\begin{array}{cc} 0 & e^r \\ -2\,\s(u) e^{-r} & 0 \end{array}\right).\label{drinsok}
\end{align}
We emphasize that in these boundary conditions, $\s(u)$ is an arbitrary function which is allowed to vary; this describes the freedom in $A$ at the displayed order in $e^{-r}$. We should check that the action for this quantity agrees with expectations from JT gravity. First, we can fix the constant in (\ref{BA}) by requiring that (\ref{1stconstmet}) is consistent with the large $r$ behavior of (\ref{constmet}), which requires $\phi = \frac{\gamma}{\epsilon} = \gamma \frac{e^r}{2}$. This fixes $c = 2\gamma$, so the boundary term becomes
\be
I = \gamma \int du\, \Tr\left(A_u^2\right) = -\gamma\int du\,\s(u)
\ee
which is the same Schwarzian action one gets from the standard treatment of JT.

\subsubsection{The boundary wiggles and symplectic form}
We can now give a $BF$ theory description of the boundary wiggles and compute the measure from the symplectic form (\ref{sympform}). In Chern-Simons theory, the boundary WZW modes arise from would-be gauge transformations that act nontrivially on the boundary. In $BF$ theory we are in the same situation: the physical mode describing the boundary wiggles is a transformation by a would-be gauge transformation that does not vanish at infinity, but instead limits to a function $\Theta(u,r)$ with some asymptotic $r$ dependence and arbitrary $u$ dependence.

In principle $\Theta(u,r)$, as an $sl(2,\mathbb{R})$ element, contains three independent functions. However, imposing that the transformation $A \rightarrow A + d\Theta + [A,\Theta]$ should preserve the condition (\ref{drinsok}) allows us to solve for the large $r$ behavior of two of the functions in terms of the third, so that the most general transformation has the large $r$ behavior
\be\label{limitingPhi}
\Theta(u,r) \rightarrow \left(\begin{array}{cc} \tfrac{1}{2}\varepsilon'(u) & \tfrac{1}{2}e^{r}\varepsilon(u) \\ -e^{-r}\left[\text{Sch}(u) \varepsilon(u) + \varepsilon''(u)\right] & -\tfrac{1}{2}\varepsilon'(u) \end{array}\right)
\ee
for some function $\varepsilon(u)$. This preserves (\ref{drinsok}) but induces the transformation
\be
\s(u) \rightarrow \s(u) + \varepsilon'''(u) + \varepsilon(u)\s'(u) + 2\varepsilon'(u)\s(u).
\ee
This transformation is the correct transformation of the Schwarzian derivative $\s(f(u),u)$ under an infinitesimal reparametrization acting on the right: $f(u) \rightarrow f(u + \varepsilon(u))$. This identifies the mode (\ref{limitingPhi}) with an infinitesimal reparametrization, introducing a small wiggle.

We are finally in a position to evaluate the measure for the wiggles, by evaluating the symplectic form 
\be\label{reduceto}
\Omega(\delta_1 A,\delta_2 A) = 2\alpha\int_{\mathcal{M}} \text{Tr}(\delta_1 A\wedge \delta_2 A)
\ee
on a pair of configurations where $\delta_i A = d\Theta_i + [A,\Theta_i]$ and $\Theta_i$ have the limiting behavior (\ref{limitingPhi}). For any formal gauge transformations $\Theta_i$, the integrand in (\ref{reduceto}) is a total derivative, and a short calculation shows that the above can be written as a boundary integral
\be
\Omega(\delta_1 A,\delta_2 A) = 2\alpha \int_{\partial \mathcal{M}} \Tr\big(\Theta_1(d\Theta_2 + [A,\Theta_2])\big).
\ee
Now inserting the limiting form (\ref{limitingPhi}) and integrating by parts in $u$, assuming that all functions including $\s(u)$ are periodic, one finds
\be
\Omega(\delta_1 A,\delta_2 A) = \alpha \int_0^\beta du\bigg[\varepsilon_1'(u)\varepsilon''_2(u) - \s(u)\bigg(\varepsilon_1(u)\varepsilon_2'(u) - \varepsilon_1'(u)\varepsilon_2(u)\bigg)\bigg]
\ee
which can also be written as
\be\label{schform}
\Omega = \frac{\alpha}{2} \int_0^\beta du\bigg[d\varepsilon'(u)\wedge d\varepsilon''(u) - 2\,\s(u)\,d\varepsilon(u)\wedge d \varepsilon'(u)\bigg].
\ee
This is a multiple of the same (Kirillov-Kostant-Souriau) symplectic form that was assumed for the Schwarzian path integral in \cite{Stanford:2017thb}. Here we have derived it from the bulk theory. Also, we have related the coefficient of the symplectic form for boundary wiggles to the coefficient of the Weil-Petersson form (\ref{WPform}), because both are special cases of the form (\ref{sympform}).

\subsubsection{The path integral over the boundary wiggles}
We will now evaluate the path integral over the boundary wiggles for the cases that are relevant for our problem. There are two cases, corresponding to a wiggly boundary on the disk, and a wiggly boundary at the big end of a hyperbolic ``trumpet'' geometry as shown below. 
\be\label{disktrumpetfig}
\includegraphics[width=.6\textwidth,valign =c]{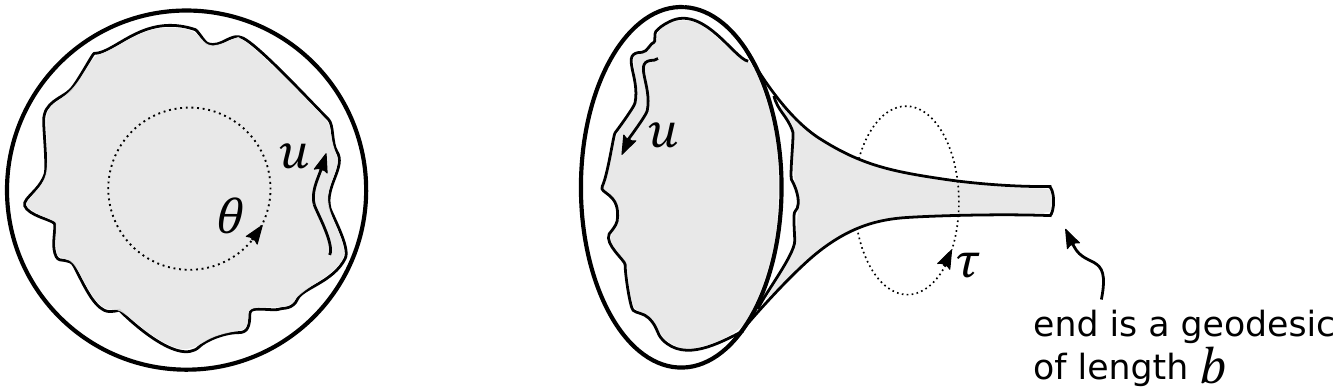}
\ee
A useful simplification is that both path integrals are one-loop exact \cite{Stanford:2017thb}, so we can evaluate them exactly by just doing the path integral for small fluctuations.

We start with the disk. In this case, it is convenient to use the following coordinates for the hyperbolic disk
\be
ds^2 = d\rho^2 + \sinh^2(\rho)d\theta^2.
\ee
The wiggly boundary can be described by giving $\theta(u)$, where $u$ is a rescaled proper length coordinate along the boundary, running from zero to $\beta$. The other coordinate $\rho(u)$ is determined by the condition that the induced metric for the boundary is $g_{uu} = \frac{1}{\epsilon^2}$. The JT action reduces to the boundary extrinsic curvature term (\ref{nvx}) in this case, which is $-\gamma \int du\, \text{Sch}(\tan\frac{\theta}{2},u)$ \cite{Maldacena:2016upp}. Evaluating the Schwarzian derivative explicitly and integrating by parts, one finds that the path integral we want is
\be\label{sdisk1}
Z_{\text{Sch}}^{\text{disk}}(\beta) = \int \frac{d\mu[\theta]}{SL(2,\mathbb{R})}\exp\left[-\frac{\gamma}{2}\int_0^\beta du \left(\frac{\theta''^2}{\theta'^2} - \theta'^2\right)\right].
\ee
The measure $d\mu[\theta]$ means the measure induced by the symplectic form (\ref{schform}). We are dividing by $SL(2,\mathbb{R})$ for the following reason: the hyperbolic disk has an $SL(2,\mathbb{R})$ isometry group. Acting with an isometry on the left panel of (\ref{disktrumpetfig}) moves the shaded droplet around in the hyperbolic space, but doesn't change the geometry of the shaded region. So to avoid overcounting, we should integrate over wiggly boundaries only up to $SL(2,\mathbb{R})$ equivalence \cite{Maldacena:2016upp}.

Using the one-loop exactness of this integral, we can get the exact answer by doing the Gaussian integral for small fluctuations about the classical solution
\be
\theta(u) = \frac{2\pi}{\beta}\left(u + \varepsilon(u)\right).
\ee
At quadratic order, the action three modes $\varepsilon = 1, e^{\pm \frac{2\pi}{\beta} i u}$. These zero modes correspond to linearized $SL(2,\mathbb{R})$ transformations of the classical solution, and we can implement the quotient in (\ref{sdisk1}) by not integrating over these modes. So we integrate over functions parametrized by
\be\label{vareps}
\varepsilon(u) = \sum_{|n|\ge 2} e^{-\frac{2\pi}{\beta} i n u}\left(\varepsilon_n^{(R)} + i \varepsilon_n^{(I)}\right)
\ee
where in order for $\varepsilon(u)$ to be real, the real and imaginary parts satisfy $\varepsilon_n^{(R)} = \varepsilon_{-n}^{(R)}$ and $\varepsilon_{n}^{(I)} = -\varepsilon_{-n}^{(I)}$. One can view the independent variables are $\varepsilon_n^{(R)}$ and $\varepsilon_{n}^{(I)}$ for positive $n\ge 2$. It is straightforward to insert (\ref{vareps}) into (\ref{schform}), along with the saddle point value of the Schwarzian $\s(\tan\frac{\pi u}{\beta},u) = \frac{2\pi^2}{\beta^2}$ and get the symplectic form in terms of these variables
\begin{align}
\Omega &= 2\alpha\frac{(2\pi)^3}{\beta^2}\sum_{n\ge 2}(n^3-n)d\varepsilon_{n}^{(R)}\wedge d\varepsilon_n^{(I)}.
\end{align}
After also working out the action in (\ref{sdisk1}) to quadratic order in $\varepsilon$, one finds that the properly normalized path integral is
\begin{align}
Z_{\text{Sch}}^{\text{disk}}(\beta) &= e^{\frac{2\pi^2\gamma}{\beta}}\prod_{n\ge 2}2\alpha \frac{(2\pi)^3}{\beta^2}(n^3-n)\int d\varepsilon_n^{(R)}d\varepsilon_n^{(I)}e^{-(2\pi)^4\gamma\frac{(\varepsilon_n^{(R)})^2 + (\varepsilon_n^{(I)})^2}{\beta^3}(n^4-n^2)}\\
&=e^{\frac{2\pi^2\gamma}{\beta}}\prod_{n\ge 2}\frac{\alpha\beta}{\gamma\, n} = \frac{1}{\alpha^{3/2}}\frac{\gamma^{3/2}}{(2\pi)^{1/2}\beta^{3/2}}e^{\frac{2\pi^2\gamma}{\beta}}.\label{Zdisk}
\end{align}
In the last step, we regularized the product using e.g.~zeta function regularization or by writing the product as an exponential of a sum of logs, introducing a smooth cutoff in the sum, and then discarding the divergent piece.

Next we consider the trumpet. There is actually a one-parameter family of trumpet geometries, labeled by the length of the geodesic at the small end of the trumpet. We take this length to be $b$. The relevant geometry can be obtained by a piece of hyperbolic space in the coordinates
\be
ds^2 = d\sigma^2 + \cosh^2(\sigma)d\tau^2, \hspace{20pt} \tau\sim \tau+ b.
\ee
The periodic identification of $\tau$ breaks the $SL(2,\mathbb{R})$ symmetry of the hyperbolic plane down to $U(1)$ translations in $\tau$. The wiggly boundary is described by a function $\tau(u)$, and in this case the boundary action becomes $-\gamma\int du\, \s(e^{-\tau},u)$.\footnote{One way to derive this is to find the relation between the $\tau$ and $\theta$ coordinates at the boundary of the regular hyperbolic disk. This is $\cos\theta = \tanh(\tau)$, which implies $\tan\frac{\theta}{2} = e^{-\tau}$. The standard formula for the Lagrangian $\s(\tan\frac{\theta}{2},u)$ is then equal to $\s(e^{-\tau},u)$.} After writing this out explicitly and integrating by parts, we find
\be\label{strumpet1}
Z_{\text{Sch}}^{\text{trumpet}}(\beta,b) = \int \frac{d\mu[\tau]}{U(1)}\exp\left[-\frac{\gamma}{2}\int_0^\beta du \left(\frac{\tau''^2}{\tau'^2} + \tau'^2\right)\right].
\ee
This is also a one-loop exact integral \cite{Stanford:2017thb}, so we can get the exact answer by studying the Gaussian integral over small fluctuations about the saddle point:
\be
\tau(u) = \frac{b}{\beta}(u + \varepsilon(u)).
\ee
One can again decompose $\varepsilon$ into modes and work out the symplectic form, this time inserting into (\ref{schform}) the classical value $\s(e^{-\frac{ub}{\beta}},u) = -\frac{b^2}{2\beta^2}$. The path integral becomes
\begin{align}
Z_{\text{Sch}}^{\text{trumpet}}(\beta,b) &= e^{-\frac{\gamma b^2}{2\beta}}\prod_{n\ge 1}2\alpha \frac{(2\pi)^3}{\beta^2}\left(n^3+\tfrac{b^2}{(2\pi)^2}n\right)\int d\varepsilon_n^{(R)}d\varepsilon_n^{(I)}e^{-(2\pi)^4\gamma\frac{(\varepsilon_n^{(R)})^2 + (\varepsilon_n^{(I)})^2}{\beta^3}\left(n^4+\tfrac{b^2}{(2\pi)^2}n^2\right)}\notag\\
&=e^{-\frac{\gamma b^2}{2\beta}}\prod_{n\ge 1}\frac{\alpha\beta}{\gamma n} = \frac{1}{\alpha^{1/2}}\frac{\gamma^{1/2}}{(2\pi)^{1/2}\beta^{1/2}}e^{-\frac{\gamma}{2}\frac{b^2}{\beta}}.\label{Ztrumpet}
\end{align}

\subsection{Putting it together: the contribution of genus \texorpdfstring{$g$}{g} in JT gravity}
We are now in a position to put the pieces together -- the Weil-Petersson volume of the bulk moduli space and the path integral over the wiggly boundary -- and compute the genus $g$ partition function with $n$ boundaries, $Z_{g,n}(\beta_1,...,\beta_n)$, introduced in (\ref{zgn}). The formula is simply
\begin{align}\notag
Z_{0,1}(\beta) &= Z_{\text{Sch}}^\text{disk}(\beta)\\
Z_{0,2}(\beta_1,\beta_2) &= \alpha\int_0^\infty b db Z_{\text{Sch}}^\text{trumpet}(\beta_1,b)Z_{\text{Sch}}^\text{trumpet}(\beta_2,b)\label{finalZgn}\\
Z_{g,n}(\beta_1,...,\beta_n) &= \alpha^n\int_0^\infty b_1 db...\int_0^\infty b_n db_n  V^\alpha_{g,n}(b_1,...,b_n) Z_{\text{Sch}}^\text{trumpet}(\beta_1,b_1)...Z_{\text{Sch}}^\text{trumpet}(\beta_n,b_n)\notag
\end{align}
where we have written the first two (which are special cases) separately. 

Let's explain this formula for the case $Z_{g,1}(\beta)$. In this case, the geometries that we integrate over have a trumpet connected to the rest of the space across a minimal geodesic of some length $b$ that should be integrated over:
\be
\includegraphics[width=.35\textwidth,valign =c]{figures/together.pdf}
\ee
For fixed $b$, the path integral factorizes into an integral over the moduli of the internal part of the surface, and an integral over the wiggly boundary of the trumpet. Together, these factors give
\be
V^\alpha_{g,n}(b) Z_{\text{Sch}}^\text{trumpet}(\beta,b),
\ee
where we have written $V_{g,n}^\alpha$ with $\alpha$ explicit to make clear the dependence on the normalization of the Weil-Petersson form (\ref{WPform}). All that remains is to integrate over $b$. In fact, there is both the length $b$ and a twist parameter $\tau$, and the measure is again the one that follows from the Weil-Petersson form (\ref{WPform}). For this case, since twisting by $b$ leaves the surface invariant, the twist is bounded between zero and $b$, so the total measure for the length parameter is
\be
\int_{\tau} \Omega = \alpha db\int_0^b d\tau = \alpha b db
\ee
and the final answer is
\be
Z_{g,1}(\beta) = \alpha\int_0^\infty b db\, V^\alpha_{g,n}(b) Z_{\text{Sch}}^\text{trumpet}(\beta,b).
\ee
The logic for $n$ boundaries is similar.

Note that the $\alpha$ dependence of the entire expression is proportional to 
\be
\alpha^n\cdot \alpha^{3g+n-3}\cdot \alpha^{-n/2} = \alpha^{\frac{3}{2}(2g+n-2)} = \alpha^{-\frac{3}{2}\chi}.
\ee
Here the first factor comes from the $n$ explicit factors of $\alpha$ in (\ref{finalZgn}), and the second factor comes from the fact that $V_{g,n}$ involves an integral over $3g+n-3$ pairs of internal variables, each carrying a factor of $\alpha$ form the form (\ref{WPform}). The final factor comes from the $\alpha$ dependence of the $n$ factors of (\ref{Ztrumpet}). Since the total $\alpha$ dependence is of the form $\alpha^{-\frac{3}{2}\chi}$, we can set $\alpha$ to one by a shift of $S_0$ by $\frac{3}{2}\log(\alpha)$.\footnote{Note that although the above calculation does not apply to the special cases $(g,n) = (0,1)$ or $(0,2)$, it is separately also true for those cases that the $\alpha$ dependence is $\alpha^{-\frac{3}{2}\chi}$.}

\subsubsection{Example: $Z_{0,2}$}
Although it is a somewhat special case from the perspective of the above formulas, it is interesting to work out the formula for $Z_{0,2}$ explicitly. This corresponds to the case where the two asymptotic boundaries are connected by a ``double trumpet''
\be
\includegraphics[width=.2\textwidth,valign =c]{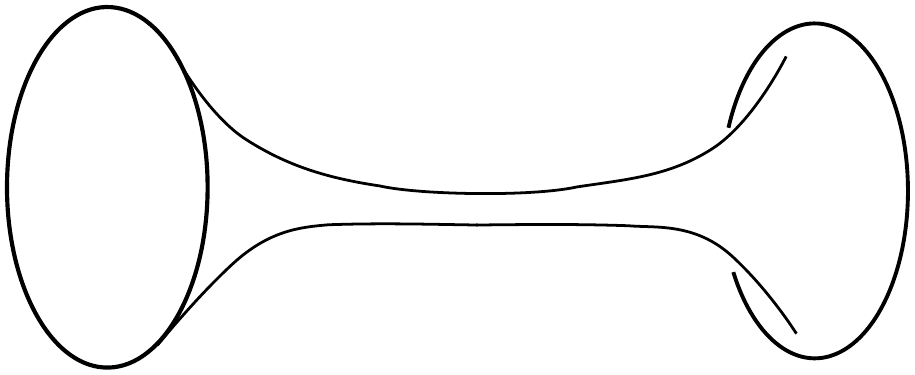}
\ee
Explicitly, in this case one has
\begin{align}\label{trumpetsee}
Z_{0,2}(\beta_1,\beta_2) &= \int_0^\infty b db\left(\frac{\gamma^{1/2}}{(2\pi)^{1/2}\beta_1^{1/2}}e^{-\frac{\gamma}{2}\frac{b^2}{\beta_1}}\right)\left(\frac{\gamma^{1/2}}{(2\pi)^{1/2}\beta_2^{1/2}}e^{-\frac{\gamma}{2}\frac{b^2}{\beta_2}}\right)\\
&=\frac{\sqrt{\beta_1\beta_2}}{2\pi(\beta_1+\beta_2)}.\label{yue}
\end{align}
If we continue $\beta_1 \rightarrow  \beta_1 + i t$ and $\beta_2 \rightarrow \beta_2 - it$, and take $t \gg \beta$, we get the linearly growing ``ramp'' contribution to the spectral form factor
\be
Z_{0,2}(\beta+it,\beta-it)\rightarrow \frac{1}{2\pi}\frac{t}{\beta_1+\beta_2}.
\ee
The coefficient here agrees with the prediction of random matrix theory, but more is actually true. The whole function (\ref{yue}) agrees with random matrix theory. To see this, we compute the contribution to the correlator of resolvents (assuming $E_1,E_2<0$)
\begin{align}
R_{0,2}(E_1,E_2) &= \int_0^\infty d\beta_1 d\beta_2 \frac{\sqrt{\beta_1\beta_2}}{2\pi(\beta_1+\beta_2)}e^{\beta_1E_1 + \beta_2E_2} \\
&= \frac{1}{4(-E_1)^{1/2}(-E_2)^{1/2}\left[(-E_1)^{1/2} + (-E_2)^{1/2}\right]^2},\label{pertSing}
\end{align}
which agrees with the universal answer (\ref{pairdouble}) for double-scaled matrix integrals.

This expression can be continued to positive energy either through the upper half plane to $E+i\epsilon$ or through the lower half plane to $E-i\epsilon$. The difference between these continuations gives the density of states (\ref{eq:discres}), and one finds (using an obvious notation)
\be
\langle \rho(E_1)\rho(E_2)\rangle_0 = \frac{R(++) + R(--) - R(+-) - R(-+)}{(-2\pi i)^2} = \frac{-1}{(2\pi)^2}\frac{E_1+E_2}{E_1^{1/2}E_2^{1/2}(E_1-E_2)^2}
\ee
Note that this expression is singular as $E_1$ approaches $E_2$. This is not true for higher genus contributions to the same correlator. This is easy to check for the genus one contribution, by doing the integral with the explicit function $V_{1,2}$ in (\ref{WPvolsTable}).

\subsubsection{Aside: these are not classical solutions}
Except for the disk, none of the path integral configurations that we considered above are full solutions to the equations of motion of JT theory. The equation of motion of $\phi$ is satisfied, since $R +2 = 0$. But the equation of motion for the metric implies that $\epsilon^{\mu\nu}\partial_\nu\phi$ is a Killing vector \cite{Banks:1990mk}, and except for the disk and double trumpet, the manifolds we consider here do not have isometries. For the double trumpet, a separate argument rules out a solution \cite{Maldacena:2018lmt}.

The basic point is that because of the $e^{-b^2/4\beta}$ factor in the trumpet partition function, there is pressure pushing $b$ to be small \cite{Hawking:1987mz}, which prevents us from finding a fully on-shell configuration. Of course, this is not a problem in the present context, since the JT theory is simple enough that we can simply do the integral over $b$.

\subsection{Correspondence with the matrix integral recursion relation}
The agreement of $R_{0,2}$ above with random matrix theory is an example of a more general fact, which is that all of the $Z_{g,n}$, taken together, give a solution to Eynard's recursion relation applied to double scaled matrix integrals (\ref{topRecur}).

As a first step, we should understand what the spectral curve function $y(z)$ is for this case. This can be obtained from the genus zero expression for the density of states as in (\ref{spec2}). The genus zero density of states for the JT theory is obtained by writing $Z_{0,1}$ as an integral with some density of states
\be
Z_{0,1}(\beta) = \int_0^\infty dE \rho_0(E) e^{-\beta E}.
\ee
In our case, $Z_{0,1} = Z_{\text{Sch}}^{\text{disk}}$ was given in (\ref{Zdisk}). After adjusting $S_0$ to make $\alpha = 1$, one can check that this answer is reproduced by the function
\be
\rho_0(E) = \frac{\gamma}{2\pi^2}\sinh(2\pi\sqrt{2\gamma E}),
\ee
which implies
\be\label{specCurve}
y(z) = \frac{\gamma}{2\pi}\sin(2\pi \sqrt{2\gamma}\,z).
\ee

We have already seen that $R_{0,2}$ has the right form for a double-scaled matrix integral, so it remains to check the generic $R_{g,n}$ case. More precisely, we will check that $W_{g,n}$ defined in (\ref{nwy}) satisfy (\ref{topRecur}). The quantities $W_{g,n}$ are in general related to $Z_{g,n}$ by an integral transform
\be
W_{g,n}(E_1,...,E_n) = 2^n z_1...z_n \int_0^\infty d\beta_1...d\beta_n e^{-(\beta_1z_1^2 + ... +\beta_n z_n^2)}Z_{g,n}(\beta_1,...,\beta_n).
\ee
For the JT gravity case (\ref{finalZgn}), after inserting the explicit expression for $Z_{\text{Sch}}^{\text{trumpet}}$ in (\ref{Ztrumpet}), one can do the integral over $\beta$ inside the integral over $b$. Then, using the integral
\be
2z\int_0^\infty d\beta e^{-\beta z^2} \frac{\gamma^{1/2}}{(2\pi)^{1/2}\beta^{1/2}}e^{-\frac{\gamma}{2}\frac{b^2}{\beta}} = (2\gamma)^{1/2}e^{-\sqrt{2\gamma}\, bz},
\ee
one finds that
\be
W_{g,n}(z_1,...,z_n) = (2\gamma)^{n/2}\int_0^\infty b_1 db_1 e^{-\sqrt{2\gamma}\,b_1 z_1}...\int_0^\infty b_n db_n e^{-\sqrt{2\gamma}\,b_nz_n}V_{g,n}(b_1,...,b_n).
\ee
If we take $\gamma = \frac{1}{2}$, then we land on the case discussed near eq.~(\ref{elt}). Eynard and Orantin showed that Mirzakhani's recursion relation implies that these functions indeed satisfy the recursion relation (\ref{topRecur}) for the spectral curve $y = \frac{\sin(2\pi z)}{4\pi}$, which is just (\ref{specCurve}) for $\gamma = \frac{1}{2}$. 

The case for general $\gamma$ follows trivially from this. One simply shows that the recursion relation (\ref{topRecur}) behaves correctly under rescalings of energy. Concretely, if $W_{g,n}(z_1,...,z_n)$ satisfy (\ref{topRecur}) with a given $y(z)$ then so do $(2\gamma)^{n/2}W_{g,n}(\sqrt{2\gamma}z_1,...,\sqrt{2\gamma} z_n)$ with $y(z) \rightarrow (2\gamma) y(\sqrt{2\gamma}z)$.

\section{Connection to minimal string theory}\label{sec:minimalString}
In the last section, we saw that the sum over topologies in JT gravity reproduces the genus expansion of a particular double-scaled matrix integral. This is reminiscent of previous connections between two-dimensional gravity and matrix integrals \cite{David:1984tx,Ambjorn:1985az,Kazakov:1985ds,Kazakov:1985ea},\footnote{For reviews see \cite{Ginsparg:1993is,DiFrancesco:1993cyw}.} which were motivated by thinking of the double-line graphs of matrix perturbation theory \cite{tHooft:1973alw,Brezin:1977sv} as a discretization of an integral over surfaces. This becomes precise in the double-scaled limit \cite{Brezin:1990rb,Douglas:1989ve,Gross:1989vs}, and we will find evidence that the correspondence identified above between JT gravity and matrix integrals is related to a further limit of these models.

Let's first briefly review the older story. In one presentation, one regards the sum over surfaces as a type of noncritical string theory called the ``minimal string,'' where the worldsheet theory is a minimal model CFT on a fluctuating geometry described by Liouville theory. Minimal model CFTs are labeled by a pair of relatively prime integers $(p,p')$, and for general values of $(p,p')$, the minimal string is dual to a multi-matrix integral. But for the special case of $(2,p)$, with $p$ an odd positive integer, one finds a single Hermitian matrix integral, and we will focus on this case. The $(2,p)$ minimal models are non-unitary, and have central charge $c = -3p + 13 - \frac{12}{p}$. In order to get a $c = 0$ worldsheet theory, the minimal model is combined with the usual $c = -26$ ghosts, and a Liouville theory \cite{Polyakov:1981rd}\footnote{For reviews see \cite{Seiberg:1990eb,Ginsparg:1993is}.} of central charge $3p + 13 + \frac{12}{p}$. This can be accomplished by taking the Liouville theory \footnote{Note that in the following action the only matter field operator appearing is the identity operator, coupling to the cosmological constant.    To achieve this in the matrix realization of the nonunitary minimal string the negative dimension matter fields must be fine-tuned to zero.   This is referred to as the  ``conformal background'' \cite{Moore:1991ir}.}
\be
I = \frac{1}{4\pi}\int d^2 x\sqrt{\hat g}\left((\hat{\nabla}\varphi)^2 + (b^{-1}+b)\varphi\hat{R} + 4\pi\mu e^{2b\varphi}\right), \hspace{20pt} b = \sqrt{\frac{2}{p}}.
\ee
The dictionary between this theory and the matrix integral is similar to what we discussed for JT gravity. In particular, matrix ensemble averages of $\langle \text{Tr}(e^{-\ell_1H})...\text{Tr}(e^{-\ell_n H})\rangle$ are given by the worldsheet path integral over surfaces with $n$ boundaries. The boundary conditions are fixed finite lengths $\ell_1,...,\ell_n$ for the Liouville sector, and the identity Cardy state for the minimal model.

Assuming that the minimal string is dual to {\it some} matrix model, one can determine the correct one by computing the disk partition function, interpreting it as the leading contribution to $\langle \text{Tr}(e^{-\ell H})\rangle$, and reading off $\rho_0(E)$.\footnote{More precisely, to compute $\langle \text{Tr}(e^{-\ell H})\rangle$, one should consider a disk with a marked boundary, meaning that a base point on the boundary has been chosen, and we do not consider translations of the boundary time coordinate as a gauge symmetry. The difference is a factor of $\ell$: if we do not pick a base point on the boundary, we would be computing $\frac{1}{\ell}\langle \text{Tr}(e^{-\ell H})\rangle$.} This computation was first done by \cite{Moore:1991ir}, using a minisuperspace approximation to Liouville, which was shown to be exact, as in \cite{Edwards:1991jx}. It was later redone in \cite{Seiberg:2003nm}, using formulas from the bootstrap solution of the Liouville boundary problem \cite{Fateev:2000ik}. In this approach, one uses, as an intermediate step, the FZZT boundary condition, which corresponds to an unmarked boundary in Liouville with fixed boundary cosmological constant $\mu_B$. For this problem, expressions for one-point functions are known \cite{Fateev:2000ik}. By an integral transform, one can change from FZZT boundary conditions to a marked boundary with fixed length $\ell$. The resulting one-point function of the cosmological constant operator with these boundary conditions is given in Eq.~(2.44) of \cite{Fateev:2000ik} as 
\be
W_b(\ell)= (\text{const.}) \mu^{\frac{1-b^2}{2b^2}}K_{\frac{1-b^2}{b^2}}(\kappa \ell), \hspace{20pt} \kappa^2 = \frac{\mu}{\sin(\pi b^2)}.
\ee
This one-point function can be interpreted as the derivative with respect to the cosmological constant $\mu$ of the disk partition function, which is then given by integrating:
\be
W(\ell) = \int d\mu \,W_b(\ell) = (\text{const.})  \frac{1}{\ell}\mu^{\frac{1}{2b^2}}K_{\frac{1}{b^2}}(\kappa \ell).\label{Wofell}
\ee

To determine the relevant matrix ensemble, we set $b = \sqrt{2/p}$ and interpret (\ref{Wofell}) as $\langle \text{Tr}(e^{-\ell H})\rangle$, and then compute the inverse Laplace transform to obtain the leading density of eigenvalues. Before doing this, it is convenient to adjust a boundary counterterm to multiply (\ref{Wofell}) by a factor of $e^{\kappa \ell}$, which will have the effect of setting the ground state energy to zero. Then, using
\be
\frac{e^{s}}{s}K_{\frac{1}{b^2}}(s) = b^2\int_0^\infty dt\, e^{-st}\sinh\left(\frac{1}{b^2}\text{arccosh}(1+t)\right)
\ee
one finds that
\be
e^{\kappa \ell}W(\ell) = \int_0^\infty dE \rho_0(E) e^{-\ell E}
\ee
with
\begin{align}
\rho_0(E) &= (\text{const.})\sinh\left[\frac{p}{2}\text{arccosh}\left(1 + \frac{E}{\kappa}\right)\right]\\
&= \sqrt{E}\left(a_0 + a_1 E + a_2 E^2 + ... + a_{\frac{p-1}{2}}E^{\frac{p-1}{2}}\right).
\end{align}
This gives the leading density of eigenvalues in the matrix model dual to the $(2,p)$ minimal string. In the second line we wrote a schematic form for the function in the case of interest where $p$ is an odd integer. The coefficients $a_k$ are positive numbers.

We would now like to understand the relationship of this to JT gravity.\footnote{We are grateful to Nathan Seiberg for emphasizing this question.} For large $p$, the central charge of the Liouville theory is large, and the fluctuations about the classical solution will be small. The classical solution of Liouville theory is hyperbolic space, and by working in units with $4\pi\mu = \frac{1}{b^2}$, we can arrange that the radius of curvature of the classical solution is one, so $R = -2$, as in our conventions for JT gravity. It seems plausible that this theory could be related to JT gravity. However, since both are supposed to be dual to matrix integrals, we can compare them by comparing the $\rho_0(E)$ function that specifies the matrix integral.  And, indeed, it is easy to check that if we take $p\rightarrow \infty$ with $E = \frac{2\pi}{p}E_{\text{JT}}$ held fixed, we land on the JT gravity spectral curve:
\be
\sinh\left[\frac{p}{2}\text{arccosh}\left(1 + \frac{E}{\kappa}\right)\right]\rightarrow \sinh\left(2\pi \sqrt{E_{\text{JT}}}\right).
\ee
The agreement of these spectral curves suggests that JT gravity is in some sense the $p\rightarrow \infty$ limit of the $(2,p)$ minimal string.\footnote{A more complete description of this correspondence will be forthcoming in \cite{seibergStanford}.} We will use this correspondence below to import intuition from the minimal string on the role of FZZT and ZZ branes in Liouville theory.

\section{Nonperturbative effects}\label{sec:nonperturbative}

The effects we have discussed so far in this paper are perturbative from the perspective of a matrix integral. More precisely, they are perturbative in the expansion parameter $1/L$, or $e^{-S_0}$ in the double-scaled case. From the perspective of JT gravity, or the collective field description of the SYK model, these are nonperturbative effects, because the quantity $e^{-S_0}$ is itself nonperturbatively small relative to the $G_N$ or $\frac{1}{N}$ expansion:
\be
\text{gravity: } \, e^{-S_0} = e^{-\frac{\#}{G_N}},\hspace{60pt} \text{SYK: }\,  e^{-S_0} = e^{-\#\, N}.
\ee
In matrix integrals these perturbative effects come from smooth  fluctuations around the large $L$ density of states $\rho_0(E)$.   The perturbative series describing these fluctuations are typically divergent, with the coefficient at genus $g$ behaving like $(2g)!$, indicating the existence of nonperturbative effects of order $\exp(-cL)$, or $\exp(-ce^{S_0})$ in the double scaled case (here $c$ can be complex).  

Indeed, below we will derive the following formula, which summarizes the leading perturbative (i.e.~genus zero) and leading nonperturbative effects in the density of eigenvalues:
\be\label{densityoneloopintro}
\langle \rho(E) \rangle \approx \begin{cases} e^{S_0} \rho_0(E)-\frac{1}{4\pi  E} \cos\Big(2\pi  e^{S_0} \int_0^E dE' \rho_0(E') \Big) & E>0 
\vspace{5pt}
\cr
  \frac{1}{-8 \pi E} \exp\Big(\hspace{-2pt}-\hspace{-2pt}2e^{S_0} \int_{0}^{-E} dx\, y\left(\sqrt{x}\right)\Big)& E<0.
  \end{cases}
\ee 
Let's explain some of the qualitative features of this formula. 
One feature that is easy to understand concerns the density of eigenvalues in the ``classically forbidden'' region $E <0$. This is zero to all orders in perturbation theory, and the main nonzero contribution comes from the exponentially small probability of a single eigenvalue being located at $E<0$. The weighting for such a configuration is controlled by the ``effective potential'' $V_{\text{eff}}(E)$ that the eigenvalue feels: the sum of the actual potential $V(E)$ and the Vandermonde repulsion from the remaining eigenvalues.  In the forbidden region we have from (\ref{effpot})
\begin{align}
V'_{\text{eff}}(E) &= L V'(E) - 2 e^{S_0} R_{0,1}(E) = -2e^{S_0}y(\sqrt{-E}) \\
V_{\text{eff}}(E)  &= 2e^{S_0} \int_{0}^{-E} dx\, y\left(\sqrt{x}\right)
\end{align}
and so in the forbidden region \eqref{densityoneloopintro} can be rewritten as
\be
\langle \rho(E)\rangle\approx \frac{1}{-8\pi E}\exp\left(-V_{\text{eff}}(E)\right) ~.\label{densityforbidden}
\ee
This simple picture only makes sense if $V_{\text{eff}}(E) >0$ in the forbidden region, indicating stability of the tree level distribution $\rho_0(E)$.    We will assume this for now to simplify our presentation.  For JT gravity it actually isn't true, as we will discuss in section \ref{instab}.

\begin{figure}[t]
 \centering
\includegraphics[scale=0.5]{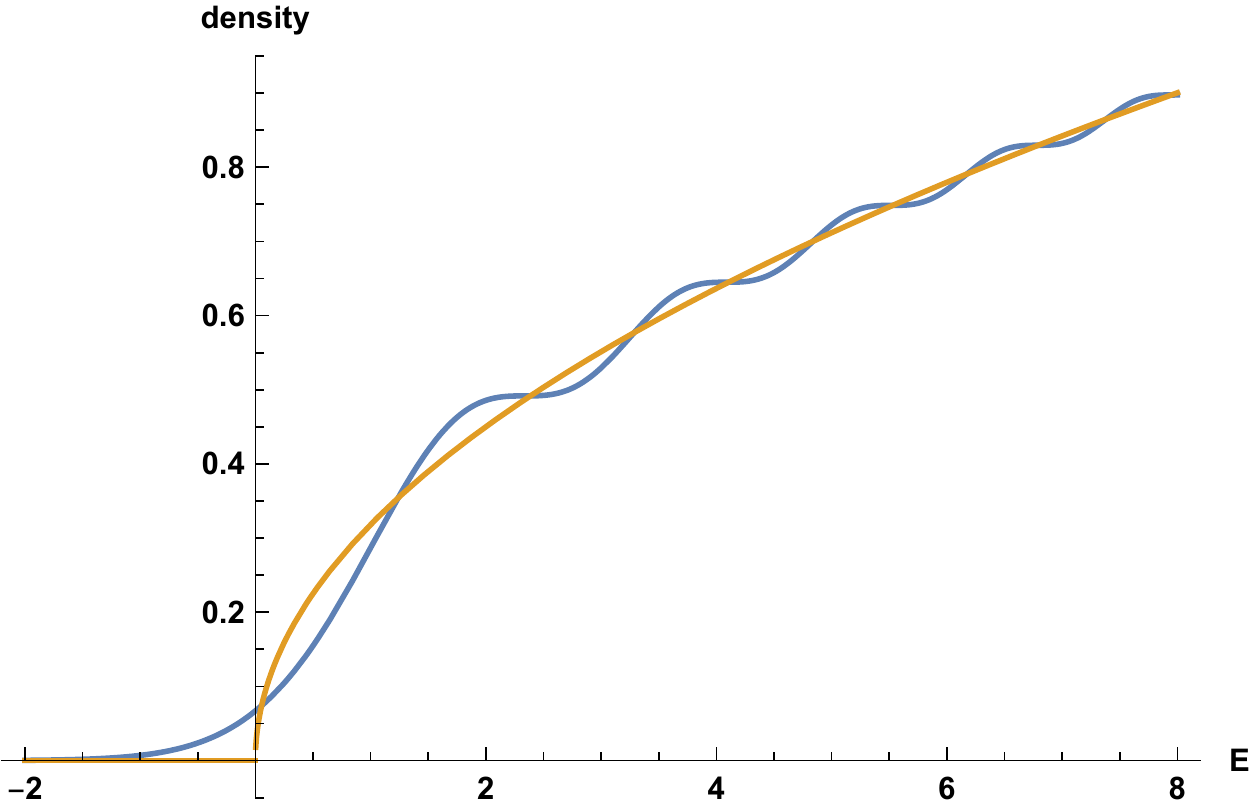}
\caption{The exact eigenvalue density for the Airy case \eqref{exactairy} (blue, dark), and the tree-level answer $\rho_0^{\text{total}}(E) = \frac{e^{S_0}}{\pi}\sqrt{E}$ (orange, light) are plotted for $e^{S_0} = 1$. Note the oscillations in the ``allowed'' region $E>0$ and the small but nonzero tail in the ``forbidden'' region $E<0$.}\label{fig:Airydensgraph}
\end{figure}

The oscillating behavior of (\ref{densityoneloopintro}) in the allowed region is in some respects more dramatic. 
The nonperturbative cosine term  in (\ref{densityoneloopintro}) is rapidly oscillating, but is not particularly small, of order one in powers of $e^{S_0}$. This is smaller than the tree level term $e^{S_0}\rho_0$, but it is larger than the first perturbative (genus one) correction, which we haven't displayed in (\ref{densityoneloopintro}), but which would be proportional to $e^{-S_0}$. So the nonperturbative correction isn't small, but it is rapidly oscillating as a function of $E$, on the scale of the separation between individual eigenvalues.  This is clearly visible in Figure \ref{fig:Airydensgraph}. 

For integrated observables that are sufficiently smooth in energy, such a term will give small effects. But in special observables they can be important. For example, oscillating nonperturbative terms in the two-eigenvalue correlation function are needed in order to explain the ``plateau'' in the spectral form factor. More generally, the oscillating nonperturbative term can be understood as a consequence of the underlying discreteness of the energy levels, as the oscillations are on the scale of the separation between adjacent levels.

The kinds of nonperturbative effects discussed here have been given a ``bulk'' interpretation as D-branes in minimal string theories described by double scaled matrix integrals.   One eigenvalue instantons correspond to ``ZZ branes,'' an analog of D-instantons.   The allowed region oscillations are probed by ``FZZT branes,'' an analog of probe D0 flavor branes.

In this section, we will give a heuristic derivation of (\ref{densityoneloopintro}) and then interpret it in the context of JT gravity by analogy to the minimal string identification. This is particularly interesting since in that context, the effects are doubly-nonperturbative from the perspective of ordinary gravitational perturbation theory in $G_N$.

\subsection{The determinant}
In order to discuss nonperturbative effects in matrix integrals, it is convenient to start out by working with determinants instead of resolvents.\footnote{We are grateful to Juan Maldacena for this suggestion.}   In these quantities, the nonperturbative effects contribute at leading order, and are therefore easier to examine.  In particular, we define the quantity
\be
\psi(E)\equiv \det(E-H) e^{-\frac{L}{2} V(E)}.\label{detV}
\ee
As we will discuss later, this operator can be interpreted as the insertion of a ``brane'' \cite{Aganagic:2003qj,Maldacena:2004sn}.

A useful general fact is that the expectation value of $\psi(E)$ in the matrix integral is equal to $P_L(E)e^{-\frac{L}{2}V(E)}$, where $P_n(E)$ is the degree $n$ monic polynomial from the family of polynomials that are orthogonal with respect to the measure $e^{-L V(E)}$. In the double-scaled limit, the polynomial index $n$ is replaced by a continuous parameter $n/L$, and the orthogonal functions satisfy a pair of linear differential equations. For large $e^{S_0}$, these equations can be solved using WKB methods, and a ``one loop'' solution at large $e^{S_0}$ is \cite{Moore:1990mg,Maldacena:2004sn}
\be\label{genwkb}
\langle \psi(E)\rangle \sim \frac{A}{z^{1/2}}\exp\left[e^{S_0}\int_0^z dE(z')y(z')\right] + \frac{B}{z^{1/2}}\exp\left[-e^{S_0}\int_0^z dE(z')y(z')\right].
\ee
where $E = -z^2$. This is written in terms of the spectral curve coordinate $y$ (\ref{spec2}). The coefficients $A,B$ are arbitrary so far, and as we will see, they can depend on the angle of $E$ in the complex plane. 
\subsubsection{Airy}
A useful example to have in mind is the case where $\rho_0 = \frac{1}{\pi}\sqrt{E}$ and the corresponding spectral curve is $y^2 = -E$. In this case, it is possible to write the exact answer for   $\langle\psi(E)\rangle$ by representing it as the  effect of integrating  out an additional ``matter field'' in the fundamental representation. This motivates  the interpretation  of  $\psi(E)$  as a probe ``flavor'' brane.\footnote{See for example \cite{Maldacena:2004sn}.     This kind of construction is used extensively in the quantum chaos literature.  For reviews see \cite{Efetov:1997fw,2005hep.ph....9286S,2009NJPh...11j3025M,haake2010quantum}.}

We begin with the Gaussian matrix potential $V(x)= \frac{2}{a^2} x^2$, which has the leading order density of states given by the Wigner semicircle:
\be
\rho_0^{\text{total}}(E) = \frac{e^{S_0}}{\pi}\sqrt{\frac{a^2-E^2}{2a}}, \hspace{20pt}  -a<E<a.
\ee
For this to be normalized, $L= e^{S_0} (a/2)^{3/2}$. For this theory, one can write the determinant $\det(E-H)$ as an integral over Grassmann vectors $\chi_i$ and $\overline{\chi}_i$ and perform the Gaussian integral over $H$ to arrive at an integral over the Grassmann variables that only depends on the combination $\sum_i \overline{\chi}_i \chi_i$,
\be
\langle \det(E-H)\rangle= \int d\chi_i d \overline{\chi}_i e^{ E \overline{\chi}_i \chi_i -\frac{a^2}{8 L} (\overline{\chi}_i \chi_i)^2}.
\ee
We can introduce an auxiliary variable $s$ so that $\chi_i$ and $\overline{\chi}_i$ can be simply integrated out,
\begin{align}\label{hermiteint}
\langle \det(E-H)\rangle&=\sqrt{\frac{2 L}{\pi a^2}} \int ds\, d\chi_i d \overline{\chi}_i  e^{(E+i s) \overline{\chi}_i \chi_i - \frac{2L}{a^2} s^2}
\cr
& = \sqrt{\frac{2 L}{\pi a^2}} \int ds (E+is)^L e^{- \frac{2L}{a^2} s^2}.
\end{align}
The final line is an integral representation for the Hermite polynomial $(\frac{a^2}{8L})^{L/2}H_L(\sqrt{2L}E/a).$ The Hermite polynomials are the orthogonal polynomials for the Gaussian measure, so this is an example of the general result mentioned above. We see that in this case $\langle \psi(E) \rangle$ is a harmonic oscillator wavefunction, and so satisfies a Schrodinger equation (with $\hbar = \frac{1}{L}$). For this simple potential the WKB estimates described in \eqref{genwkb} follow from a standard WKB analysis of of this Schrodinger equation.\footnote{We will not use the Schrodinger equation perspective directly, but it is a convenient way to obtain the leading exponential behavior of (\ref{genwkb}). See e.g.~ section 4.3 of \cite{Dijkgraaf:2018vnm} for recent discussion.} 
To obtain the theory $\rho_0(E) = \frac{1}{\pi}\sqrt{E}$, we study this Gaussian theory near the lower edge of the eigenvalue distribution. This corresponds to studying the wavefunction near its left classical turning point. To do this, we replace $E_{\text{old}} = -a + E_{\text{new}}$ so that the support of $\rho_0$ is $0<E_{\text{new}}<2a$. Then we send $a \rightarrow \infty$ holding $E_{\text{new}} = E$ fixed. One can take this limit in the integral representation (\ref{hermiteint}).

After multiplying by $e^{-\frac{L}{2}V(E)}$ as in (\ref{detV}), one finds\footnote{We change variables to $s \rightarrow -\frac{ia}{2} + \sqrt{\frac{a}{2}}e^{-S_0/3}s$ before taking $a\rightarrow \infty$. Also, as the overall normalization in $\langle \psi(E)\rangle$ can be rescaled by a constant shift of the potential, we will choose things so that the constant of proportionality in (\ref{aiint}) is one.}
\be\label{aiint}
\langle \psi(E)\rangle \propto \frac{1}{2\pi} \int_{-\infty}^{\infty} ds \;e^{\frac{i}{3} s^3 +i \xi s} = \text{Ai}(\xi), \hspace{20pt} \xi = -e^{\frac{2S_0}{3}}E.
\ee 
The leading asymptotics of the Airy function on the real axis can be obtained from a standard saddle point analysis of this integral. There are two saddle points, and in the allowed region $\xi < 0$, the defining contour is deformable to the union of the two associated steepest-descent contours. In the forbidden region, $\xi>0$, the defining contour is deformable to the steepest descent contour associated to only one of the saddles. This leads to the leading large $\xi$ asymptotics
\be
\text{Ai}(\xi) \sim \begin{cases}\frac{1}{\sqrt{\pi}(-\xi)^{1/4}}\cos\left(-\tfrac{\pi}{4} + \tfrac{2}{3}(-\xi)^{3/2}\right)& \xi<0\\
 \frac{1}{2\sqrt{\pi}\xi^{1/4}}\exp\left(-\tfrac{2}{3}\xi^{3/2}\right)&\xi >0.
\end{cases}\label{airyasymp}
\ee
This is consistent with the general formula (\ref{genwkb}) applied to the case $y^2 = -E$, and it determines the $A,B$ coefficients in both the forbidden and allowed regions.

In this paper we are concerned with matrix integrals whose spectral curve approaches the Airy curve $y^2 = -E$ at small $E$, up to a rescaling. We will determine the $A,B$ coefficients in general by demanding that the results join onto the Airy results for small $E$. This prescription is consistent with rigorous results for unscaled matrix models \cite{MR1702716}.

\subsubsection{Disks and Cylinders}
We now describe another technique for obtaining semiclassical formulas like \eqref{genwkb}.\footnote{See for example \cite{Aganagic:2003qj,Kutasov:2004fg,Bergere:2009zm,Dijkgraaf:2018vnm}.}  This technique is not rigorous, but it does have two advantages.  First, it makes direct contact with the quantities used in topological recursion.   Second, and more importantly, it is directly connected to the D-brane interpretation of these nonperturbative effects.   

The essence of the technique is very simple.   It is based on the identity
\be\label{dettotrlog}
\det(E-H) = \exp\Big(\Tr \log (E-H)\Big).
\ee
We would like to take the expectation value of the RHS in the matrix ensemble, using the general formula
\be\label{expX}
\langle e^X\rangle = \exp\left(\langle X\rangle + \frac{1}{2}\langle X^2\rangle_{\text{conn.}}+...\right),
\ee
where in our case $X = \Tr\log(E-H)$. In a matrix integral, the higher connected correlators are suppressed by powers of $e^{S_0}$. So, in a leading approximation, we can keep only the first term in (\ref{expX}), and we evaluate it to leading order. This can be done by integrating the genus-zero resolvent:\be
\langle \Tr \log(E-H)\rangle_{g=0} = e^{S_0}\int^E dE' R_{0,1}(E').
\ee
This combines nicely with the explicit factor of $e^{-\frac{L}{2}V(E)}$ in (\ref{detV}). Using $e^{S_0} y = e^{S_0}R_{0,1}(E) - \frac{L}{2}V'(E)$, we have the leading formula
\be\label{treeexpdisk}
\langle \psi(E) \rangle = e^{-\frac{L}{2} V(E)}  \langle \det(E-H)\rangle \simeq \exp\left(e^{S_0}  \int^{E} dE' \,y\right),
\ee
which matches the exponential factors in \eqref{genwkb}. 

This approach is analogous to Polchinski's calculation \cite{Polchinski:1994fq} of D-brane amplitudes,  of order $e^{-1/g_s}$, as an expansion in disconnected worldsheets of  disk topology, that is in powers of $1/g_s$. We will give a corresponding brane interpretation of the effects in \eqref{treeexpdisk}. Anticipating this discussion, we will refer to the term in the exponential in (\ref{treeexpdisk}) as the disk amplitude.

For the discussion below, we will need to be more precise about the choice of branch. Because $y$ is multivalued in the $E$ plane, it is convenient to view the disk amplitude as a function of $z$, where $E = -z^2$:
\be
\text{Disk}(z) \equiv e^{S_0}\int_0^{z}y(z')dE(z').
\ee
For the Airy spectral curve $y^2 = -E$, or alternatively $y = z$, this is
\be
\text{Disk}(z) = e^{S_0}\int_0^z z \cdot (-2z dz) = -\frac{2z^3}{3}e^{S_0}.
\ee
For the JT gravity curve, it is
\be
\text{Disk}(z) = e^{S_0}\int_0^z \frac{\sin(2\pi z)}{4\pi}\cdot(-2z dz) = -\frac{\sin(2\pi z)-2\pi z\cos(2\pi z)}{8\pi^3}e^{S_0}.
\ee
In order to insert these expressions in (\ref{treeexpdisk}), we have to decide which branch of $z(E)$ to use. In principle, we have this choice in each factor of $\langle \Tr \log (E-H) \rangle_{g=0}$ that appears when we expand out \eqref{dettotrlog}.  In order to determine the branch, we need to supply additional input.   The known general form \eqref{genwkb} can be reproduced if we choose the same branch for all terms in the expansion of the exponential, and possibly also choose the other choice in another complete sum. This gives the two exponentials in \eqref{genwkb}.  Additional information has to be supplied to fix their coefficients.  In our case joining onto the Airy limit will determine this.

In order to reproduce the prefactor powers in (\ref{genwkb}), we need to include the order-one terms in (\ref{expX}). For Hermitian matrix integrals, the only term of order one comes from the connected two-point function. It will be convenient to view this as a function of separate $E_1,E_2$ arguments:
\be
\langle \Tr \log (E_1 - H) \Tr \log (E_2-H) \rangle_{\text{conn.}} =    \int^{E_1} dE_1' \int^{E_2} dE_2'  \; R_{0,2}(E_1', E_2') + {\cal O}(e^{-2 S_0}).
\ee
Anticipating the brane discussion, we will refer to the leading term on the RHS as the cylinder, and we will regard it as a function of $z$:
\begin{align}
\text{Cyl}(z_1, z_2) &\equiv \int_{\infty}^{z_1} dE(z_1')\int_{\infty}^{z_2}  dE(z_2') \; R_{0,2}(E(z_1'), E(z_2'))\label{cylLine1}\\
&=\int_{\infty}^{z_1}\int_{\infty}^{z_2}\frac{dz_1'dz_2'}{(z_1'+z_2')^2}=-\log(z_1+z_2) + \infty.
\end{align}
In the second line, we used (\ref{pairdouble}) and did the integral, assuming $z_1,z_2>0$. The infinite additive constant is independent of $z_1,z_2$, and it reflects an ambiguity in the normalization of $\psi(E)$ in the double-scaled limit. For the moment, we simply add a constant to the potential in order to cancel the $\infty$. In the better-defined ratio of determinants that we will consider below, it will cancel automatically.

In terms of the disk and cylinder quantities, for a given choice of branch, we can write the RHS of (\ref{expX}) to one-loop accuracy as a function of $z$
\be
\Psi(z) \equiv \exp\left[\text{Disk}(z) +\frac{1}{2} \text{Cyl}(z,z)\right] = \frac{1}{\sqrt{2z}}\exp\left[\text{Disk}(z)\right].\label{PsiDisk}
\ee
We expect to be able to express $\langle \psi(E)\rangle$ in general in terms of a linear combination of the two branches $\Psi(z)$ and $\Psi(-z)$. By matching to the Airy case (\ref{aiint}) and using the asymptotics of the Airy function (\ref{airyasymp}), we find that specific linear combination has to be
\be\label{psidiskannulus}
\langle \psi(E)\rangle \propto \begin{cases} \Psi(z) + \Psi(-z) &  E>0 \cr \Psi(z) & E<0. \end{cases}
\ee
In this expression, we have $z^2 = -E$ as always. Because of the $1/\sqrt{2z}$ in (\ref{PsiDisk}), we have to specify a branch for the prefactor. The correct prescription from matching to Airy is to take $\Psi(e^{i\frac{\pi}{2}}\sqrt{E}) + \Psi(e^{-i\frac{\pi}{2}}\sqrt{E})$ on the first line, and $\Psi(\sqrt{-E})$ on the second line.

\subsection{The inverse determinant}
We would like to understand the origin of nonperturbative effects in more familiar quantities like the average density, the pair density correlation function, or in resolvents.   The orthogonal polynomial approach to matrix integrals allows one to write all such quantities in terms of orthogonal polynomials \cite{eynard2015random}.  So in principle having determined the orthogonal polynomial from the $\langle \det(E-H) \rangle$ we have all that we need.   But we would like to attempt a more direct, if more formal, calculation.  

An important intermediate tool is the insertion of
\be
\tilde\psi(E) \equiv \frac{e^{\frac{L}{2}V(E)}}{\det(E-H)}.
\ee
The operator  $\tilde{\psi}(E)$ is interpreted as the insertion of  a ``ghost brane,'' which is a term used to mean an object that totally cancels a brane if it is inserted at the same point.\footnote{See for example \cite{Aganagic:2003qj,Bergere:2009zm,Dijkgraaf:2018vnm}. In some of these papers this object is referred to as an antibrane. We are using the term ghost brane in the sense of \cite{Okuda:2006fb}.} The ghost brane is a more complicated object than the brane. Before double scaling   $\langle 1/\det(E-H) \rangle$ is the Hilbert transform\footnote{More precisely $\langle 1/\det(E-H) \rangle_{L+1}$ is the Hilbert transform of $e^{-LV(E)}\langle \det(E-H)\rangle_{L}$.}  of $\langle \det(E-H) \rangle$ with respect to the measure $e^{-LV(E)}$ \cite{eynard2015random}. This procedure defines two independent functions, depending on whether the energy argument is above or below the real axis. Of course, both functions can be continued in the complex $E$ plane, and they define two separate entire functions. We will refer to them as $\langle \tilde\psi_\pm(E)\rangle$, where the plus subscript means that the formula gives the actual ghost brane answer for $E$ in the upper half-plane, and the minus subscript means that it agrees in the lower half-plane.

The double scaled version of $\langle\tilde{\psi}(E)\rangle$ also obeys a differential equation \cite{2009arXiv0909.0854B} which yields a semiclassical expansion  of the same general form as \eqref{genwkb}.   To understand the more subtle determination of coefficients in this case we turn again to the Airy example and then proceed to a formal disk and cylinder calculation.

\subsubsection{Airy}
We start with the Airy case, with spectral curve $y^2 = -E$. One can compute $\langle\tilde\psi\rangle$ using the same technique we used for $\langle\psi\rangle$. Instead of introducing Grassmann flavor fields we introduce bosonic fields $\phi_i,\bar\phi_i$ to implement $1/\det(E-H)$. After mimicking the procedure that led us to the Airy integral, and adjusting the arbitrary overall normalization, we find (see \ref{appendixantibrane})
\be
\langle \widetilde{\psi}_\pm(E) \rangle = \int_{\mathcal{C}_\pm} dr \;e^{\frac{1}{3} r^3-\xi r}, \hspace{20pt} \xi = -e^{\frac{2}{3}S_0}E.\label{eq:antibraneairyintegral}
\ee
As explained in appendix \ref{appendixantibrane}, the contour $\mathcal{C}_\pm$ starts at $r=-\infty$ and ends at $r= e^{\pm i\frac{\pi}{3}}\cdot\infty$. By comparing to the contour integral definitions of the Airy function, we see that the exact answer for this integral is
\be\label{airyantibrane}
\langle \tilde{\psi}_\pm(E) \rangle =\pi \text{Bi}(\xi) \pm i\pi\text{Ai}(\xi).
\ee
As in the case of the brane, the overall normalization of this quantity is not significant.

\subsubsection{Disks and Cylinders}
We would now like to write a candidate general formula in terms of the disk and cylinder function. We use the same formula (\ref{expX}), but now with $X = -\Tr\log(E-H)$. At one-loop order, this is equal to 
\be
\tilde{\Psi}(z)\equiv \exp\left[ -\text{Disk}(z) +\frac{1}{2} \text{Cyl}(z,z)\right] = \frac{1}{\sqrt{2z}}\exp\left[-\text{Disk}(z)\right]\label{PsiDisk2}
\ee

At one loop, one would expect to be able to write $\langle \tilde\psi\rangle$ in terms of a linear combination of the two branches $\tilde\Psi(\pm z)$. Matching to (\ref{airyantibrane}) determines the linear combination, up to an overall constant multiple, as
\be
\langle \tilde{\psi}_\pm(E) \rangle \sim 
\begin{cases} \tilde{\Psi}(z)& E>0
\cr 
\tilde{\Psi}(z) + \frac{1}{2}\tilde\Psi(-z)& E<0.  \end{cases}\label{eq:antibranediskannulus}
\ee
In these expressions, we mean that $z$ is continued through the lower half-plane for $\tilde\psi_+$ and through the upper half plane for $\tilde\psi_-$. So in the upper line we have $\tilde{\Psi}(e^{\mp i\frac{\pi}{2}}\sqrt{E})$ and in the lower line we have $\tilde{\Psi}(\sqrt{-E}) + \frac{1}{2}\tilde{\Psi}(e^{\mp i\pi}\sqrt{-E})$. We expect (\ref{eq:antibranediskannulus}) to be valid for a general spectral curve that limits to the Airy curve near the origin.

\subsection{The resolvent}
We will now combine the brane and ghost brane computations in order to get (\ref{densityoneloopintro}). 
The key formula for this section relates the resolvent to the derivative of a dipole of a brane and ghost brane\footnote{See for example \cite{Efetov:1997fw,2009NJPh...11j3025M,haake2010quantum,Bergere:2009zm}.}
\be
\Tr \frac{1}{E-H} =  \partial_E \frac{\det(E-H)}{\det(E'-H)}\bigg|_{E'\rightarrow E}~.
\ee
Though the above way of writing this relation will be more suitable for calculations, the following, more trivial-looking formula may also be helpful to think about
\be
\Tr \frac{1}{E-H} =  \Tr \left(\frac{1}{E-H}\right) \frac{\det(E-H)}{\det(E'-H)}\bigg|_{E'\rightarrow E}~.\label{eq:resolventdettrivial}
\ee
The introduction of these determinants may seem unnecessary. However, if we include the ratio of determinants for $E\neq E'$ and calculate using the formal method of the previous sections we find in addition to the perturbative series for the resolvent there are nonperturbative pieces that survive in the limit $E'\rightarrow E$.

\subsubsection{Airy}
As always, we start with the Airy case $y^2 = -E$. Modifying the techniques for the brane and ghost brane, we show in appendix \ref{appendixdipole} that
\begin{align}
\big\langle\psi(E)\tilde{\psi}_\pm(E')\big\rangle &=-
\int_{\mathcal{C_\pm}} \frac{dr ds}{2\pi} (r+is)\exp\Big[ \tfrac{1}{3}r^3+\tfrac{i}{3}s^3 +e^{\tfrac{2S_0}{3}}(E' r- i E s)\Big]\label{eq:detoverdetintegral}\\ 
&= \pi e^{-\frac{2S_0}{3}}(\partial_E-\partial_{E'})\text{Ai}(\xi)\big(\text{Bi}(\xi')\pm i \text{Ai}(\xi')\big), \hspace{20pt} (\xi,\xi') = -e^{\frac{2S_0}{3}}(E,E').\notag
\end{align}
By contrast to the brane and the ghost brane computations, this dipole quantity has a well-defined and meaningful normalization. Taking another derivative with respect to $E$, using the definition of the Airy function $\text{Ai}''(\xi)=\xi \text{Ai}(\xi)$, and setting $E'\rightarrow E$, we find an exact formula for the resolvent:
\be
\langle R^{\pm}(E)\rangle - \frac{L V'(E)}{2}= \pi e^{\frac{2 S_0}{3}} \left[-\text{Ai}'(\xi) \big(\text{Bi}'(\xi)\pm i \text{Ai}'(\xi)\big)+ \xi \text{Ai}(\xi) \big(\text{Bi}(\xi)\pm i\text{Ai}(\xi)\big)\right].
\ee
Note that like the ghost brane, the resolvent defines two separate entire functions, as discussed in \cite{Maldacena:2004sn}. Taking the difference between the two along the real axis, we recover (\ref{exactairy}).

One point to note is the following. The saddle point structure of the integral (\ref{eq:detoverdetintegral}) is the same as for the product of the integrals for $\langle \psi(E)\rangle$ and $\langle \tilde{\psi}(E')\rangle$. This means that the asymptotics of $\langle\psi(E)\tilde{\psi}(E')\rangle$ is to leading order the same as the asymptotics of $\langle \psi(E)\rangle \langle \tilde{\psi}(E')\rangle$. At first sight this may seem problematic; $\langle\psi(E)\tilde{\psi}(E')\rangle$ should be exactly one as $E'\rightarrow E$, while the product $\langle \psi(E)\rangle \langle \tilde{\psi}(E)\rangle$ is not equal to one. The one-loop correction away from the product structure of (\ref{eq:detoverdetintegral}) is critical for ensuring the correct behavior at $E'\rightarrow E$, as we will see in a moment.

\subsubsection{Disks and cylinders}
We would now like to reproduce the asymptotics of this formula with a formal disk-and-cylinder approach. 
To do this, we write
\be
\frac{\det(E-H)}{\det(E'-H)} = \exp\bigg[\int_{E'}^E dE'' R(E'')\bigg]
\ee 
and then apply the formula (\ref{expX}). A nice feature of the ratio of determinants is that although the determinant and the inverse determinant do not by themselves have a natural normalization, their product does. This is because the arbitrary constant in the potential cancels out. And, correspondingly, the total one-loop factor
\be
C(z,z') \equiv \frac{1}{2}\Big(\text{Cyl}(z,z) + \text{Cyl}(z',z')\Big) - \text{Cyl}(z,z') = \log\frac{z+z'}{2\sqrt{zz'}}
\ee
is finite without any subtraction of infinite terms. By analogy to the previous formulas (\ref{PsiDisk}) and (\ref{PsiDisk2}), we expect to write a one-loop formula for the brane-ghost brane pair in terms of the function
\be
\Psi(z;z') = \frac{z+z'}{2\sqrt{z z'}}\exp\Big(\text{Disk}(z)-\text{Disk}(z')\Big).\label{braneGhostBranePsi}
\ee
We separate the arguments by a semicolon to emphasize that this function is not symmetric. The first argument refers to the determinant, and the second to the inverse determinant.

We find that the exact Airy expression can be matched to the leading asymptotics 
\be
\big\langle \psi(E)\tilde{\psi}_{\pm}(E')\big\rangle \approx \begin{cases} \Psi(z;z') + \Psi(-z;z')& E,E'>0
\cr
\Psi(z;z') + \frac{1}{2}\Psi(z;-z')& E,E'<0.
  \end{cases}\label{eq:braneantibraneoneloop}
\ee
The branch prescriptions for the two arguments are the same as in the respective determinant and inverse determinant cases. Explicitly,
\be
\big\langle \psi(E)\tilde{\psi}_{\pm}(E')\big\rangle \approx \begin{cases} \Psi(e^{i\frac{\pi}{2}}\sqrt{E};e^{\mp i\frac{\pi}{2}}\sqrt{E'}) + \Psi(e^{-i\frac{\pi}{2}}\sqrt{E};e^{\mp i\frac{\pi}{2}}\sqrt{E'})& E,E'>0
\cr
\Psi(\sqrt{-E};\sqrt{-E'}) + \frac{1}{2}\Psi(\sqrt{-E};e^{\mp i\pi}\sqrt{-E'})& E,E'<0.
  \end{cases}\label{eq:braneantibraneoneloop2}
\ee
Unlike in the previous cases, the normalization here is meaningful. As a simple check, we can take the limit $E' = E$. Using (\ref{braneGhostBranePsi}), we find that
\be
\Psi(z,z) = 1, \hspace{20pt}\Psi(z,-z) = 0,\label{cancelsPerfectly}
\ee
so (\ref{eq:braneantibraneoneloop}) reduces to one in the limit $E' = E$. This is expected because $\psi(E)\tilde\psi(E) = 1$ exactly.

Now, to compute the resolvent we take the derivative $\partial_E = -\frac{1}{2z}\partial_z$ and then set $E'\rightarrow E$. Again using (\ref{braneGhostBranePsi}), and also using that $\text{Disk}(z)$ is odd in $z$, one finds
\be
\partial_z \Psi(z;z')\Big|_{z'=z} = \partial_z\text{Disk}(z),\hspace{20pt} \partial_z \Psi(z;z')\Big|_{z'=-z} = \frac{1}{2\sqrt{-z^2}}\exp\big[2\,\text{Disk}(z)\big].
\ee
This is an interesting expression. In the coincidence limit when the arguments are on the same sheet, the derivative of the brane-ghost brane function (\ref{braneGhostBranePsi}) reduces to a simple object, the derivative of the disk. But when the $z$ arguments are on opposite sheets, we get a nontrivial expression with an exponentiated disk amplitude. This is the source of the nonperturbative effects in this way of calculating the resolvent. 

Multiplying by $-\frac{1}{2z}$ to convert the $z$ derivative to an $E$ derivative, and then plugging into (\ref{eq:braneantibraneoneloop2}) and taking care with the branches, we find 
\be
\langle R^\pm(E) \rangle \approx \begin{cases} \mp e^{S_0}y(i\sqrt{E})\pm\frac{i}{4 E} \exp\big[\pm 2 \text{Disk}(i\sqrt{E}) \big] & E>0
\cr
 e^{S_0} y(\sqrt{-E}) \mp \frac{i}{-8 E} \exp\big[2\text{Disk}(\sqrt{-E})\big] & E<0.
  \end{cases}\label{eq:resolventoneloop}
\ee
Given this expression for the resolvent, we can find the one-loop approximation to the density of eigenvalues by using $R^+ - R^- = -2\pi i \rho$. Using $ y(i\sqrt{E})= i \pi \rho_0(E)$ for $E>0$, we find
\be\label{densityoneloop}
\langle \rho(E) \rangle \approx \begin{cases} e^{S_0} \rho_0(E)-\frac{1}{4\pi  E} \cos\big(2\pi  e^{S_0} \int_0^E dE' \rho_0(E') \big) & E>0 
\cr
  \frac{1}{-8 \pi E} \exp\big(-2e^{S_0} \int_0^{-E} dx\, y(\sqrt{x})\big)& E<0.
  \end{cases}
\ee
as claimed in (\ref{densityoneloopintro}).

Although it is more complicated, one can follow a similar procedure to calculate a nonperturbative correction to the two-resolvent correlator, and extract a generalization of the sine kernel formula for the pair correlation function $\langle \rho(E)\rho(E')\rangle$. In the appendix \ref{appendixsinekernel}, we find that by writing each resolvent as a derivative of a ratio of determinants and calculating to one-loop order, we find for $E, E'>0$
\be\label{pairdensityoneloop}
\langle \rho(E)\rho(E')\rangle \supset \frac{\cos\big(2 \pi  \int_E^{E'} dE'' \rho_0^{\text{total}}(E'')\big)}{2\pi^2 (E-E')^2}, \hspace{20pt} |E-E'|\ll 1.
\ee
Adding the contribution from the leading perturbative cylinder geometry between the two resolvents and approximating the integral in the above formula to leading order in $E-E'$, we find the familiar sine kernel expression for the pair correlation function. While the nonperturbative correction and cylinder are separately divergent as $E'\rightarrow E$, this expression is finite in that limit (although there is also a delta function contact term, see (\ref{sineApp}) for more detail). 

\subsection{Connection to JT gravity}
We would like to understand how to interpret the above computations in the context of JT gravity. We will start by thinking about $\langle \psi(E)\rangle$, using the representation of the determinant as $\det(E-H) = \exp(\Tr\log(E-H))$. Expanding out the exponential, one has to compute expectation values of arbitrary powers of $\Tr\log(E-H)$. These are related by an integral transform to an expectation value of products of partition functions $Z(\beta_1)...Z(\beta_n)$, which we know how to compute in JT gravity. As described before, for $\langle Z(\beta_1)...Z(\beta_n)\rangle$, one sums over surfaces with $n$ boundaries with lengths proportional to $\beta_1,...,\beta_n$.

The boundary condition required to compute $\Tr\log(E-H)$ is a type of fixed energy boundary condition and it can be obtained as an integral transform of the fixed length boundary condition that defines $Z(\beta) = \Tr\,e^{-\beta H}$. Naively, the relation is
\be
\Tr\log(E-H) = -\int_{\epsilon}^\infty \frac{d\beta}{\beta}e^{\beta E}Z(\beta) + \text{const}.\label{trLogE}
\ee
This expression is quite natural from the perspective of quantum gravity. The integral over $\beta$ is an integral over the length of the boundary, and the $e^{\beta E}$ weighting amounts to a boundary cosmological constant. The measure factor of $\frac{1}{\beta}$ is familiar in string theory. It means that we consider the boundary to have no marked point, so configurations that differ by a translation of the boundary time coordinate should not be counted independently.\footnote{The integral to compute $\Tr\frac{1}{E-H}$ has no such factor, and corresponds to a marked boundary.} In the context of the minimal string, the boundary condition implied by (\ref{trLogE}) is simply the FZZT boundary condition in Liouville, with $\mu_B \propto -E$.

The following is a technical remark. Eq.~(\ref{trLogE}) works for contributions with genus $g>0$, but for the disk topology, the integral has an exponential divergence at $\beta = 0$. This reflects the divergence in $\Tr\log(E-H)$ in a double-scaled matrix integral. In the quantity $\psi(E)$, this is cured by subtracting the bare potential $\frac{L}{2} V(E)$. In JT gravity, a prescription that effectively subtracts this is as follows. The function $Z(\beta)$ is a double-valued function, due to a branch point at $\beta = 0$. The two branches differ by a minus sign, so we can rewrite the integrand in (\ref{trLogE}) as one-half of the difference of the two branches. The integral can then be interpreted as the discontinuity across a branch cut. A way to define the integral in general is to deform the contour slightly to pass around the branch point at the origin, avoiding the singularity. This prescription is equivalent to the naive one for $g>0$ and it also gives the correct answer (subtracting the potential) for $g = 0$. In order to converge at infinity, the integral has to be done along a ray that is oriented in an appropriate half-plane, depending on the phase of $E$.

\begin{figure}[t]
\begin{center}
\includegraphics[width=.3\textwidth]{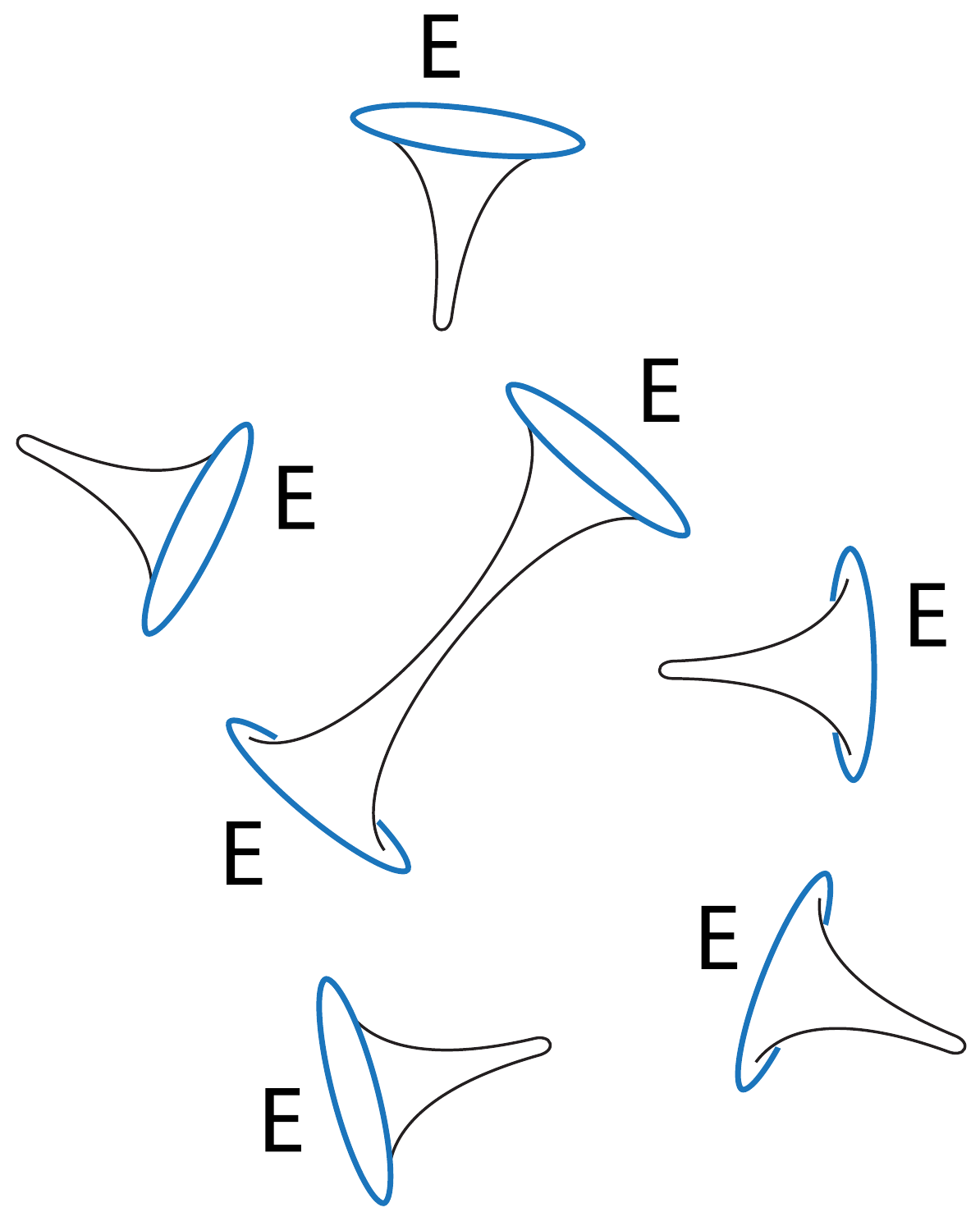}
\end{center}
\caption{{\small An example of a geometry that contributes to $\langle \psi(E)\rangle$.}}
\label{fig:branegeometries}
\end{figure}

In any case, an insertion of $\Tr\log(E-H)$ amounts to a boundary in JT gravity of a particular kind. To compute $\langle \psi(E)\rangle$ we sum over configurations with an arbitrary number of such boundaries. These boundaries can be connected to each other or disconnected. See figure \ref{fig:branegeometries} for an example configuration.
 So $\psi(E)$ is fundamentally a ``many universe'' quantity, whose computation involves disconnected spacetime geometries.\footnote{In the introduction we used the term ``baby universe'' to refer to a disconnected spatial universe.  Here we are referring to disconnected spacetimes, which would be the histories of such ``baby universes.''}

This is analogous to the mathematics of D-branes as described in \cite{Polchinski:1994fq}. The $\Tr\log(E-H)$ boundary condition is interpreted as the worldsheet boundary associated to a D-brane. $\langle\psi(E)\rangle$ is then the D-brane partition function, involving a sum over many disconnected worldsheets ending at the same type of boundary condition. In this analogy, the D-brane is labeled by the argument $E$, and the multivaluedness reflects the branched structure of the semiclassical space of D-branes \cite{Maldacena:2004sn}. In the Liouville minimal string case, the relevant D-branes are called FZZT branes \cite{Fateev:2000ik}, and they are defined by a boundary condition similar to (\ref{trLogE}). So the mathematics is the same but the interpretation is different: we view the two-dimensional geometry as spacetime, rather than a string worldsheet.\footnote{It would be interesting to explore the D-brane interpretation of $\langle \psi(E)\rangle$ in JT gravity further, including understanding the ``target space'' in which this D-brane lives.  The description outlined in Section \ref{sec:minimalString} of JT gravity as a limit of the minimal string might provide a way of addressing this issue.}

The computation of $\langle \tilde\psi\rangle$ is similar to that of $\langle\psi\rangle$: it is also a many-universe quantity. One obvious difference is a minus sign associated to each boundary, which comes from writing $\det(E-H)^{-1} = \exp(-\Tr\log(E-H))$. A less obvious difference is that the sum over branches has to be treated differently than for $\langle \psi\rangle$. We do not have a derivation of this branch prescription from JT gravity, but we can follow the rules outlined in the matrix integral discussion above.

The calculation of the ``dipole'' partition function $\langle \psi(E)\tilde{\psi}_\pm (E')\rangle$ is also similar. For a given choice of branches, we sum over all geometries with any number of brane or ghost brane boundaries. In the disk-and-cylinder approximation, the disconnected spacetimes exponentiate to form the combination $\Psi(z,z')$. For coincident $E' = E$, the ghost brane perfectly cancels the brane, thanks to (\ref{cancelsPerfectly}).

\begin{figure}[t]
\begin{center}
\includegraphics[width = .3\textwidth]{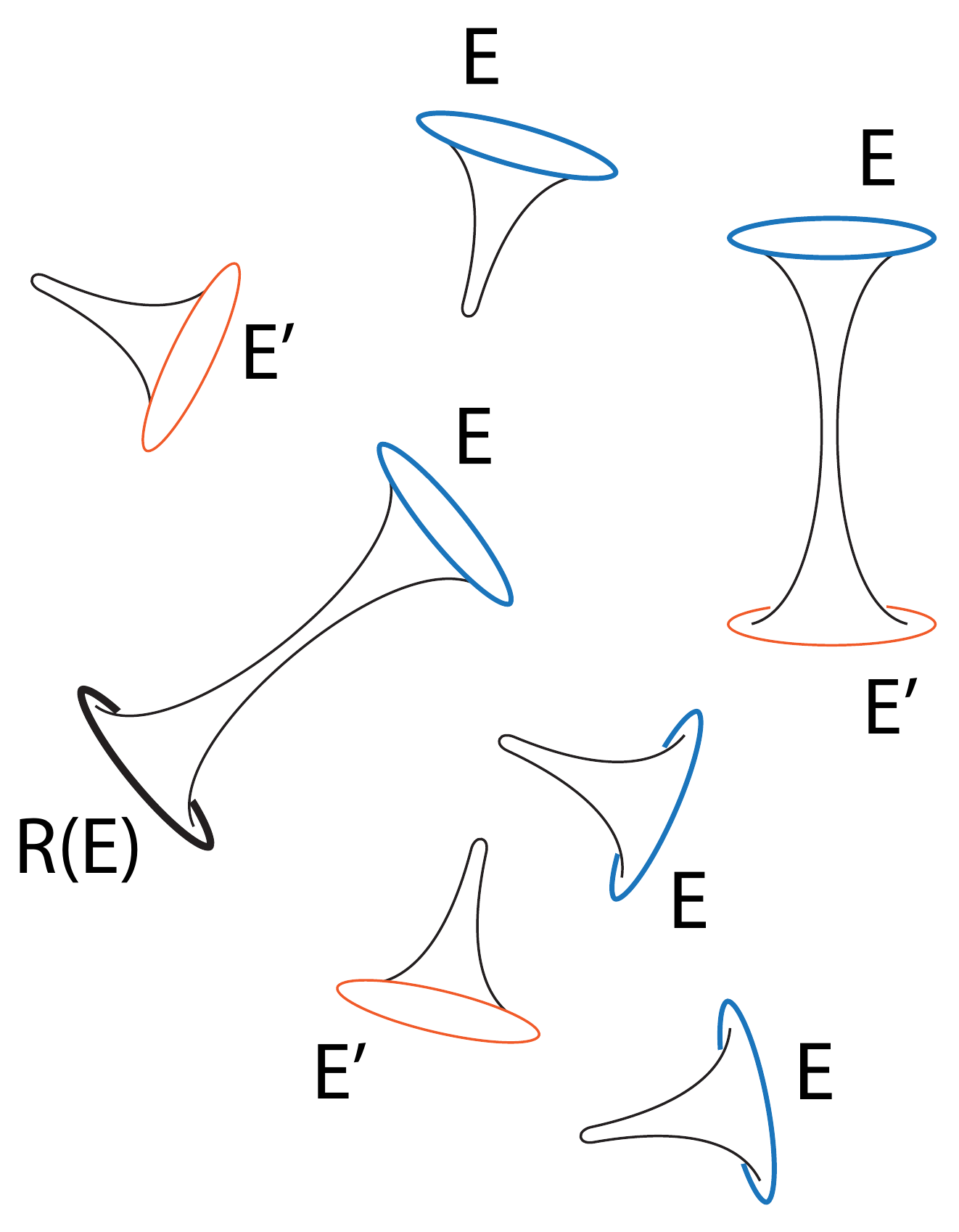}
\end{center}
\caption{{\small An example of a geometry that leads to the nonperturbative correction to the resolvent. The black (dark) boundary corresponds to the insertion of a resolvent operator, $\Tr \frac{1}{E-H}$. The blue (medium) boundaries correspond to the insertion of a brane operator, $\Tr \log (E-H)$.  The orange (light) boundaries correspond to the insertion of a ghost brane operator, $-\Tr \log (E'-H)$.}}
\label{fig:resolventbrane}
\end{figure}

The JT interpretation of the resolvent is more interesting. In (\ref{eq:resolventoneloop}), we got the resolvent by differentiating the dipole partition function, and then setting $E' = E$. For the present discussion, it is convenient to think about this procedure using the formula (\ref{eq:resolventdettrivial}):
\be
R(E) =  \lim_{E'\rightarrow E}R(E)\psi(E)\tilde{\psi}(E').\label{eq:resolventdettrivial2}
\ee
To calculate this, we sum over geometries with a single resolvent boundary in addition to the brane and ghost brane boundaries, see figure \ref{fig:resolventbrane}.\footnote{A key point is that the resolvent boundary must be on the same branch as the brane boundary; we do not sum over separate branch choices for the resolvent. This can easily be seen to follow from the requirement that we match with the version of the calculation where we take the $E$ derivative of the dipole partition function.} In the sum over geometries, there is a class in which the resolvent does not connect to any of the brane boundaries. These sum up to the factorized expectation value $\langle R^\pm(E) \rangle \langle \psi(E)\tilde{\psi}_\pm(E')\rangle$. Taking the limit $E'\rightarrow E$, the brane and ghost brane cancel and we recover the perturbative series for the resolvent.

The leading correction to factorization comes from a geometry where the resolvent is connected to one of the branes by a cylinder diagram. By a calculation similar to the one in (\ref{cylLine1}), we find the cylinder between the resolvent and the brane is
\be
\langle R(E)\Tr\log(E'-H)\rangle_{\text{conn.}}= \frac{1}{2z(z+z')}.
\ee
The cylinder to the ghost brane is minus the same thing. Combining these two, and with the right sum over branches, we have at leading order
\be
\langle R(E)\psi(E_1)\tilde{\psi}(E_2)\rangle - \langle R(E)\rangle\langle\psi(E_1)\tilde{\psi}(E_2)\rangle = \sum_{\text{branches}}\frac{z_2-z_1}{2z(z+z_1)(z+z_2)}\Psi(z_1,z_2).\label{explicitCyl}
\ee
Let's first consider this function for the case where $E_1 = E_2 \neq E$. Then the explicit cylinder factor vanishes if $z_1,z_2$ are on the same sheet, and from (\ref{braneGhostBranePsi}) the function $\Psi$ vanishes if $z_1,z_2$ are on opposite sheets, so we find zero. However, when we obtain the resolvent as a limit of the derivative of the brane-ghost brane dipole, we are effectively setting $E = E_1$ before taking the limit $E_2 = E_1$. In this case the answer still vanishes if $z_1,z_2$ are on the same sheet, but if they are on opposite sheets, a factor in the denominator of (\ref{explicitCyl}) cancels the zero in $\Psi(z,-z)$ and we get a nonzero answer, proportional to $\exp[2\,\text{Disk}(z)]$. In this term, the sum over an infinite number of disconnected spacetimes remains!

This surprising result would be natural if we were studying a different problem, a string theory where the worldsheet is described by JT gravity --  the ``JT string.''   Then the infinite number of spacetimes would just be the infinite number of worldsheets connected to a D-brane, as in the minimal string theory calculations that motivated the above analysis.   
But this result is highly unnatural from the point of view of gravity, where the resolvent is a ``single-spacetime'' object, at least at the perturbative  level of the sum over geometries.  Could it be that there is a different gravitational theory that does not include the nonperturbative effects discussed above?

The gravitational sum over geometries is divergent.   If we make the plausible assumption that it is a resurgent asymptotic series then, as we will see in  section \ref{sec:largeorder}, the perturbation series itself contains information about nonperturbative effects. That information is consistent with the results of the  D-brane approach.  
So, perhaps the right question is whether there is another method of calculating these effects that is natural from the single-spacetime gravitational point of view -- although it might be less efficient than the D-brane method that is natural in the JT string.   SYK model considerations, discussed in section \ref{sec:openquestions}, hint at the existence of such a method.    Then the question would become, what does the existence of these two  different approaches mean for quantum gravity?

\subsection{Nonperturbative instability and ZZ brane}\label{instab}
Up to this point we have made the simplifying  assumption that
$V_{\text{eff}}(E) >0$ in the forbidden region, meaning that the most likely position for an eigenvalue is in the allowed region where $\rho_0(E)$ is supported.    But in some cases $V_{\text{eff}}(E)$ can become negative, indicating that the density $\rho_0(E)$ is not the dominant contribution to the matrix integral.   Another eigenvalue density, perhaps with multiple regions of support, could dominate.

\begin{figure}[t]
 \begin{center}
\includegraphics[width=.75\textwidth]{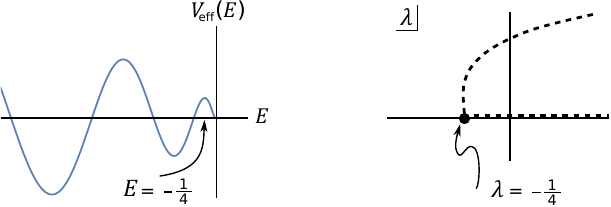}
\caption{{\small At left we show a plot of the effective potential for the JT spectral curve in the forbidden region. At right we sketch an example of an integration contour for the eigenvalues, corresponding to a steepest-ascent path for $V_{\text{eff}}$ that leaves the real axis at $\lambda = -\frac{1}{4}$.}}\label{fig:forbiddenRegion}
\end{center}
\end{figure}

In fact, for  JT gravity the situation is more extreme.  The JT spectral curve, $y = \frac{\sin(2\pi z)}{4\pi}$ gives the  effective potential 
\be
V_{\text{eff}}(E) = \frac{e^{S_0}}{4\pi^3}\left[\sin(2\pi\sqrt{-E}) - 2\pi\sqrt{-E}\cos(2\pi\sqrt{-E})\right], \hspace{20pt} E<0.\label{effectivePot}
\ee
We give a plot of this function in figure \ref{fig:forbiddenRegion}.
As we enter the forbidden region, the potential initially rises, reaching a local maximum value at $E = -\frac{1}{4}$, before decreasing and becoming negative.  It then oscillates repeatedly, with regions that are more and more negative.  
 The consequence is that this model is nonperturbatively ``unstable.''\footnote{A similar situation applies in the $(2,p)$ minimal string: there the potential oscillates a finite number of times, passing through zero $\frac{p-1}{2}$ times before eventually going to either positive infinity (a metastable situation)  or negative infinity (an unstable situation).} 
 
 To give the model a nonperturbative definition, we must choose a contour other than the real axis for the integral over eigenvalues. We choose a contour so that the integral converges and so that the perturbative series which involves expanding around the $\rho_0(E)$ saddle point is actually asymptotic to the true value of the integral.   For example the contour can extend along the real axis from positive infinity down to $E = -\frac{1}{4}$, and then follow this steepest descent contour either upwards or downwards into the complex $E$ plane and out to infinity, see figure \ref{fig:forbiddenRegion}.
Any linear combination of these contours (suitably normalized) would also produce the same asymptotic series and be a viable nonperturbative completion of the theory, so such a completion is not unique.   

A special role is played by the saddle point at $E = -\frac{1}{4}$.  In the matrix integral literature, such saddle points are known as one-eigenvalue instantons. In the minimal string, they are related to boundary conditions in Liouville known as ZZ branes. We will study this in matrix integral language in  the next section. First, we will give a somewhat tentative JT gravity interpretation.

An important point is that in the minimal string, ZZ branes are interpreted in a fundamentally different way than FZZT branes. As we have seen, FZZT branes correspond to observables that one can insert into the matrix integral partition function. By contrast, ZZ branes are non-optional: they are like instantons that occur as part of the matrix integral, and will contribute a small amount to any computation.

Let's first study the effect of a ZZ brane on the matrix integral partition function, $\mathcal{Z}$. As discussed in section \ref{secondCompact}, $\mathcal{Z}$ has an asymptotic series given by the sum over closed surfaces. The ZZ brane correction amounts to allowing surfaces to end in a particular boundary associated to the ZZ brane. This boundary condition can be described in the minimal string as a difference of two FZZT boundary conditions \cite{Martinec:2003ka}:
\be\label{ZZFZZT}
|\text{ZZ}\rangle = |\text{FZZT}(z_{\text{ZZ}})\rangle - |\text{FZZT}(-z_{\text{ZZ}})\rangle
\ee
Here we are using the notation that $|\text{FZZT}(z)\rangle$ is the boundary state associated to an insertion of $\Tr\log(E-H) - \frac{L V(E)}{2}$, with the understanding that $z$ encodes both the energy via $z^2 = -E$ and also the choice of branch. This can be rewritten as 
\be\label{intBranches}
|\text{ZZ}\rangle = \int_{-z_{\text{ZZ}}}^{z_{\text{ZZ}}}dE(z')|R(z')\rangle
\ee
where $|R(z)\rangle$ is the boundary state associated to an insertion of $R(E) - \frac{L V'(E)}{2}$. The $z$ coordinate of the ZZ brane is determined by the condition that it be a stationary point of the effective potential, so that $y(z) = 0$. At such points, the two branches of the spectral curve meet, and the integral in (\ref{intBranches}) can be understood as a closed contour on the spectral curve, see \cite{Seiberg:2003nm}. Concretely, for the ZZ brane associated to the local maximum of $V_{\text{eff}}$ at $E = -\frac{1}{4}$, we have $z_{\text{ZZ}} = \frac{1}{2}$.

In the matrix integral partition function, we propose to include surfaces that end on the ZZ brane by taking
\be
\mathcal{Z} \rightarrow \mathcal{Z}\cdot\left[1 + (\text{const.})\cdot e^{2\text{Disk}(z_{\text{ZZ}})}\right].\label{ZZpartfn}
\ee
Let's explain this formula. First of all, the term involving the $1$ is just the original series for the matrix partition function. The correction represents the possibility that a ZZ brane is ``present.''\footnote{A familiar subtlety in this type of expression is that the series without the ZZ brane is asymptotic, with a nonperturbative ambiguity that is of the order of the correction. Depending on the Borel contour, we could view the correction term as just arising from the resummation of the perturbative series. However, as with other instantons, the ZZ brane gives a semiclassical interpretation to this correction.} In this term, any number of surfaces can connect to the ZZ brane: the exponential factor is the disk amplitude associated to the boundary state (\ref{ZZFZZT}), after we remember that Disk$(-z)= -$Disk$(z)$. Note that Disk$(z_{\text{ZZ}})<0$.  Formally, the constant should include the one-loop contribution that comes from the exponentiated cylinder diagram connecting the ZZ brane to itself. Using (\ref{intBranches}), this cylinder diagram is
\be
\int_{-z_{\text{ZZ}}}^{z_{\text{ZZ}}} dE(z_1)dE(z_2)R_{0,2}(E(z_1),E(z_2)) = \int_{-z_{\text{ZZ}}}^{z_{\text{ZZ}}} \frac{dz_1 dz_2}{(z_1+z_2)^2}
\ee
but the integral appears to be infinite or ambiguous. It may be possible to resolve this directly, but we will leave the constant arbitrary, and see that it can be matched to a matrix integral computation below. We will find that the constant is imaginary, and that it depends on the contour choice described above.

Let's now discuss the contribution of the ZZ brane to the resolvent. To discuss this correction, we do not need to represent the resolvent as a brane-ghost brane dipole. We can simply treat it in the naive way as a single boundary. At leading order, the contribution of the ZZ brane reflects the possibility of this resolvent boundary connecting to the ZZ brane boundary via a cylinder diagram. This leads to 
\be
\langle R(z)\rangle \rightarrow \langle R(z)\rangle + \frac{z_{\text{ZZ}}}{z}\frac{1}{-z^2 + z_{\text{ZZ}}^2} \cdot(\text{const.})\cdot e^{2\text{Disk}(z_{\text{ZZ}})}.\label{ZZansFirst}
\ee
In this expression, the constant and exponential factors are the same as in (\ref{ZZpartfn}). The prefactor is simply the cylinder diagram between a resolvent boundary and the ZZ state (\ref{intBranches}):
\be
\int_{-z_{\text{ZZ}}}^{z_{\text{ZZ}}}dE(z')R_{0,2}(E(z),E(z')) = \int_{-z_{\text{ZZ}}}^{z_{\text{ZZ}}}\frac{-2z'dz'}{4z z'(z+z')^2} = \frac{z_{\text{ZZ}}}{z}\frac{1}{-z^2+z_{\text{ZZ}}^2}.\label{ZZans}
\ee
We see that indeed the expression has a pole when the energy argument $E$ of the resolvent agrees with the energy associated to the ZZ brane, $-z_{\text{ZZ}}^2 = -\frac{1}{4}$. This is consistent with the interpretation of the ZZ brane as an eigenvalue sitting at the local maximum of the potential.

Finally, we emphasize that the ZZ brane effects are numerically very small, because $2\,\text{Disk}(z_{\text{ZZ}}) = -\frac{e^{S_0}}{4\pi^2}$, so the contribution is double-exponentially suppressed. However, these effects are significant because they encode the large orders behavior of the asymptotic series. We will discuss this next.

\subsection{Large order behavior}\label{sec:largeorder}
In this section we will use the density of eigenvalues in the forbidden region to predict the large genus asymptotics of the Weil-Petersson volumes, following a similar approach for other problems described by topological recursion in \cite{Marino:2006hs, Marino:2007te,Marino:2012zq}. In doing this, we are assuming that the matrix integral dual to JT gravity is resurgent, so that it can be analyzed by Borel transform techniques. We will start by reviewing these.

In general, given a formal series
\be\label{defA}
A(u) \simeq \sum_{n=0}^\infty a_k u^k
\ee
where the coefficients grow asymptotically as $a_k\sim k!$, one can define a related convergent series called the Borel transform of $A$:
\be
\widehat{A}(t) = \sum_{k = 1}^\infty \frac{a_k}{\Gamma(k)}t^{k-1}.\label{hat}
\ee
This sum will have a finite radius of convergence under our growth assumption for $a_k$. One can then define a third function, sometimes called the Borel sum of $A$, or inverse Borel transform of $\widehat{A}$, by
\be\label{ibt}
\widetilde{A}(u) = a_0 + \int_{\mathcal{C}}e^{-t/u}\widehat{A}(t)dt.
\ee
Here, the contour $\mathcal{C}$ starts at the origin and ends at infinity and is chosen so that the integral is convergent. By substituting in (\ref{hat}) and doing the $t$ integral, one can show that $\widetilde{A}$ has an asymptotic series that matches $A$. Therefore $\widetilde{A}$ is a candidate nonperturbative completion of $A$, although in general there will be multiple choices depending on the contour $\mathcal{C}$.

The closest singularity to the origin of the Borel transform $\widehat{A}(t)$ encodes the large orders asymptotics of the original series $a_k$. To see this, one can write a dispersion relation for $a_k$ by starting with a contour integral around the origin
\be\label{vanishes}
a_k = \frac{\Gamma(k)}{2\pi i}\oint_0 \frac{dt}{t^{k}}\widehat{A}(t)
\ee
and then deforming the contour outwards in the complex $t$ plane. For large $k$, the closest singularity to the origin will dominate the answer.

\subsubsection{Large Genus Asymptotics of $V_{g,0}$}
We would like to apply (\ref{vanishes}) to predict the large genus behavior of Weil-Petersson volumes. We start by considering the perturbative series for the matrix partition function $\mathcal{Z}$ for JT gravity, which involves a sum over closed surfaces of arbitrary genus. As discussed in section \ref{secondCompact}, the partition functions of these closed surfaces evaluate to the Weil-Petersson volumes with no boundary. So the matrix integral free energy has a perturbative series
\be
\mathcal{F}(e^{-S_0}) = \log\mathcal{Z} \simeq \sum_{g = 0}^\infty e^{(2-2g)S_0}V_{g,0}.\label{pertClosed}
\ee
The genus zero and one cases have to be defined specially, but we are interested in the large genus behavior, so this will not be important. The series (\ref{pertClosed}) is asymptotic due to the $(2g)!$ growth of the volumes $V_{g,0}$. In order to apply the Borel transform formalism, we should think in terms of a series indexed by $k = 2g-2$ instead of $g$, so that the growth is $k!$. Explicitly, one can write
\be
\mathcal{F}(e^{-S_0}) \simeq \sum_{k = -2}^\infty f_k e^{-k S_0}, \hspace{20pt} f_k = \begin{cases}0 & k \text{ odd}\\
V_{\frac{k+2}{2},0} & k \text{ even}.\end{cases}
\ee

Now, we switch perspective to the matrix integral. We start with a conventional matrix integral with finite $L$. Then $\mathcal{Z}$ is an integral over $L$ eigenvalues. We can can organize this into integrals over the allowed region $E>0$ and the forbidden region $E<0$:
\begin{align}\label{faintegral}
\mathcal{Z}&= \int_A d\lambda_1\int_A d\lambda_2 \dots \int_A d\lambda_L \; \mu(\{\lambda_i\})+ L \int_F d\lambda_1 \int_A d\lambda_2\dots \int_A d\lambda_L\; \mu(\{\lambda_i\})+\dots
\cr
& =\mathcal{Z}^{(0)}+\mathcal{Z}^{(1)}+\dots
\end{align}
Here $\mu(\{\lambda_i\})$ denotes the measure, including both the Vandermonde and the potential. We will refer to $\mathcal{Z}^{(1)}$ as the ``one-instanton'' contribution to $\mathcal{Z}$. The nonperturbative expression for the free energy will be 
\be
\mathcal{F} = \log\mathcal{Z} = \log\mathcal{Z}^{(0)} + \frac{\mathcal{Z}^{(1)}}{\mathcal{Z}^{(0)}} + ...
\ee
where $\frac{\mathcal{Z}^{(1)}}{\mathcal{Z}^{(0)}}$ is the one-instanton contribution to the free energy. From (\ref{faintegral}), we can recognize this quantity simply as
\be\label{torecognize}
\frac{\mathcal{Z}^{(1)}}{\mathcal{Z}^{(0)}} = \int_F d\lambda\,\langle\rho(\lambda)\rangle
\ee
after integrating over the other eigenvalues $\lambda_2,...,\lambda_L$. This generalizes in a trivial way to a double-scaled theory.

The main idea is to recognize the integral over $\lambda$ in (\ref{torecognize}) as a contribution to an inverse Borel transform integral for the free energy $\mathcal{F}$
\be
\mathcal{F}(e^{-S_0}) = \int_{\mathcal{C}} dt \,\exp(-e^{S_0}t)\,\widehat{\mathcal{F}}(t)\label{torecognize2}
\ee
with some mapping between $t$ and $\lambda$. This relationship can be fixed by interpreting the exponential term in the density of eigenvalues, given by (\ref{densityforbidden}), as $\exp(-e^{S_0}t)$. In other words $t = \frac{V_{\text{eff}}(\lambda)}{e^{S_0}}$. To determine the large genus asymptotics via (\ref{vanishes}), we need to know $\widehat{\mathcal{F}}(t)$ near the closest singularity to the origin. This singularity is associated to the saddle point in the effective potential at $\lambda = -\frac{1}{4}$. Expanding (\ref{effectivePot}) near this point, and matching (\ref{torecognize}) to (\ref{torecognize2}), we find
\be\label{branchpoint}
t \approx \frac{1}{4\pi^2} - \frac{1}{2}(\lambda+\tfrac{1}{4})^2, \hspace{20pt} \widehat{\mathcal{F}}(t) \approx \frac{1}{2\pi\sqrt{2(\frac{1}{4\pi^2}-t)}}.
\ee
So, in particular, $\widehat{\mathcal{F}}(t)$ has a branch point singularity at $t = \frac{1}{4\pi^2}$.

We now apply (\ref{vanishes}). Since the perturbative series for $\mathcal{F}$ is an even function of $e^{-S_0}$, the same will be true for $\hat{\mathcal{F}}$, and the integral will vanish for odd $k$. For even $k$, we will get the contribution from the branch point described in (\ref{branchpoint}), together with an equal contribution from the reflection under $t\rightarrow -t$. For large $g$, this leads to
\be
V_{g,0} \approx 2\cdot\frac{\Gamma(2g{-}2)}{2\pi i}\int_{\frac{1}{4\pi^2}}^\infty\frac{dt}{t^{2g-2}} \text{disc}\left[\widehat{\mathcal{F}}(t)\right]\approx \frac{(4\pi^2)^{2g-\frac{5}{2}}}{2^{1/2}\pi^{3/2}}\Gamma(2g-\tfrac{5}{2}).\label{predictV0}
\ee
In the final step, we substituted in (\ref{branchpoint}) and did the integral. This expression is expected to be accurate at order one, which means that it will have multiplicative corrections of the form $(1 + \frac{a_1}{g} + \frac{a_2}{g^2}+...)$. And, indeed, at this level of precision (\ref{predictV0}) matches a conjecture due to Zograf \cite{zograf2008large}.\footnote{Parts of this conjecture have been established rigorously \cite{1999math......2051M,mirzakhani2013growth,mirzakhani2011towards}.}

\subsubsection{Large genus asymptotics of $V_{g,1}(b)$}
We now move on to discuss the large genus asymptotics of $V_{g,1}(b)$, for which we will find a new formula. We start by defining a generating function of the volumes for $g\ge 1$:
\be
V(b,e^{-S_0}) \simeq \sum_{g = 1}^\infty e^{(1-2g)S_0}V_{g,1}(b) = \sum_{n = 0}^\infty e^{-nS_0}v_n(b).
\ee
Here $v_n = 0$ for even $n$ and $v_n(b) = V_{\frac{n+1}{2},1}(b)$ for odd $n$. In terms of this quantity, the expectation value of the resolvent is given by (dropping the genus zero piece)
\be
\langle R(-z^2)\rangle =-\frac{1}{2z}\int_0^\infty \hspace{-5pt}b\,db \,V(b,e^{S_0}) e^{-z b}.\label{toinvert}
\ee
In appendix \ref{appB}, we compute the nonperturbative contribution to the resolvent that results from integrating one eigenvalue in the classically forbidden region. The result is
\be
\langle R(E)\rangle^{(1)} = \int_F d\lambda \frac{\langle\rho(\lambda)\rangle}{E-\lambda} \sqrt{\frac{\lambda}{E}}.\label{resolventToBeDerived}
\ee
One can use this to get a formula for the nonperturbative contribution to $V(b,e^{-S_0})$ by inverting (\ref{toinvert}) by inverse Laplace transform. This leads to
\begin{align}
V^{(1)}(b,e^{-S_0})&=\frac{2}{b}\int_F d\lambda\, \langle\rho(\lambda)\rangle \sinh(\sqrt{-\lambda}b)\\
&=\int_F\frac{dz}{2\pi}\frac{\sinh(bz)}{bz}e^{-e^{S_0}t(z)}.\label{interpretint}
\end{align}
In the second line, we substituted in the formula for the density of eigenvalues in the forbidden region (\ref{densityforbidden}) and changed variables to $\lambda = -z^2$. The function $t(z)$ is
\be
t(z) = 2\int_0^{z^2}\hspace{-5pt}dx\, \frac{\sin(2\pi\sqrt{x})}{4\pi} = \frac{\sin(2\pi z) - 2\pi z\cos(2\pi z)}{4\pi^3}.\label{tandzeq}
\ee

\begin{figure}[t]
\begin{center}
\includegraphics[width=.55\textwidth]{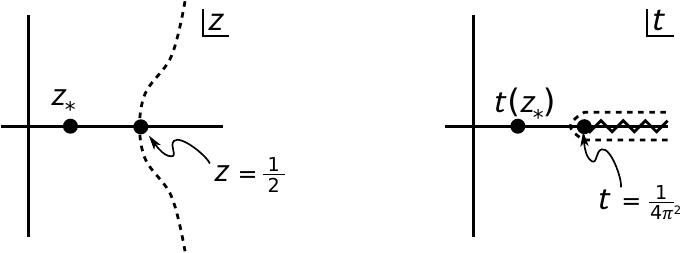}
\caption{{\small The contour that computes the integral of the discontinuity of $\widehat{V}(b,t)$ is shown on the $t$ plane and the $z$ plane. The contour can be deformed to pass through the saddle point $z_*$ which is somewhere between $z = 0$ (for $b\gg g$) and $z = \frac{1}{2}$ (for $b\ll g$).}}\label{fig:zandt}
\end{center}
\end{figure}

Similarly to the discussion of the free energy, we would like to interpret the integral in (\ref{interpretint}) as an inverse Borel transform $V(b,e^{-S_0}) = \int dt e^{-e^{S_0}t}\widehat{V}(b,t)$ and thus determine $\widehat{V}$. In making this correspondence, the mapping between $t$ and $z$ will be (\ref{tandzeq}). Then, to predict the large-orders behavior we need to compute (for odd $n$)
\begin{align}
V_{\frac{n+1}{2},1}(b)&\approx 2\cdot \frac{\Gamma(n)}{2\pi i}\int_{\frac{1}{4\pi^2}}^\infty\frac{dt}{t^{n}}\text{disc}\left[\widehat{V}(b,t)\right].\label{rewriteint}
\end{align}
This integral is over a contour in the $t$ plane that surrounds the branch cut along the positive $t$ axis. In the free energy case, we computed $\widehat{\mathcal{F}}(t)$ approximately by inverting (\ref{tandzeq}) near the singularity at $t = \frac{1}{4\pi^2}$, which maps to $z = \frac{1}{2}$. However, in the present case we will need to know the behavior for more general values of $t$. Because it is difficult to invert (\ref{tandzeq}) in general, it turns out to be more convenient to work in terms of the $z$ variable. In terms of this coordinate, one finds that the integral (\ref{rewriteint}) becomes
\be
V_{g,1}(b)\approx \frac{\Gamma(2g-1)}{\pi i}\int_{\mathcal{C}}\frac{dz}{2\pi}\frac{\sinh(bz)}{bz}\frac{1}{t(z)^{2g-1}}.\label{volPrediction}
\ee
We expect this formula to be correct at large $g$ for arbitrary $b$. More precisely, we expect multiplicative corrections $(1 + a_1/g + a_2/g^2+...)$ where the coefficients $a_1,a_2,...$ are bounded functions of $b$. The contour $\mathcal{C}$ in the $z$ plane is sketched in figure \ref{fig:zandt}. For large $g$ where the formula should be correct, the answer is dominated by a saddle point with $0<z<\frac{1}{2}$. For small values of $b/g$, the saddle point is close to $z = \frac{1}{2}$, and for large values of $b/g$, it is close to the origin.

\begin{figure}[t]
\begin{center}
\includegraphics[width=.55\textwidth]{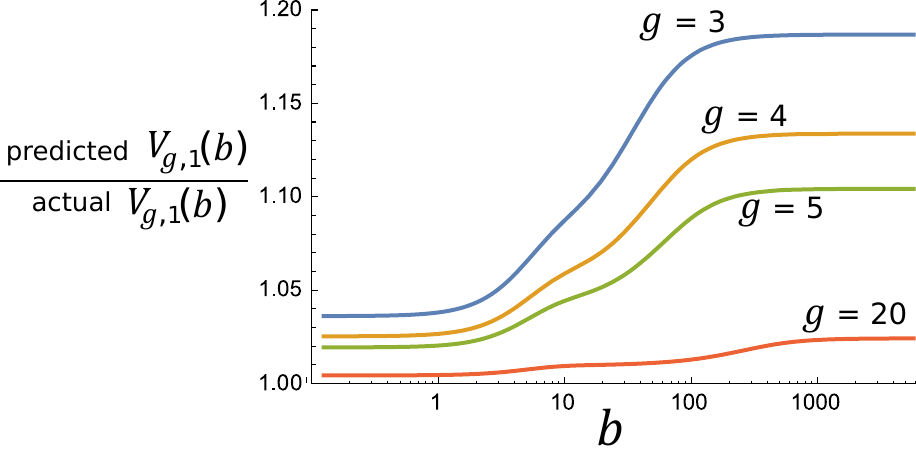}
\caption{{\small We compare the formula (\ref{volPrediction}) to exact results provided to us by Peter Zograf \cite{ZografData}.}}\label{volfig}
\end{center}
\end{figure}

 We will provide a few checks of this formula.
\begin{enumerate}
\item If $b\ll g$, then the saddle point is very close to $z = \frac{1}{2}$. We can approximate $t(z)\approx \frac{1}{4\pi^2} - \frac{1}{2}(z-\frac{1}{2})^2$ and do the integral for large $g$, finding
\be
V_{g,1}(b) \approx \frac{4(4\pi^2)^{2g-\frac{3}{2}}}{(2\pi)^{3/2}}\Gamma(2g-\tfrac{3}{2})\frac{\sinh(\frac{b}{2})}{b},\hspace{20pt} g\gg b, \hspace{10pt} g\gg 1.\label{smallb}
\ee
This formula follows from Conjecture 1 and Conjecture 2 of \cite{zograf2008large}, after translating from intersection numbers to volumes using
\be
V_{g,1}(b) = \sum_{k = 0}^{3g-2}\frac{(2\pi^2)^{3g-2-k}}{k!(3g-2-k)!}\left(\frac{b^2}{2}\right)^k \int_{\overline{\mathcal{M}}_{g,1}} \kappa_1^{3g-2-k}\psi_1^k.
\ee
\item At the other extreme, where $b\gg g\gg 1$, the saddle point is close to the origin. This time we can expand $t(z) \approx \frac{2 z^3}{3}$ and find
\be
V_{g,1}(b) \approx \left(\frac{3}{2}\right)^{2g-1}\frac{\Gamma(2g-1)}{\Gamma(6g-2)}\frac{b^{6g-4}}{2\pi}, \hspace{20pt} b\gg g\gg 1.\label{largeb}
\ee
This agrees at large $g$ with the exact formula for $\int \psi_1^{3g-2} = \langle \tau_{3g-2}\rangle$ in (5.31) of \cite{Itzykson:1992ya}.
\item At the level of the factorial dependence, it agrees with the conjecture in (26) of \cite{Maloney:2015ina}.
\item Peter Zograf provided us with exact results for $V_{g,1}(b)$ up to $g = 20$. In figure \ref{volfig} we plot the ratio of (\ref{volPrediction}) divided by the exact answer for a few different values of $g$. The agreement seems to be reasonable for all $b$, and certainly improving as we increase $g$.
\item Using this same data, we attempted to extrapolate in $1/g$. For each value of $b$, we modeled the ratio of (\ref{volPrediction}) to the exact answer as $a_0 + a_1/g + a_2/g^2+a_3/g^3+a_4/g^4$. We fit the coefficients using Zograf's data from $g = 14,...,20$, and thus extracted a prediction $a_0$ for the ratio at $g = \infty$. We tried two methods in the extrapolation. In method one, we held $b$ fixed as we varied $g$. In method two, we held $B = b/g$ fixed. The two methods can be compared approximately using the maximum value $g = 20$. Method one works better for $b \lesssim 2g$ and method two works better for $b\gtrsim 2g$. We found that the maximum over $b$ of the minimum of the two extrapolated errors  $|a_0-1|$ was around $3\times 10^{-6}$, and the maximum of the maximum was around $0.004$. Repeating the analysis with $g = 9,...,15$, the max of the min error is roughly ten times higher.
\end{enumerate}

These checks give us some confidence in our understanding of nonperturbative effects in the matrix integral dual to JT gravity. As a side note, it would be possible to use a similar method to predict the large $g$ behavior of $V_{g,n}(b_1,...,b_n)$. It is also interesting to consider the possibility that the large genus limit of any expansion described by topological recursion can be obtained in a similar way, by using the one-eigenvalue sector of a formal matrix integral with spectral curve $y$, see \cite{Marino:2006hs, Marino:2007te,Marino:2012zq} for some related work.

\section{Discussion}\label{sec:Discussion}

\subsection{Disordered but \texorpdfstring{$k$}{k}-local theories like SYK}
The SYK model is a type of sparse random matrix ensemble, and there is numerical evidence that certain aspects of the model are well described by random matrix statistics. However, it is not a random matrix ensemble in the sense of section \ref{secMatrix}, and so in general we do not expect the JT gravity results in this paper to match the behavior of SYK.

One reason is the following. Varying the SYK couplings $J_{a_1...a_q}$ will ``rattle around'' the fine-grained structure of the spectrum, as in a true random-matrix ensemble. However, because the number of variable parameters is relatively small, of order $N^q$, one also expects there to be some ``global'' features of the spectrum with relatively large $1/N^q$ fluctuations. These features include the overall normalization of the Hamiltonian, as well as other low-order moments \cite{Cotler:2016fpe,Gharibyan:2018jrp,VerbaarschotTalk}. They lead to relatively large connected correlations in $\langle Z(\beta_1)Z(\beta_2)\rangle$ of order $1/N^q$. Taken at face value, these represent a large departure from random matrix statistics, and one without a clear JT gravity interpretation. However, we note that these modes can be reduced or eliminated by ``unfolding'' the spectrum sample-by-sample \cite{VerbaarschotTalk}, so we will not dwell on this point.

There is another reason. It seems possible that the collective field description of SYK involves analogs of the JT geometries discussed in this paper. For example, the ramp saddle point \cite{Saad:2018bqo} is analogous to the double trumpet considered here. However, even for these contributions, one expects differences from JT gravity. As a crude model, one can consider the effect of free bulk matter fields on the JT gravity computation. The measure for the genus $g$ moduli space will now include the matter determinant as well as the Weil-Petersson measure. At higher genus this leads to a dramatic effect: the portion of moduli space with long thin tubes will be exponentially divergent, due to the negative Casimir energy on the small circle. This is essentially the tachyon divergence familiar from string theory.

In the full SYK theory we expect that this divergence will be preempted by a first order Hawking-Page transition, very similar to the situation discussed by Maldacena and Qi \cite{Maldacena:2018lmt}. There, before a cylinder becomes long and thin, it disconnects into two disks, describing disconnected Euclidean black holes. This mechanism could effectively remove the dangerous part of moduli space, but in doing so it will change the JT gravity answer significantly.

So, in general, we do not expect the JT gravity results in this paper to match those of SYK, or indeed any $k$-local theory. An important exception is for the short-ranged spectral correlators that are subject to random matrix universality (probed e.g.~by the late-time spectral form factor). These are described by the universal ramp saddle point of \cite{Saad:2018bqo}, together with D-brane effects for the plateau. As we have seen, the D-brane contribution  in JT is determined by the disk and cylinder amplitude.   The matter field correction to the disk contribution determines the correction to the leading order density of states.   For short range spectral correlations the cylinder is far from the transition point and so the matter fields should not change its energy dependence.\footnote{ The matter field corrections to the cylinder do  limit the range of agreement of eigenvalue pair correlations with random matrix theory to energy differences less than a ``Thouless energy'' of order $1/\log N$ \cite{Saad:2018bqo}.}   So we expect a close relative of the the D-brane effects described in JT to explain the sine kernel result with a modified SYK density of states. This clearly deserves more work.

\subsection{Non-disordered theories}
So far all the comments in this paper have concerned {\it averaged} systems.    Some of the deepest questions in this subject concern the  fine-grained behavior of the energy eigenvalues of gauge/gravity dual systems that are not averaged --  large $N$ Super Yang-Mills theory (SYM) is the canonical example.  In trying to apply the ideas of this paper to such a system one encounters an immediate problem. Consider for example the double-trumpet:
\be\label{wormholepic2}
\includegraphics[width=.2\textwidth,valign =c]{figures/wormholeExample.pdf}
\ee
This contributes to the connected correlator $\langle Z(\beta_1)Z(\beta_2)\rangle - \langle Z(\beta_1)\rangle\langle Z(\beta_2)\rangle$. But in a non-disordered theory, $Z(\beta)$ is a fixed function rather than an observable in an ensemble, and such a correlator would not make sense \cite{Maldacena:2004rf,ArkaniHamed:2007js}.

We are not sure what to say about this, but there is a well understood example of quantum chaos that may be instructive.  This is the semiclassical quantum dynamics of classically chaotic systems with a few degrees of freedom.\footnote{For a review see \cite{2009NJPh...11j3025M}, especially the ArXiv version.} These systems are studied using the Feynman path integral representation (or actually its refinement known as the Gutzwiller trace formula) for $\Tr e^{-iHT/\hbar}$ as a sum over periodic paths $a$:
\begin{equation}
\Tr e^{-iHT/\hbar} \sim \sum_{a } e^{\frac{i}{\hbar} S_a} ~.
\end{equation}
 Here $S_a$ is the classical action.  To study the spectral form factor one uses two copies,
 \begin{equation}\label{sfforbit}
 \Tr e^{iHT/\hbar} ~\Tr e^{-iHT/\hbar} \sim \sum_{a, b} e^{\frac{i}{\hbar} S_a} e^{- \frac{i}{\hbar} S_b} ~.
 \end{equation}
  Note that this factorizes into the product of the result for each copy. Semiclassically the long time behavior is determined by long periodic classical orbits.  In a chaotic system  the details of such orbits are extremely complicated and the resulting behavior in \eqref{sfforbit}  is erratic.\footnote{The results will resemble Figure 10 in \cite{Cotler:2016fpe}.}

But this signal simplifies after averaging, for instance over a time window.   For long orbits the phases in the orbit sum in \eqref{sfforbit} are very large and after any appreciable averaging most terms  in the double sum over $a, b$ cancel.   The only surviving terms for times in the ramp region are those with $a=b$ up to a time translation.  For these terms the action precisely cancels.  This pairing is Berry's  ``diagonal approximation'' \cite{1985RSPSA.400..229B}, and gives the ramp. This pattern is a close analog of the double cone \cite{Saad:2018bqo}, which is a continuation of the geometry (\ref{wormholepic2}). The two systems, decoupled without averaging, become correlated into a connected ``geometry'' after averaging.   Factorization is destroyed by suppressing the $a \neq b$ terms.\footnote{The  potential for averaging to make  such connected geometries, and the link to Coleman's ideas, were pointed out in \cite{Maldacena:2004rf}.}

\begin{figure}[t]
\begin{center}
\includegraphics[width = .375\textwidth]{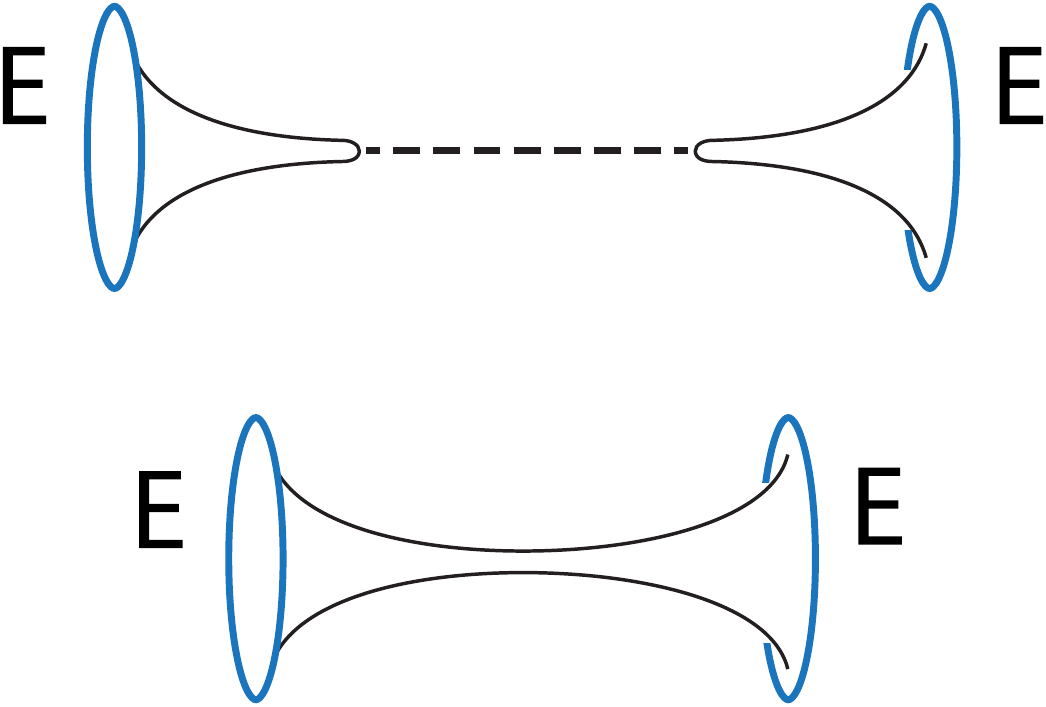}
\end{center}
\caption{{\small  A cartoon of the way that averaging over erratic phases might result in the wormholes present in Figure \ref{fig:branegeometries}.}}
\label{fig:averagingconnection}
\end{figure}

The role of essentially random phases in creating these connections  calls to mind the work of Coleman \cite{Coleman:1988cy,Coleman:1988tj} on Euclidean wormholes.\footnote{This work was motivated by \cite{Giddings:1987cg,Lavrelashvili:1987jg,Hawking:1987mz}. See also,  \cite{Banks:1988je,Klebanov:1988ka,Giddings:1988wv,Klebanov:1988eh,Polchinski:1994zs}.} Here Euclidean wormholes result from integration over  random couplings.   Operator insertions with Gaussian random coefficients on the same or different ``universes'' are paired by Wick contraction and fattened by an OPE  into wormholes.   In our setup we might imagine ``forming'' the wormholes in Figure \ref{fig:branegeometries} by averaging over erratic phases as in Figure \ref{fig:averagingconnection}.    Figure \ref{fig:branegeometries} would be an open ``many-universe'' analog of the  closed ``many-universe''  configurations that are summed up in this approach \cite{Coleman:1988tj,Klebanov:1988ka,Giddings:1988wv,Klebanov:1988eh}.     This sum also produces  results doubly exponential in $1/G_N$, for analogous reasons.

Perhaps we have a choice about the bulk description of spectral statistics for non-disordered systems: an unaveraged description with simple topology but with exceedingly detailed and complicated information about microstates, like the individual orbits $a$, and their intricate and rapidly fluctuating phases; or an averaged description made up of simpler  geometrical objects.  But the price for this simplicity  seems to be third quantization, wormholes, and branes.

\subsection{Other open questions}\label{sec:openquestions}
\begin{enumerate}
\item What is the result for the cylinder amplitude that connects a ZZ brane to itself?
\item Is it possible to recover the description of the matrix integral in terms of eigenvalues, starting with the bulk JT gravity description? A possible hint is the McGreevy-Verlinde proposal \cite{McGreevy:2003kb} for the $c = 1$ string. As refined in \cite{Klebanov:2003km} this proposal asserts that the $L$ unstable  ZZ branes, combined with the tachyonic stretched strings between them, form the matrix of the matrix quantum mechanics.
\item Is there another way to get the nonperturbative effects without summing over disconnected geometries? SYK may give a useful perspective. The determinant approach from section \ref{sec:nonperturbative} should work, but it involves infinitely many SYK replicas as an intermediate step. As an alternative, it should be possible in principle to get the full spectral form factor (including the plateau) by doing the exact $G,\Sigma$ path integral with two replicas. Can the plateau be obtained this way in practice?
\item We have emphasized the doubly exponential character of the brane effects, of order $\exp(ce^{1/G_N})$ or $\exp(c e^{N_{\text{SYK}}})$. This suggests  two layers of asymptotic expansion. Consider the SYK spectral form factor expressed as a two replica $G, \Sigma$ integral. Away from the Schwarzian limit where certain quantities become one loop exact, we expect to have an asymptotic expansion in $1/N$.  Presumably its Borel transform has singularities corresponding to various $e^{-N}$ effects.   Perhaps the higher genus surfaces present in the JT limit persist in some fashion and determine some of these effects. If this is the case then these $e^{-N}$ effects themselves will form another asymptotic series, whose Borel plane singularities will contain the doubly exponential effects. Can we find a simpler problem with a similar two-layer structure?\footnote{This is unlike the more familiar pattern, where the sum over $e^{-N}$ instantons is convergent, see e.g.~\cite{ Marino:2008ya}. There is a field theory example where such phenomena occur \cite{2017PhRvD..96i6022A},   due to IR ``renormalon'' effects.}

\item Another natural set of observables in JT gravity are correlators of probe bulk matter fields.  Some preliminary remarks about higher genus corrections to them were made in \cite{Saad:2018bqo} and a systematic treatment has been initiated in \cite{Blommaert:2019hjr}.  One interesting aspect of these observables is that they probe matrix elements as well as energies and so give information about the random matrix character of eigenstates, i.e.~ETH. Higher order OTOCs may be especially interesting to study in this regard.

\item Are there theories of quantum gravity in higher dimensions that are dual to ensemble averages of boundary theories, rather than specific boundary theories?

\item In cases where the boundary theory is fixed, what prevents us from including the contribution of connected geometries? See \cite{Maldacena:2004rf,ArkaniHamed:2007js}. These papers considered actual solutions to the bulk equations of motion. By contast, the gravity configurations considered in this paper are not solutions: there is no on-shell configuration for the dilaton (more physically, the action is not stationary with respect to $b$, as in \cite{Hawking:1987mz}). If we are willing to consider such geometries in higher dimensions, the problem could be even worse than envisioned in \cite{Maldacena:2004rf,ArkaniHamed:2007js}.


\item The disk and cylinder geometries are the ingredients required to determine the ramp and plateau.  These seem to have a natural extension to higher dimensions: the Euclidean black hole solution and, when continued to Minkowski signature, the double cone portion of the  eternal AdS-Schwarzschild geometry \cite{Saad:2018bqo}.  Do these provide an explanation of spectral statistics in suitably averaged higher dimensional gauge/gravity dual systems?

\item The branched structure of the spectral curve was important for getting the nonperturbative effects. What is the meaning of this spectral curve in gravity?

\end{enumerate}

We will close the paper with the following comment. In an ensemble average of quantum systems, unitarity of time evolution is not as dramatic as for an individual system (e.g.~there are no recurrences). But it still means something. For example, it implies that the spectral form factor cannot decay to zero at late time. So Maldacena's version of the black hole information problem \cite{Maldacena:2001kr} can be formulated for disordered theories. 

Concretely, an analog of Maldacena's puzzle for disordered theories is to explain in bulk language the late time behavior of the spectral form factor. For the case where the bulk dual is JT gravity, the nonperturbative effects discussed in this paper provide a type of answer.\footnote{The ideas discussed in \cite{Polchinski:1994zs} may be related.} Of course, we did not derive the effects from bulk reasoning: in section \ref{sec:nonperturbative} we were essentially reinterpreting the rules of a matrix integral in JT language. A clearer bulk understanding of these rules, or perhaps a different version of them, would be valuable.

\section*{Acknowledgements}
We are grateful to Alex Altland, Clay Cordova, Alexei Kitaev, Juan Maldacena, Alex Maloney, Greg Moore, Nathan Seiberg,  Jacobus Verbaarschot,  Edward Witten, and Zhenbin Yang for discussions.  We are particularly grateful to Edward Witten for pointing out the connection between JT gravity and Weil-Petersson volumes, and their connection to topological recursion, and to Peter Zograf for providing us with exact results for $V_{g,1}(b)$.

PS is partially supported by a Fellowship from the Ashok and Gita Vaish Charitable Trust.  PS and SS are supported in part by NSF grant PHY-1720397.  DS is supported by Simons Foundation grant 385600. Parts  of this work were performed at the Aspen Center for Physics, which is supported by National Science Foundation grant NSF-PHY-1607611, and at KITP, which is supported by National Science Foundation grant  NSF PHY-1748958.

\appendix

\section{Some details on nonperturbative effects}
\subsection{The Airy ghost brane and dipole}
\subsubsection{Integral representation for ghost brane}\label{appendixantibrane}

We begin with the Gaussian model\footnote{For related work in this context see \cite{Okuyama:2018gfr}.} described by $V(H) = \frac{2}{a^2}H^2$.  Instead of introducing Grassmann flavor fields we introduce bosonic fields $\phi_i$ to implement $1/\det(E-H)$.\footnote{The opposite statistics of these fields motivates the name ``ghost brane.''  Another point of view derives from the field theory on the spectral curve discussed in \cite{Moore:1990mg,Aganagic:2003qj,Dijkgraaf:2018vnm}.} Performing the Gaussian integral over $H$, we find an integrand that depends only on $\sum_i \overline{\phi}_i \phi_i$. We then introduce an auxiliary variable $r$ that renders the integral over the $\phi_i$ and $\overline{\phi}_i$ Gaussian, allowing us to integrate them out. The result is the integral
\be
\langle \det(E\pm i \epsilon-H)^{-1} \rangle = \sqrt{\frac{2L}{\pi a^2}} \int_{-\infty}^\infty dr (E\pm i\epsilon+r)^{-L} e^{-\frac{2L}{a^2} r^2}.\label{eq:antibranegaussian}
\ee
\begin{figure}[t]
\begin{center}
\includegraphics[width=.3\textwidth]{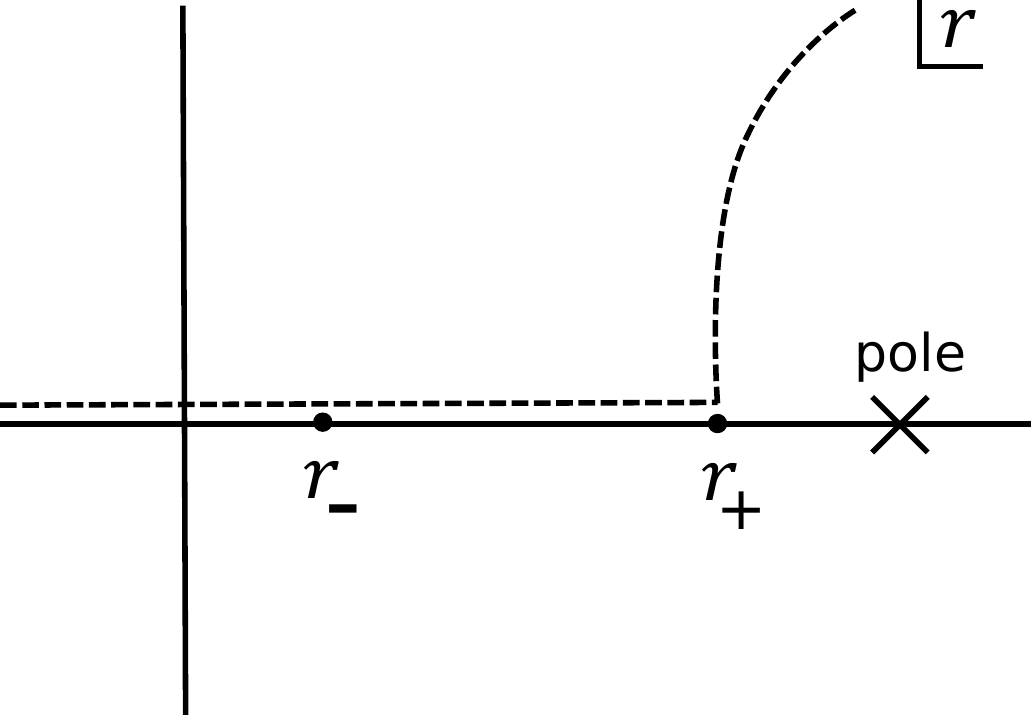}\hspace{20pt}\includegraphics[width=.3\textwidth]{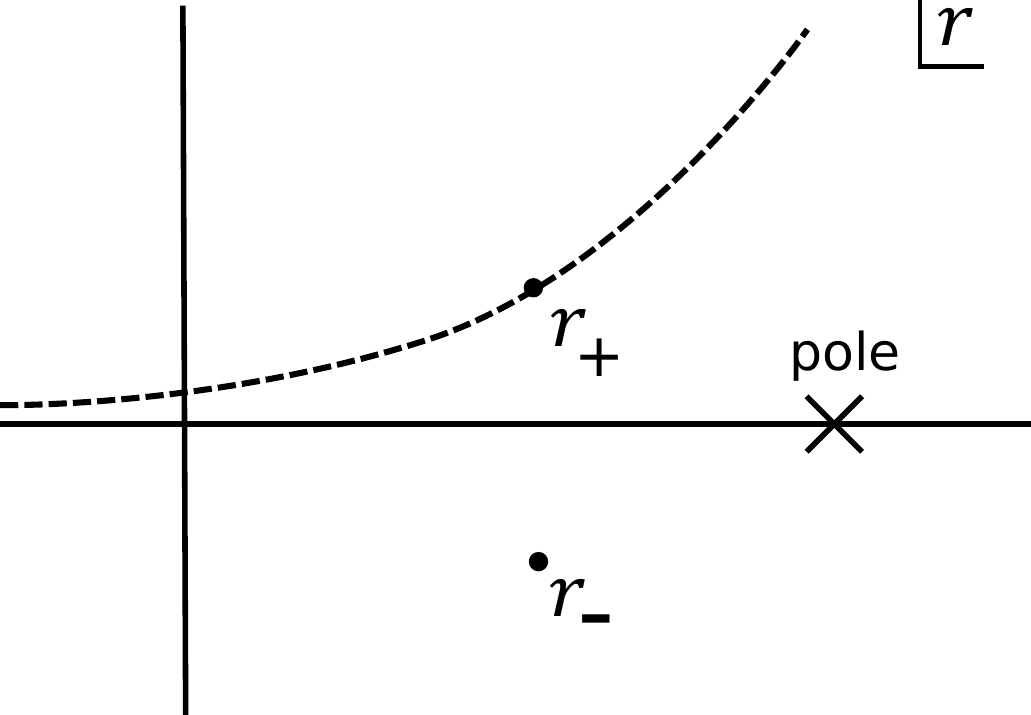}
\end{center}
\caption{Here we display the saddle points and integration contour for the integral (\ref{eq:antibranegaussian}). The left diagram corresponds to $E<-a$ and the right diagram corresponds to $E>-a$. The contours are drawn for the case with $+i\epsilon$, so that the $r$ contour runs above the pole.}
\label{fig:AntibraneContour}
\end{figure}
Before we discuss the saddle point structure of this integral, we will verify that this integral representation reproduces the expected discontinuity across the real axis. First, we shift $r\rightarrow r\mp i\epsilon$ so that the the pole is on the real axis, and the contour for $r$ runs either just above or just below the real axis. The discontinuity of the integral is the integral over the difference of the two contours. The two contours can be joined at infinity to form a loop, which we can shrink to surround the pole at $r=-E$. We recognize the resulting integral as the familiar contour integral representation for the $L-1$'th Hermite Polynomial. Up to a constant of proportionality, we find
\be
\text{disc}\big[\langle \tilde{\psi}(E)\rangle_{L}\big] \propto i \langle \psi(E) \rangle_{L-1}.
\ee
Here we are defining the $L$ and $L-1$ ensembles as described around (\ref{eq:measures}).

Now we move on to a saddle point analysis of (\ref{eq:antibranegaussian}). There are two saddle points, $r=r_\pm$. We will focus on the case $E<0$ because we are ultimately interested in double-scaling near $E = -a$. Then the behavior of the integral depends on how $E$ compares to $-a$: 
\begin{enumerate}
\item For $E<-a$, the two saddle points are on the real axis. The steepest descent contour for $r_-$ runs along the real axis, while the steepest descent contour for $r_+$ passes vertically through $r_+$. We may deform the defining contour into two portions of these steepest descent contours. The first segment comes in from $r=-\infty$, passes through $r_-$, then reaches $r_+$. Here we go either upwards or downwards along the steepest descent contour for $r_+$. Which way we go is determined by the $i\epsilon$ prescription in (\ref{eq:antibranegaussian}), which requires the $r$ contour to go either below or above the pole, see figure \ref{fig:AntibraneContour}. Schematically,
\be
\langle \det(E\pm i \epsilon-H)^{-1} \rangle\sim  e^{- L I(r_-)} \pm \frac{i}{2}  e^{- L I(r_-)},\hspace{20pt} E<-a.
\ee
Here $e^{-L I(r_\pm)}$ is the integrand evaluated on the saddle point $r_\pm$. In the above expression we have ignored the contributions of one-loop factors except to emphasize the $\frac{i}{2}$.

\item For $E>-a$, the defining contour can be deformed into just one of the steepest descent contours. Again, the choice is determined by the $i\epsilon$ prescription. The leading order saddle point approximation gives
\be
\langle \det(E \pm i \epsilon-H)^{-1} \rangle\sim e^{ \pm i\frac{\pi}{4} - L I(r_{\pm})},\hspace{20pt} -a<E<a.
\ee 
\end{enumerate}

So far, we have discussed the standard Gaussian matrix integral. To get the double-scaled theory, we multiply (\ref{eq:antibranegaussian}) by $e^{\frac{L}{2}V(E)}$, shift $E\rightarrow E-a$, shift $r\rightarrow \frac{a}{2}+\sqrt{\frac{a}{2}} e^{-\frac{S_0}{3}} r$, and take the limit $a\rightarrow \infty$, holding fixed $e^{S_0} = (2/a)^{3/2} L$. This leads to (\ref{eq:antibraneairyintegral}), up to a constant of proportionality. This limit scales away the pole in (\ref{eq:antibranegaussian}), but it leaves a remnant: the integrand grows along the positive real axis. Because of this, the integration contour has to go to infinity in the right half-plane at a finite angle either above or below the real axis, ending at $e^{\pm i\frac{\pi}{3}}\cdot\infty$. In taking the double-scaled limit, this choice matches onto the choice of whether the contour goes above or below the pole on the real axis, and is therefore determined by the $\pm i\epsilon$ prescription.

\subsubsection{Integral representation for dipole}\label{appendixdipole}

We start again with the potential $V(H)=\frac{2}{a^2} H^2$. By writing the ratio of determinants as an integral over Grassmann vectors $\chi_i, \;\overline{\chi}_i$ and bosonic vectors $\phi_i,\; \overline{\phi}_i$, we find an integral that depends only on the matrix
\be
\hat{S} = \begin{pmatrix} \sum_i \overline{\phi}_i \phi_i &  \sum_i \overline{\phi}_i \chi_i \cr \sum_i  \overline{\chi}_i \phi_i &  \sum_i \overline{\chi}_i \chi_i \end{pmatrix}.
\ee
Defining the supertrace $\text{Str} (\hat{M})\equiv \hat{M}_{11}- \hat{M}_{22}$, we have

\be
\bigg\langle \frac{\det(E-H)}{\det(E'\pm i\epsilon-H)}\bigg\rangle = \int \prod_{i=1}^L \bigg[\frac{d\chi_id\overline{\chi}_i d\overline{\phi}_i d\phi_i}{2\pi}\bigg] \exp\bigg[ \frac{a^2}{8L} \text{Str} (\hat{S}^2) - \overline{\phi}_i \phi_i (E'\pm i\epsilon) +\overline{\chi}_i \chi_i E\bigg].
\ee
We can introduce an auxiliary matrix $S$ which has bosonic diagonal elements and Grassmann off-diagonal elements to render the integral over the $\chi_i, \;\overline{\chi}_i$ and $\phi_i,\; \overline{\phi}_i$ Gaussian,
\be
S= \begin{pmatrix} S_{11} & S_{12} \cr S_{21} & i\; S_{22}\end{pmatrix}.
\ee
Defining the matrix $\mathcal{E}= \text{diag}(E'\pm i\epsilon, E)$ and performing the Gaussian integral we are led to an integral just over the matrix $S$,
\be
\bigg\langle \frac{\det(E-H)}{\det(E'\pm i\epsilon-H)}\bigg\rangle = \int \frac{dS_{11}dS_{22}dS_{12}dS_{21}}{2\pi} \exp\bigg[ -L \;\text{Str} \log(\mathcal{E}+S) - \frac{2 L}{a^2} \text{Str}(S^2)\bigg].
\ee

To take the double scaled limit we multiply this by $\exp[ L \frac{V(E')-V(E)}{2}]$, shift $E, E'$ by $a$, shift the matrix $S\rightarrow \frac{a}{2}+ \sqrt{\frac{a}{2}} S$, then take the limit $a\rightarrow \infty$ keeping $e^{S_0}$ fixed. We find
\be
\big\langle\psi(E)\tilde{\psi}(E'\pm i\epsilon)\big\rangle = \int_{\mathcal{C_\pm}}  \frac{dS_{11}dS_{22}dS_{12}dS_{21}}{2\pi}\;\exp\Big[e^{S_0}  \text{Str}\big(\tfrac{1}{3}S^3+\mathcal{E} S\big) \Big].
\ee
The contours for $S_{11}, S_{22}$ are the same as the contours for $r,s$ in the respective ghost brane and brane computations. To obtain the formula (\ref{eq:detoverdetintegral}), we simply integrate out the Grassmann variables $S_{12}$ and $S_{21}$. This leads to
\be
\big\langle\psi(E)\tilde{\psi}(E'\pm i\epsilon)\big\rangle = -e^{S_0}\int_{\mathcal{C_\pm}}  \frac{dS_{11}dS_{22}}{2\pi}(S_{11}+i S_{22})\exp\Big[e^{S_0} \big(\tfrac{1}{3}S_{11}^3+\tfrac{i}{3}S_{22}^3 + E'S_{11} - i E S_{22}\big) \Big].
\ee 
After relabeling $S_{11}= e^{-\frac{S_0}{3}} r$, $S_{22}=  e^{-\frac{S_0}{3}} s$, we find (\ref{eq:detoverdetintegral}). Note that for this Airy case, apart from the factor of $(S_{11}+iS_{22})$, this is the product of the brane and ghost brane answers. This correction term can be obtained by differentiating with respect to $E,E'$.

This calculation is a simple application of a central  technique in the modern treatment of quantum chaos \cite{Efetov:1997fw,2005hep.ph....9286S,2009NJPh...11j3025M,haake2010quantum,Verbaarschot:2004gj,2006JPhA...3913191G}.   In this approach correlators of arbitrary numbers of determinants and inverse determinants are represented by  higher dimensional versions of $\hat{S}$.   The integral over  $\hat{S}$ is then treated as a nonlinear sigma model whose target is embedded in a  superspace.   Its ``pion''  perturbative excitations produce the long range energy correlations.   A nonperturbative saddle point, the analog of a baryon, produces the D-brane effects.  In this context the saddle point is called the ``Altshuler-Andreev'' instanton \cite{andreev1995spectral}.  

This approach offers a framework to demonstrate  random matrix universality.   ``All'' one needs to do is show that the system being studied produces the necessary pattern of symmetry breaking in the effective sigma model.   It would be interesting to explore this approach in the gravity context.

\subsection{Density pair correlation function}\label{appendixsinekernel}
We want to calculate the density pair correlation function $\langle \rho(E_1) \rho(E_2)\rangle$ for $|E_1-E_2|\ll 1$, $E_1, E_2>0$. To do so, we compute the resolvent correlator using the formula
\be
\big\langle R^\pm(E_1) R^{\pm'}(E_2) \big\rangle = \partial_{E_1}\partial_{E_2} \big\langle  \psi(E_1)\psi(E_2) \tilde{\psi}_\pm(E_3)\tilde{\psi}_{\pm'}(E_4)\big\rangle \big|_{E_1=E_3, \;E_2=E_4}\label{appb1}
\ee
and then extract the density correlator using
\be
(-2\pi i)^2\langle \rho \rho\rangle = \langle R^+R^+\rangle + \langle R^- R^-\rangle - \langle R^+R^-\rangle - \langle R^{-}R^+\rangle.\label{rhotoRes}
\ee
This computation involves many different cases, and we will simplify our task somewhat by restricting to the pieces that are singular as $E_2\rightarrow E_1$. The one-loop function, including the exponentiated disks and cylinders, is
\be
\Psi(z_1,z_2;z_3,z_4) = \frac{(z_1+z_3)(z_1+z_4)(z_2+z_3)(z_2+z_4)}{4\cdot\sqrt{z_1 z_2 z_3 z_4}(z_1+z_2)(z_3+z_4)}e^{\text{Disk}(z_1)+\text{Disk}(z_2)-\text{Disk}(z_3)-\text{Disk}(z_4)}.
\ee
The branch prescription is determined by the prescription for the dipole (\ref{eq:braneantibraneoneloop2}). In the allowed region, the rule is to sum over both branches for the brane operators, but to take only the branch $z = e^{\mp i\frac{\pi}{2}}\sqrt{E}$ for the ghost brane operator associated to $R^\pm$. In all cases, we will evaluate the contribution to the resolvent, which is
\be
\partial_{E_1}\partial_{E_2}\Psi = \frac{-1}{2z_1}\partial_{z_1}\frac{-1}{2z_2}\partial_{z_2}\Psi(z_1,z_2;z_3,z_4)\label{Psi4}
\ee
where $z_1^2 = z_3^2 = -E_1$ and $z_2^2 = z_4^2 = -E_2$ with a choice of branch in the square root that has to be specified. The singular terms arise from the action of derivatives on the prefactor (cylinder) expressions, so we do not need to differentiate the disk functions. We will now go through the cases.
\begin{enumerate}
\item First, we consider the computation of $\langle R^+R^+\rangle$, so $z_3 = e^{-i\frac{\pi}{2}}\sqrt{E_1}$ and $z_4 = e^{-i\frac{\pi}{2}}\sqrt{E_2}$. The computation for $\langle R^-R^-\rangle$ can be treated similarly. There are four branch choices. In the two cases $(z_1,z_2) = \pm (z_3,z_4)$, one finds no singular contribution. In the case $(z_1,z_2) = (-z_3,z_4)$, one finds 
\be
\langle R^+R^+\rangle \supset \frac{i}{4\sqrt{E_1E_2}(E_1-E_2)}e^{2\text{Disk}(i\sqrt{E_1})}.
\ee
This is singular at $E_1 = E_2$, but the singularity cancels when we include the term from $(z_1,z_2) = (z_3,-z_4)$, which is obtained from the above by switching $1\leftrightarrow 2$. So, in summary, thre are no singular terms in $\langle R^+R^+\rangle $ or $\langle R^-R^-\rangle$.
\item Next, we consider $\langle R^+R^-\rangle$, so $z_3 = e^{-i\frac{\pi}{2}}\sqrt{E_1}$ and $z_4 = e^{i\frac{\pi}{2}}\sqrt{E_2}$. The case $\langle R^-R^+\rangle$ follows from complex conjugation, or by interchanging $1\leftrightarrow 2$.
\begin{enumerate}
\item If both $z_1 = z_3$ and $z_2 = z_4$, we find the singular term
\be
\langle R^+R^-\rangle \supset \frac{-1}{4\sqrt{E_1E_2}(\sqrt{E_1}-\sqrt{E_2})^2}.\label{pertResTerm}
\ee
This is simply the singularity that is present for the resolvent in the perturbative theory, coming from the double trumpet (\ref{pertSing}).
\item If $z_1 = -z_3$ and $z_2 = z_4$, we find
\be
\langle R^+R^-\rangle \supset \frac{-i}{4\sqrt{E_1E_2}(E_1-E_2)}e^{2\text{Disk}(i\sqrt{E_1})}.\label{getscombined}
\ee
\item If $z_1 = z_3$ and $z_2 = -z_4$, we find
\be
\langle R^+R^-\rangle \supset \frac{-i}{4\sqrt{E_1E_2}(E_1-E_2)}e^{-2\text{Disk}(i\sqrt{E_2})}.\label{getscombined2}
\ee
\item If $z_1 = -z_3$ and $z_2 = -z_4$, we find
\be
\langle R^+R^-\rangle \supset \frac{1}{16E_1E_2}\frac{(\sqrt{E_1}+\sqrt{E_2})^2}{(\sqrt{E_1}-\sqrt{E_2})^2}e^{2\text{Disk}(i\sqrt{E_1})-2\text{Disk}(i\sqrt{E_2})}\label{nonPertResTerm}
\ee
\end{enumerate}
So, we see that for $\langle R^+R^-\rangle$ and $\langle R^-R^+\rangle$, all terms are singular. The term (\ref{pertResTerm}) has a double pole at $E_1 = E_2$, but no single pole. The terms (\ref{getscombined}) and (\ref{getscombined2}) have single poles at $E_1 = E_2$, but the coefficient is rapidly oscillating as a function of energy. For our purposes, the most interesting term is (\ref{nonPertResTerm}). This contains a double pole that cancels the double pole in (\ref{pertResTerm}), and it also contains a single pole with a coefficient proportional to the derivative of the disk amplitude.
\end{enumerate}

One can now add the contributions for $\langle R^+R^-\rangle$ and $\langle R^-R^+\rangle$ together to compute $\langle \rho\rho\rangle$, as in (\ref{rhotoRes}). In doing so, it is important to treat the poles carefully, giving a small $\pm$ imaginary part to the energy argument of $R^\pm$. With this prescription, the poles discussed above lead to $\delta(E_1-E_2)$ contact terms in $\langle \rho \rho\rangle$. One finds that the term (\ref{nonPertResTerm}) and its analog in $\langle R^-R^+\rangle$ lead to the contribution
\begin{align}\label{eq:paircorrelationnonperturbative}
 \langle \rho(E_1)\rho(E_2) \rangle \supset& \frac{1}{32 \pi^2 E_1 E_2} \frac{(\sqrt{E_1}+\sqrt{E_2})^2}{(\sqrt{E_1}-\sqrt{E_2})^2} \cos\big[2 \pi e^{S_0}\hspace{-5pt} \int_{E_2}^{E_1}\hspace{-5pt} dE \rho_0 (E) \big]\\&\hspace{20pt}+ \rho^{\text{total}}_0(E_1)\delta(E_1{-}E_2).\notag 
 \end{align}
In writing this expression, we used that
\be
\partial_E\text{Disk}(i\sqrt{E}) = i\pi e^{S_0}\rho_0(E) = i\pi\rho_0^{\text{total}}(E).
\ee
The pole from (\ref{getscombined}) and (\ref{getscombined2}) corrects the coefficient of the delta function in this expression slightly, by adding the leading oscillating correction to the density of eigenvalues (\ref{densityoneloop}). The perturbative term (\ref{pertResTerm}) adds a term that cancels the double pole in the first line of (\ref{eq:paircorrelationnonperturbative}).

Away from $E_1,E_2=0$, we may simplify our expressions to find the universal sine kernel formula for the connected pair correlation function in regions of large density. Expanding the integral $\int_{E_2}^{E_1} dE \rho_0(E) \approx \rho_0(E_2) (E_1-E_2)$, keeping only the singular parts as $E_2\rightarrow E_1$ in (\ref{eq:paircorrelationnonperturbative}), and subtracting a double pole to account for the perturbative term (\ref{pertResTerm}), we find that for non-coincident points,
\begin{align}
\langle \rho(E_1)\rho(E_2)\rangle_{\text{conn}.} &\supset  - \frac{1}{2\pi^2 (E_1-E_2)^2}+ \frac{1}{2 \pi^2 (E_1-E_2)^2} \cos\big[ 2\pi e^{S_0} \rho_0(E_2) (E_1-E_2)\big]
\cr
&= - \frac{\sin^2\big[\pi e^{S_0} \rho_0(E_2) (E_1-E_2)]}{\pi^2 (E_1-E_2)^2}, \hspace{20pt} E_1,E_2>0, \;|E_1{-}E_2|\ll 1.
\end{align}
The full sine-kernel approximation to the pair correlation also includes the leading factorized piece $\rho_0^{\text{total}}(E_1)\rho_0^{\text{total}}(E_2)$. In our present setup, this arises from a nonsingular but large contribution to $\langle R^+R^+\rangle$ and $\langle R^-R^-\rangle$ in which the derivatives in (\ref{Psi4}) act on the disk amplitudes. Including this piece, and the delta function at coincident points, one finds
\be
\langle \rho(E_1)\rho(E_2)\rangle \approx  \rho_0^{\text{total}}(E_1)\rho_0^{\text{total}}(E_2) +\rho_0^{\text{total}} \delta(E_1{-}E_2)- \frac{\sin^2\big[\pi\rho_0^{\text{total}}(E_2) (E_1{-}E_2)]}{\pi^2 (E_1{-}E_2)^2}.\label{sineApp}
\ee

\subsection{Forbidden contribution to the resolvent}\label{appB}
In this appendix, we will derive the formula (\ref{resolventToBeDerived}) from the perspective of a matrix integral. As a setup for the calculation, suppose that we divide the region of integration of the matrix eigenvalues into two pieces, an ``allowed'' region and a ``forbidden'' region, such that the probability of an eigenvalue being in the forbidden region is very small (in practice, this means that we should put the division a finite distance into the $E <0$ region). Suppose that, as a first step, we compute everything by doing the integrals over eigenvalues in the allowed region, including the integrals in the matrix partition function that normalizes expectation values. The question is: what do we have to add to the computation of the resolvent in order to accurately describe the contribution of one eigenvalue being in the forbidden region?

We will get a simple formula for double-scaled matrix integrals, but we will start out with an unscaled matrix integral with $L$ eigenvalues. It will be useful to define two un-normalized measures, one for $L$ eigenvalues and one for $L-1$:
\begin{align}
\mu_L(\lambda) &=  \Delta^2(\lambda_1,...,\lambda_L) e^{-L\sum_{j=1}^L V(\lambda_j)}\\
\tilde{\mu}_{L-1}(\lambda) &=  \Delta^2(\lambda_1,...,\lambda_{L-1}) e^{-(L-1)\sum_{j = 1}^{L-1} \tilde{V}(\lambda_j)}\label{eq:measures}
\end{align}
where $\Delta$ is the Vandermonde determinant, and
\be
(L-1)\tilde{V}(x) = L V(x).
\ee
A useful relation is that the measure for $L$ eigenvalues is a product of the measure for $L-1$ eigenvalues times the square of the determinant operator, viewed as an operator in the $(L-1)$-eigenvalue theory:
\be
\mu_L(\lambda_1...\lambda_L) = \tilde{\mu}_{L-1}(\lambda_1...\lambda_{L-1}) \psi^2(\lambda_L).
\ee
This implies, for example, that we can write a formula for the density of eigenvalues in the $L$ ensemble in terms of the determinant operator in the $L-1$ ensemble:
\be
\langle \rho(E)\rangle_L = \frac{L\mathcal{Z}_{L-1}}{\mathcal{Z}_L}\langle \psi^2(E)\rangle_{L-1}.\label{eigDet}
\ee

We can use this to write a formula for the matrix integral partition function
\be\label{expRes}
\mathcal{Z}_L = \mathcal{Z}_{L}^{(0)} + \mathcal{Z}_{L}^{(1)} + ...
\ee
Here the first term represents the contribution with all eigenvalues integrated in the allowed region:
\be
\mathcal{Z}_{L}^{(0)} = \int_A d\mu_L(\lambda).
\ee
The second term represents the contribution with one eigenvalue in the forbidden region:
\be
\mathcal{Z}_{L}^{(1)}  = L \int_A d\tilde{\mu}_{L-1}\int_F d\lambda_L \psi^2(\lambda_L) = \mathcal{Z}_{L-1}^{(0)}L\int_Fd\lambda_L\langle \psi^2(\lambda_L)\rangle_{L-1}^{(0)}.\label{usedab}
\ee
The subscript on the expectation value in the last term means that we are working in the $L-1$ eigenvalue ensemble defined by $\tilde{\mu}_{L-1}$, and we are in the zero-instanton sector, where all eigenvalues are integrated in the allowed region.

Similarly, we can write a formula for an un-normalized version of the resolvent:
\begin{align}
\int_{A+F}  \frac{d\mu_L(\lambda)}{E-\lambda_1} &= \int_A \frac{d\mu_L(\lambda)}{E-\lambda_1} + \int_F d\lambda_L \psi^2(\lambda_L)\int_Ad\tilde{\mu}_{L-1}(\lambda)\left(\frac{L-1}{E-\lambda_1} + \frac{1}{E-\lambda_L}\right).\label{torewriteas}
\end{align}
In the second term on the RHS, we are including the possibility that one of the $L$ eigenvalues is in the forbidden region. This could be any of the eigenvalues. In $L-1$ of the $L$ possible cases, this special eigenvalue is not the eigenvalue $\lambda_1$ that appears in the denominator in the LHS. In one of the $L$ cases, it is that eigenvalue. The two possibilities are represented in the two terms in parentheses.

We can rewrite the second term in (\ref{torewriteas}) as an expectation value in the $L{-}1$ ensemble:
\begin{align}
\int_{A+F}  \frac{d\mu_L(\lambda)}{E-\lambda_1} &= \int_A \frac{d\mu_L(\lambda)}{E-\lambda_1} + \mathcal{Z}_{L-1}^{(0)}\int_F d\lambda \left\langle\psi^2(\lambda)\left(R(E) + \frac{1}{E-\lambda}\right)\right\rangle_{L-1}^{(0)}.\label{torewriteas2}
\end{align}
The properly normalized resolvent is $L / \mathcal{Z}_L$ times this expression. Using (\ref{expRes}) and (\ref{usedab}), we find that at one-eigenvalue-instanton precision, this is
\begin{align}
\langle R(E)\rangle_L &= \langle R(E)\rangle_{L}^{(0)}\left(1 - \frac{\mathcal{Z}_{L}^{(1)}}{\mathcal{Z}_{L}^{(0)}}\right) + L\frac{\mathcal{Z}_{L-1}^{(0)}}{\mathcal{Z}_{L}^{(0)}}\int_Fd\lambda\left\langle \psi^2(\lambda)\left(R(E) + \frac{1}{E-\lambda}\right)\right\rangle_{L-1}^{(0)}\\
&=\langle R(E)\rangle_{L}^{(0)} + L\frac{\mathcal{Z}_{L-1}^{(0)}}{\mathcal{Z}_{L}^{(0)}}\int_F d\lambda\left[\Big\langle \psi^2(\lambda)R(E)\Big\rangle_{L-1,\text{conn.}}^{(0)} + \left\langle\frac{ \psi^2(\lambda)}{E-\lambda}\right\rangle_{L-1}^{(0)}\right]\label{combinesWithNicely}\\
&\hspace{64pt} + \frac{\mathcal{Z}_{L}^{(1)}}{\mathcal{Z}_{L}^{(0)}}\left(\langle R(E)\rangle_{L-1}^{(0)}-\langle R(E)\rangle_{L}^{(0)}\right).\label{finalLine}
\end{align}
Here we are defining $\langle \psi^2 R\rangle_{\text{conn.}} = \langle \psi^2R\rangle - \langle \psi^2\rangle\langle R\rangle$. 

So far, the discussion has been for a regular (unscaled) matrix integral. But in a double-scaled theory, we get a somewhat simpler expression. One reason is that in a double-scaled theory, the difference of expectation values on the final line (\ref{finalLine}) vanishes. Intuitively, this is because in a double-scaled matrix integral, $L$ is taken to infinity, so changing it by one has no effect on normalized quantities. Another nice feature of the double-scaled case is that we can write a simple leading-order formula for the connected correlator $\langle \psi^2 R\rangle_{\text{conn.}}$. We use
\be
\langle R(E(z))\,\Tr\log(E(z_1)-H)\rangle_{\text{conn.}} = \int_{\infty}^{z_1} dE(z')R_{0,2}(E(z),E(z')) = \frac{1}{2z(z+z_1)}
\ee
to find that
\be
\Big\langle \psi^2(E(z_1))R(E(z))\Big\rangle_{L-1,\text{conn.}}^{(0)} = \frac{1}{z(z+z_1)}\Big\langle \psi^2(E(z_1))\Big\rangle_{L-1}^{(0)}.
\ee
This term combines nicely with the $\langle \psi^2(\lambda)/(E-\lambda)\rangle$ term in (\ref{combinesWithNicely}) and we find
\begin{align}
\langle R(E)\rangle &= \langle R(E)\rangle^{(0)} + L\frac{\mathcal{Z}_{L-1}^{(0)}}{\mathcal{Z}_{L}^{(0)}}\int_F d\lambda \langle \psi^2(\lambda)\rangle_{L-1}\sqrt{\frac{\lambda}{E}}\frac{1}{E-\lambda}.
\end{align}
Finally, using (\ref{eigDet}), we find
\begin{align}
\langle R(E)\rangle=\langle R(E)\rangle^{(0)} + \int_F d\lambda\, \langle \rho(\lambda)\rangle\sqrt{\frac{\lambda}{E}}\frac{1}{E-\lambda}.
\end{align}
This expression answers the question we posed at the beginning of this appendix.

When one applies this to the contour in figure \ref{fig:forbiddenRegion}, one finds a distinct semiclassical contribution from the saddle point at $\lambda = -\frac{1}{4}$. In the leading semiclassical approximation, this gives a pole at $E = -\frac{1}{4}$. This agrees with the ZZ brane result in (\ref{ZZansFirst}), although here we get a prediction for the (imaginary!) numerical coefficient after plugging in (\ref{densityforbidden}).

{\footnotesize
\bibliography{references}
}

\bibliographystyle{utphys}

\end{document}